\documentclass[12pt,a4paper,twoside]{report}

\usepackage{amsmath,amssymb}
\usepackage{amsthm}
\newtheorem{thm}{Theorem}
\usepackage{graphicx}

\usepackage{fancyhdr}
\usepackage{braket}
\usepackage{cite}
\usepackage{hyperref}
\usepackage{lscape}

\hypersetup{
linktocpage=true,colorlinks=true,bookmarks=true,pdfborder={0 0 0},%
allcolors=blue }

\usepackage{amsfonts}
\usepackage{color}
\usepackage{bm}
\usepackage{gensymb}
\usepackage{wasysym}
\definecolor{darkgreen}{rgb}{0.0, 0.2, 0.13}

\usepackage{here}
\usepackage{url}
\usepackage{youngtab}

\newcommand{\calA}{\mathcal{A}}
\newcommand{\calU}{\mathcal{U}}
\newcommand{\calG}{\mathcal{G}}
\newcommand{\calS}{\mathcal{S}}
\newcommand{\trans}{\mathcal{T}}

\usepackage[top=40truemm,bottom=25truemm,left=25truemm,right=25truemm]{geometry}

\begin{document}

\begin{titlepage}
 \title{ {\large Doctoral Dissertation} \\ \vspace{2cm}
 {\Large Searching for an emergent $\mathrm{SU}(4)$ symmetry in real materials} \\ \vspace{0.5cm} \vspace{2cm}
 {\large A Dissertation Submitted for the Degree of Doctor of Philosophy} \\
 {\large December 2019} \\ \vspace{1.5cm}
 {\large Department of Physics, Graduate School of Science,} \\
 {\large The University of Tokyo} \\ 
 {\large Masahiko G. Yamada}
 }
 \author{}
 \date{}
\maketitle
\end{titlepage}



\addtocounter{page}{-1} 

\chapter*{Abstract}
Beauty of mathematics appears everywhere in modern condensed matter physics,
but the importance of the theory of higher-rank Lie groups has been
ignored for a long time.
The enhancement of the spin-space symmetry from 
the usual $\mathrm{SU}(2)$ to $\mathrm{SU}(N)$ with $N>2$ is promising
for finding nontrivial quantum spin liquids,
but the realization of $\mathrm{SU}(N)$ spin systems in real materials is still challenging.
Although there is a proposal in cold atomic systems,
in magnetic materials with a spin-orbital degree of freedom it is difficult
to achieve the $\mathrm{SU}(N)$ symmetry by fine tuning.
Here we propose a new mechanism by which the $\mathrm{SU}(4)$ symmetry emerges
in the strong spin-orbit coupling limit.
In $d^1$ transition metal compounds with edge-sharing anion octahedra,
the spin-orbit coupling gives rise to strongly bond-dependent and
apparently $\mathrm{SU}(4)$-breaking hopping
between the $J_\textrm{eff}=3/2$ quartets.
However, in the honeycomb structure, a gauge transformation
maps the system to an $\mathrm{SU}(4)$-symmetric Hubbard model,
which means that the system has a hidden symmetry in spite of its
large spin-orbit coupling. In the strong repulsion limit
at quarter filling, as expected in $\alpha$-ZrCl$_3,$
the low-energy effective model is the $\mathrm{SU}(4)$ Heisenberg model
on the honeycomb lattice, which cannot have a trivial
gapped ground state and is expected to host a gapless spin-orbital liquid.
In such quantum spin-orbital liquids, both the spin and orbital degrees
of freedom become fractionalized and correlated together
at low temperature due to the strong frustrated interactions between them.
Similarly to spinons in pure quantum spin liquids,
quantum spin-orbital liquids can host not only spinon excitations,
but also fermionic ``orbitalon'' excitations at low temperature, which we have
named here in distinction from orbitons in the symmetry-broken
Jahn-Teller phases.  In fact, the $\mathrm{SU}(4)$ Heisenberg model
on the honeycomb lattice is known to host such gapless exotic
excitations (spinons and orbitalons) by numerical calculations.
By generalizing this model to other three-dimensional lattices,
we also propose crystalline spin-orbital liquids
protected by the combination of
an emergent $\mathrm{SU}(4)$ symmetry and space group symmetries.

\chapter*{List of published papers}

\section*{Papers}
\begin{enumerate}
    \item Masahiko G. Yamada, Tomohiro Soejima, Naoto Tsuji, Daisuke Hirai, Mircea Dinc\u{a}, and Hideo Aoki, ``First-principles design of a half-filled flat band of the kagome lattice in two-dimensional metal-organic frameworks'', Phys. Rev. B \textbf{94}, 081102(R) (2016), as a Rapid Communication. (arXiv:1510.00164)
    \item Masahiko G. Yamada, Hiroyuki Fujita, and Masaki Oshikawa, ``Designing Kitaev Spin Liquids in Metal-Organic Frameworks'', Phys. Rev. Lett. \textbf{119}, 057202 (2017). (arXiv:1605.04471)~\cite{Yamada2017MOF}
    \item Masahiko G. Yamada, Vatsal Dwivedi, and Maria Hermanns, ``Crystalline Kitaev spin liquids'', Phys. Rev. B \textbf{96}, 155107 (2017), as Editors' Suggestion. \\
        (arXiv:1707.00898)~\cite{Yamada2017XSL}
    \item Masahiko G. Yamada, Masaki Oshikawa, and George Jackeli, ``Emergent $\mathrm{SU}(4)$ Symmetry in $\alpha$-ZrCl$_3$ and Crystalline Spin-Orbital Liquids'', Phys. Rev. Lett. \textbf{121}, 097201 (2018). (arXiv:1709.05252)~\cite{Yamada2018}
\end{enumerate}

This PhD thesis is mostly based on Paper 4, and the texts in Paper 4 were partially used in this thesis with permission of American Physical Society.
The appendix also includes a result from Paper 3.

\section*{Preprints}
\begin{enumerate}
    \item Masahiko G. Yamada, and George Jackeli, ``Magnetic and Electronic Properties of Spin-Orbit Coupled Dirac Electrons on a $(001)$ Thin Film of Double Perovskite Sr$_2$FeMoO$_6$'', arXiv:1711.08674.
    \item Masahiko G. Yamada, and Yasuhiro Tada, ``Quantum valence bond ice theory for proton-driven quantum spin-dipole liquids'', arXiv:1903.03567.
\end{enumerate}

\section*{Thesis}
\begin{enumerate}
    \item Masahiko Yamada, ``Designing various quantum spin liquids in metal-organic frameworks'', Master's thesis,
the Department of Physics, the University of Tokyo (2017).
\end{enumerate}

\vspace*{\stretch{1}}
Il s'agit de ce fait, que dans mon approche de la math\'ematique, et plus g\'en\'eralement, dans ma d\'emarche spontan\'ee \`a la d\'ecouverte du monde, la tonalit\'e de base de mon \^etre est \textbf{yin, ``f\'eminin''}\dots
Ce qui est exceptionnel par contre dans mon cas (me semble-t-il), c'est que dans ma d\'emarche de d\'ecouverte et notamment, dans mon travail math\'ematique, j'aie \'et\'e toute ma vie pleinement fid\`ele \`a cette nature originelle, sans aucune vell\'eit\'e d'y apporter des retouches ou rectificatifs, que ce soit en vertu des desiderata d'un Censeur int\'erieur (lequel de toutes fa\c{c}ons n'y a jamais vu que du feu, tellement on serait loin de soup\c{c}onner une sensibilit\'e et une approche cr\'eatrice ``f\'eminine'' dans une affaire ``entre hommes'' comme la math\'ematique!), ou par souci de me conformer aux canons de bon go\^ut en vigueur dans le monde ext\'erieur, et plus particuli\`erement, dans le monde scientifique. Il n'y a aucun doute pour moi que c'est gr\^ace surtout \`a cette fid\'elit\'e \`a ma propre nature, dans ce domaine limit\'e de ma vie tout au moins, que ma cr\'eativit\'e math\'ematique a pu se d\'eployer pleinement et sans entrave, comme un arbre vigoureux, solidement plant\'e en pleine terre, se d\'eploy\'e librement au rythme des nuits et des jours, des vents et des saisons. Il en a \'et\'e ainsi, alors pourtant que mes ``dons'' sont plut\^ot modestes, et que les d\'ebuts ne s'annon\c{c}aient nullement sous les meilleurs auspices.
\begin{flushright}
\mbox{--- Alexander Grothendieck, \textit{R\'ecoltes et Semailles}, 1985--1987.}
\end{flushright}
\vspace{\stretch{2}}
\pagebreak

\tableofcontents
\thispagestyle{plain} 

\chapter{Introduction}\label{intro}

Nontrivial quantum spin liquids (QSLs) are expected to exhibit many
exotic properties such as fractionalized
excitations~\cite{Balents2010,Savary2017},
in addition to the absence of a long-range order.
Despite the vigorous studies in the last several decades, however,
material candidates for such QSLs are still rather limited.

An intriguing scenario to realize a nontrivial QSL is by
generalizing the spin system, which usually
consists of spins representing the $\mathrm{SU}(2)$ symmetry, to $\mathrm{SU}(N)$ ``spin'' systems
with $N > 2$.
We expect stronger quantum fluctuations in $\mathrm{SU}(N)$ spin systems
with a moderate $N$, which could lead the system to an $\mathrm{SU}(N)$ QSL
even on unfrustrated, bipartite lattices, including the honeycomb
lattice~\cite{Li1998,Hermele2011,Corboz2012,Lajko2013}.

The $\mathrm{SU}(N)$ spin systems with $N > 2$
can be realized in ultracold atomic systems,
using the nuclear spin degrees of freedom~\cite{Cazalilla2014}.
In electron spin systems, however, realization of this $\mathrm{SU}(N)$ symmetry
is more challenging.
It would be possible to combine the spin and orbital
degrees of freedom, so that local electronic states are identified
with a representation of $\mathrm{SU}(N)$.
QSL realized in this context
may be called quantum spin-orbital liquids
(QSOLs) because it involves spin and orbital degrees of freedom.
Despite the appeal of such a possibility,
the actual Hamiltonian is usually not $\mathrm{SU}(N)$-symmetric,
reflecting the different physical origins of the spin
and orbital degrees of freedom.
For example, the relevance of an $\mathrm{SU}(4)$ QSOL has been discussed
for Ba$_3$CuSb$_2$O$_9$ (BCSO) with a decorated honeycomb
lattice structure~\cite{Zhou2011,Nakatsuji2012,Corboz2012}.
It turned out, however, that the estimated parameters for BCSO are
rather far from the model
with an exact $\mathrm{SU}(4)$ symmetry~\cite{Smerald2014}.
Moreover, the spin-orbit coupling (SOC) and the directional dependence of the orbital hopping usually
break both the spin-space and orbital-space $\mathrm{SU}(2)$ symmetries,
as exemplified in iridates~\cite{Jackeli2009}.
Thus, it would seem even more
difficult to realize an $\mathrm{SU}(N)$-symmetric system in real magnets with SOC.
(See Refs.~\citenum{Ohkawa1983,Shiina1997,Wang2009,Kugel2015} for proposed
realization of $\mathrm{SU}(N)$ symmetry. However, they do not lead
to QSOL because of their crystal structures.)

In this thesis, we demonstrate a novel mechanism for realizing an $\mathrm{SU}(4)$
spin system in a solid-state system with an onsite SOC.  Paradoxically, the
symmetry of the spin-orbital space can be \emph{enhanced} to $\mathrm{SU}(4)$
when the SOC is strong.
In particular, we propose
$\alpha$-ZrCl$_3$~\cite{Swaroop1964Chem,Swaroop1964Phys,Brauer1978} as the first candidate for an
$\mathrm{SU}(4)$-symmetric QSOL on the honeycomb lattice.  Its $d^1$ electronic
configuration in the octahedral ligand field, combined with the strong
SOC, implies that the ground state of the electron is described by a
$J_\textrm{eff}=3/2$ quartet~\cite{Romhanyi2017}.  In fact, the
resulting effective Hamiltonian appears to be anisotropic in the quartet
space.  Nevertheless, we show that the model is gauge-equivalent to an
$\mathrm{SU}(4)$-symmetric Hubbard model.  In the strong repulsion limit, its
low-energy effective Hamiltonian is the Kugel-Khomskii
model~\cite{Kugel1982} on the honeycomb lattice, exactly at the $\mathrm{SU}(4)$
symmetric point:
\begin{equation}
        H_\textrm{eff} = J \sum_{\langle jk \rangle} \Bigl(\bm{S}_j
\cdot \bm{S}_k+\frac{1}{4}\Bigr)\Bigl(\bm{T}_j \cdot
\bm{T}_k+\frac{1}{4}\Bigr), \label{Eq.KK_SU4}
\end{equation}
where $J>0,$ and $\bm{S}_j$ and $\bm{T}_j$ are pseudospin-$1/2$
operators defined for each site $j$.
The $\mathrm{SU}(4)$ symmetry can be made manifest by rewriting the Hamiltonian,
up to a constant shift, as
$H_\textrm{eff} = \frac{J}{4} \sum_{\langle jk \rangle} P_{jk}$,
where the spin state at each site forms the fundamental representation
of $\mathrm{SU}(4),$ and $P_{jk}$ is the operator which swaps the states at sites $j$
and $k$.  This is a natural generalization of the antiferromagnetic $\mathrm{SU}(2)$ Heisenberg model to $\mathrm{SU}(4).$

The ground state of the $\mathrm{SU}(2)$ spin-1/2 antiferromagnet on the
honeycomb lattice is simply N\'{e}el-ordered~\cite{Roger1989,Fouet2001},
reflecting the unfrustrated nature of the lattice.  On the other hand,
the $\mathrm{SU}(N)$ generalization of the N\'{e}el state by putting different
flavors on neighboring sites gives a macroscopic number of classical
ground states when $N>2$~\cite{Hermele2009,Gorshkov2010,Lajko2017},
implying its instability.
In fact, it was argued that
the $\mathrm{SU}(4)$ antiferromagnet on the honeycomb lattice has a QSOL ground
state without any long-range order~\cite{Corboz2012,Lajko2013}.

\begin{figure}
\centering
\includegraphics[width=6cm]{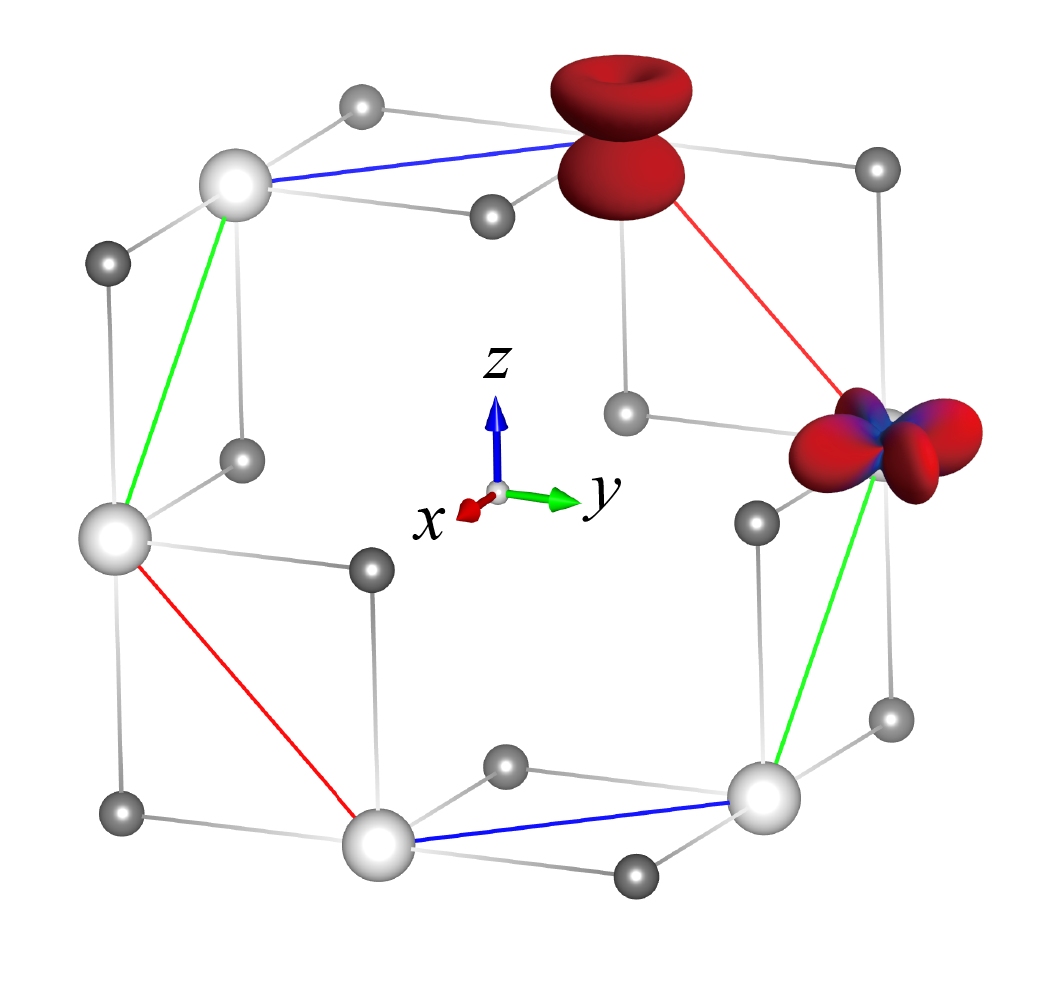}
\caption{Schematic structure of honeycomb $\alpha$-ZrCl$_3.$
White and grey spheres represent Zr and Cl atoms, respectively.}
\label{eye}
\end{figure}

In this Introduction, we first review three types of magnetic frustrations,
geometric frustration, exchange frustration, and $\mathrm{SU}(N)$ frustration
with introduction to the Lieb-Schultz-Mattis-type theorems~\cite{LSM1961}.
Next, we discuss previous methods to realize $\mathrm{SU}(N)$ spin systems
and known results for QSOLs with an $\mathrm{SU}(N)$ symmetry,
which is the central topic of this thesis.

\section{$\mathrm{SU}(2)$ spin systems and quantum spin liquids}

\begin{figure}
\centering
\includegraphics[width=14cm]{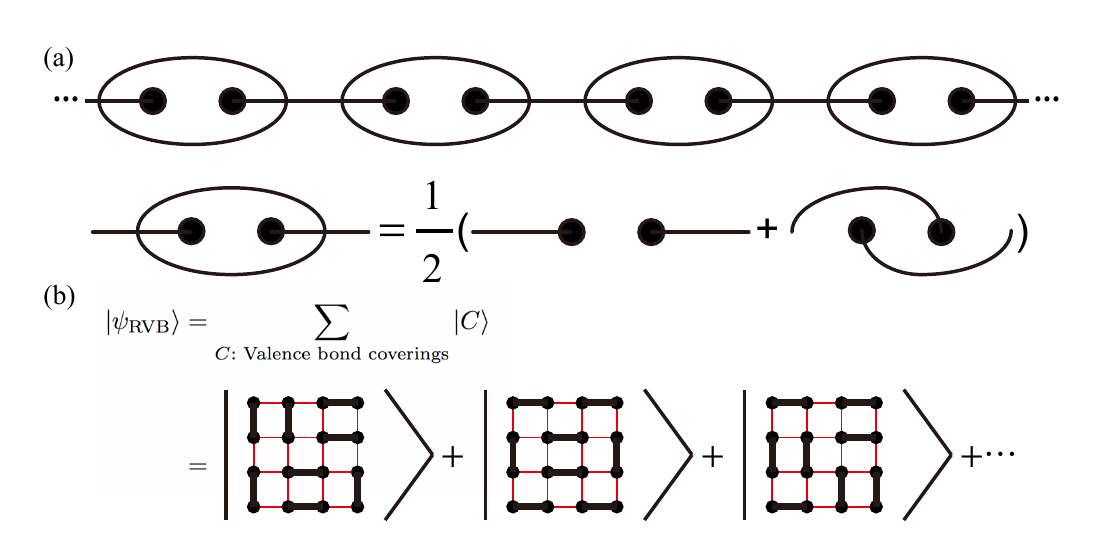}
\caption{VBS and RVB states. (a) VBS ground state of the AKLT model.
(b) RVB ansatz state, where $C$ is every possible valence bond covering on the square lattice.}
\label{vbs}
\end{figure}

Among two-dimensional (2D) $\mathrm{SU}(2)$ spin systems, a QSL
state was first proposed on the triangular lattice by P.~W.~Anderson~\cite{Anderson1973,Anderson1987,Baskaran1987}.
He proposed a symmetric ground state called resonating valence bond (RVB)
as a candidate ground state for the $\mathrm{SU}(2)$ Heisenberg model (Eq.~\eqref{heis}) on
the triangular lattice.
\begin{align}
    H_\textrm{Heisenberg} &= J\sum_{\langle jk \rangle} \bm{S}_j \cdot \bm{S}_k,\label{heis}
\end{align}
where a coupling $J = 4t^2/U$ is determined from a hopping $t$ and an interaction strength
$U > 0$ of the underlying Hubbard model.  Though the true ground state for this model
(spin-1/2) was found to have long-range ordering with a 120-degree antiferromagnetic
configuration later~\cite{Singh1992,Elstner1993}, some triangular organic/inorganic materials
are found to be QSLs in experiments~\cite{Shimizu2003,Li2015}.
Although we can still hope to explain such a spin liquid
state in the weak-coupling regime of the Hubbard model~\cite{Balents2003},
we focus on the $\mathrm{SU}(2)$ Heisenberg model in this section.

Before going on to the RVB state, we quickly discuss a valence bond solid (VBS)
state to show what a valence bond is.  A valence bond is a singlet pair of spins
and the periodic alignment of valence bonds on the lattice is called VBS.
This state is known to be a ground state of the Majumdar-Ghosh model~\cite{Majumdar1969},
or more famously of the Affleck-Kennedy-Lieb-Tasaki (AKLT) model~\cite{Haldane1983PLA,Haldane1983PRL,AKLT1987} after
the projection onto the spin-1 Hilbert space [see Fig.~\ref{vbs}(a)].
In these VBS states, spin-1/2 excitations
are confined and are not included in QSLs.  We can regard an RVB state as an disordered
version of the VBS configuration.

Though the RVB state is a bad guess for the square lattice (even for the triangular
lattice), it is instructive to investigate its property first.  In fact, it is known
that the RVB ground state is an exact ground state for a dimer model on the square lattice
at a fine-tuned point called Rokhsar-Kivelson point~\cite{Rokhsar1988},\footnote{In the
Rokhsar-Kivelson state, $\ket{C}$ are orthogonal with each other.} though it is supposed to
be unstable on the square lattice.  Just to catch a feeling, it is a good starting point to
show a form of its wavefunction.
\begin{equation}
        \ket{\psi_\textrm{RVB}} = \sum_{C: \textrm{ Valence bond coverings}} \ket{C},
\end{equation}
where $C$ is every possible valence bond covering, \textit{i.e.} the way in which
the lattice is completely covered by valence bonds (singlet pairs), and the square lattice case
is illustrated in Fig.~\ref{vbs}(b).

This RVB ground state shows the following important properties and we adopt these three
features as the definition of a QSL~\cite{Misguich2011}:
\begin{enumerate}
        \item Absence of magnetic long-range order.
        \item Absence of spontaneous symmetry breaking.
        \item Existence of fractionalized excitations.
\end{enumerate}
We note that the second one is necessary to exclude the case where the ground states are
degenerate because of spontaneous symmetry breaking.
Actually, the RVB state does not break any space group symmetries of the square lattice,
and all the correlations are apparently short-ranged, though the state itself shows
a \textit{long-range entanglement}, which is a critical feature of QSLs.
The existence of fractionalized excitations called spinons is intuitively understood
as follows.  First, we can locally excite a valence bond by changing a singlet into
a triplet.  Then, due to the superposition of all the possible coverings on
the spin-1/2 lattice model, the separation of the excited triplet pair (of spinons!)
does not cost energy.  This is contrary to VBS where the separation of the excited
pair costs energy proportional to the distance.  In this sense, we can think that
almost free spin-1/2 spinons are fractionalized excitations in the RVB state.
They are called ``fractionalized'' in the sense that they carry a spin-1/2 degree of freedom
instead of spin-1 for magnons, and are fermionic despite the fact that the system
was originally bosonic.

\begin{figure}
\centering
\includegraphics[width=12cm]{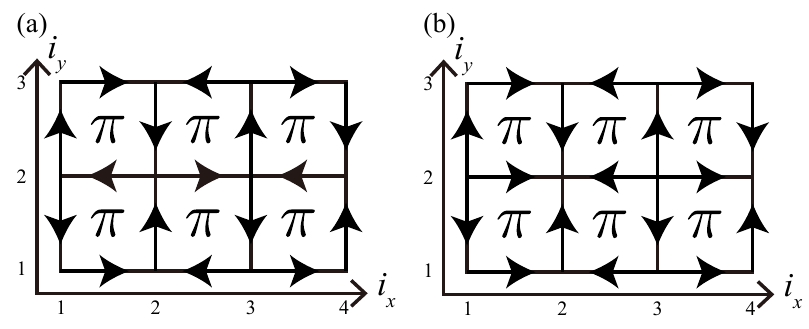}
\caption{Affleck-Marston's $\pi$-flux mean-field ansatz.  (a) Symmetric gauge.
(b) Real (or pure imaginary) gauge.
Arrows determine the direction of $\langle j\to k \rangle$ in the Hamiltonian
and $(i_x,\,i_y)$ labels each site of the square lattice.}
\label{pifluxsq}
\end{figure}

Similarly to the above mentioned RVB state,
Affleck and Marston~\cite{Affleck1988am,Marston1989} proposed the so-called $\pi$-flux ansatz
state for the square lattice QSL.  We can also regard this $\pi$-flux state as one
variation of generalized RVB states, and it obeys three definitions of QSL.
The derivation requires a large-$N$ limit, which will be discussed in Sec.~\ref{largen},
so we here only present a mean-field parton model to describe this
state.
\begin{align}
    H_\textrm{MF} &= -\chi_0\sum_{\langle j\to k \rangle,\,\sigma} (e^{\frac{i\pi}{4}} f_{j\sigma}^\dagger f_{k\sigma} + h.c.),
\end{align}
where $f_{j\sigma}^\dagger$ and $f_{j\sigma}$ are creation and annihilation operators
for spinons with a spin $\sigma = \,\uparrow,\,\downarrow$ at the $j$th site.
The condition $\sum_\sigma f_{j\sigma}^\dagger f_{j\sigma} = 1$ for each $j$ maps
the spinon representation to the original
spin model, as will be discussed in Sec.~\ref{largen}.  The direction of
$\langle j\to k \rangle$ is always determined by Fig.~\ref{pifluxsq}(a).
This is called $\pi$-flux state because a magnetic flux inside each plaquette
is always $\pi$ and spinons feel $-1$ phase factor from the Aharonov-Bohm effect.

Here we used one of the most symmetric gauges on the square lattice, and
this is why an imaginary part appears in the hopping.  Such a gauge does not necessarily
exist for other lattices, so we always use a real gauge with only $\pm 1$ in the latter part.
In such a gauge, the Hamiltonian is transformed into
\begin{align}
    H_\textrm{MF}^\prime &= -\chi_0\sum_{\langle j\to k \rangle,\,\sigma} (i f_{j\sigma}^\dagger f_{k\sigma} + h.c.),
\end{align}
where a factor $i$ is actually unnecessary, and we will omit it from now on to make
it real~\cite{Misguich2011}.  The direction of $\langle j\to k \rangle$ is always determined
by Fig.~\ref{pifluxsq}(b).  The sign has been changed to meet the $\pi$-flux condition.
It seems that assigning mean-field variables to meet the $\pi$-flux condition
(the product of the phase factors around each plaquette must be $-1$) always
breaks translation and other lattice symmetries, but all the lattice symmetries
are correctly implemented \textit{projectively} in this model and gauge degrees of
freedom ignored in this mean-field form always compensate the symmetry transformation.
Thus, there is no spontaneous symmetry breaking or long-range ordering.

One traditional way to understand this kind of phenomena in QSLs is the
projective symmetry group (PSG) theory~\cite{Wen2002}.\footnote{The naming of PSG is confusing
because the group itself is extended, not projective.  What is projective in this
theory is its representation.}
We will not review the entire theory of this framework because usually counting all
PSGs is not efficient, but we will check how it works in some specific models.
In the case of Affleck-Marston's $\pi$-flux ansatz, the gauge structure is
known to be $\mathrm{SU}(2)$ and it is a mother of many other spin liquid states.
Actually, the spectrum includes two Dirac cones~\cite{Misguich2011} and
the translation symmetry is implemented projectively.
As shown in Fig.~\ref{pifluxsq}(b), the translation along the $x$-axis (or $y$-axis)
changes the sign of hopping matrix $U_{jk}^0.$  However, this sign can be absorbed by a gauge
transformation defined by $W_i = (-1)^{i_y},$ where the coordinate of the $i$th site
is defined as $(i_x,\,i_y)$ shown in Fig.~\ref{pifluxsq}, because $W_j = -W_k$ for any
nearest-neighbor bond $\langle jk \rangle$ along the $y$-axis.
Thus, as soon as the translation changes $U_{jk}^0$ into $\tilde{U}_{jk}^0 = -U_{jk}^0,$
we can do a gauge transformation $W_i \tilde{U}_{jk}^0 W_j^\dagger = U_{jk}^0$
to recover the translation symmetry.
In spin liquids, the symmetry is usually supplemented by an additional gauge
transformation explained by PSG.  We also review PSGs of Kitaev models~\cite{Kitaev2006}
in Appendix~\ref{psg}.\footnote{The meaning of PSG
in exactly solvable Kitaev models is slightly different from that of mean-field solutions
because it describes a ``direct'' action of the symmetry on quasiparticles.
See \textit{e.g.} Appendix~F of Ref.~\citenum{Kitaev2006} for the gapped case.}

Even though the $\mathrm{SU}(2)$ Heisenberg models on the square and triangular
lattices have a long-range order at zero temperature, there is still a hope to
find a 2D lattice whose Heisenberg model hosts a QSL.  In order to kill any classical
magnetic ordering, the lattice has to have a strong geometric frustration.
One of the most important possibilities is a kagome lattice, and most studies
support the claim that the $\mathrm{SU}(2)$ Heisenberg model on the kagome lattice
has no magnetic ordering~\cite{Yan2011,Leung1992}.  Although the nature (\textit{e.g.} PSG) of the observed
QSL is still under debate, many numerical studies suggest the existence
of a gapped $Z_2$ spin liquid or a gapless $U(1)$ spin liquid in this model~\cite{Yan2011,Zhu2018}.

As for three-dimensional (3D) $\mathrm{SU}(2)$ spin systems,
a quantum spin ice state is expected on nearly $\mathrm{SU}(2)$-symmetric pyrochlore
antiferromagnets, for example.  The Heisenberg model (or more correctly an XXZ-type
model) on the pyrochlore lattice is actually related to a dimer model on the diamond
lattice (more correctly it can be mapped to a 6-vertex model on the diamond lattice),
and perturbatively realize the physics of RVB states~\cite{Ross2011,Savary2017}.  Numerically,
the 6-vertex model on the diamond lattice is shown to host a gapless $\mathrm{U}(1)$ spin liquid in the
wide range of parameters by sign-free quantum Monte Carlo (QMC) simulations~\cite{Shannon2012}.
This asymptotically shows the existence of a $\mathrm{U}(1)$ spin liquid in the Heisenberg model
on the pyrochlore lattice.

\section{Spin-orbit coupling and Kitaev spin liquids}

So far we discussed a quantum spin liquid in completely $\mathrm{SU}(2)$-symmetric
systems with geometric frustration, but another type of frustration, called
exchange frustration, can be introduced by destroying the $\mathrm{SU}(2)$ symmetry.  Usually
this can be realized in heavier elements with a strong SOC because
in $4d$ or $5d$ transition metals spin interactions become highly
anisotropic and bond-dependent.  This results in bond-dependent interactions
with noncommuting operators, and leads to frustrations between ``exchange''
Hamiltonians for each bond.  This new type of frustration can be found especially
in iridates or Ru-compounds, which would potentially realize the Kitaev
model~\cite{Kitaev2006}.  This model is interesting because it is exactly solvable
\textit{e.g.} on the honeycomb lattice.  See Appendix~\ref{psg} for more details.
The discussion here follows Ref.~\citenum{MT}.  We use a first-quantization picture
for simplicity, though we use a second-quantization picture in the rest of the thesis.

\begin{figure}
\centering
\includegraphics[width=14cm]{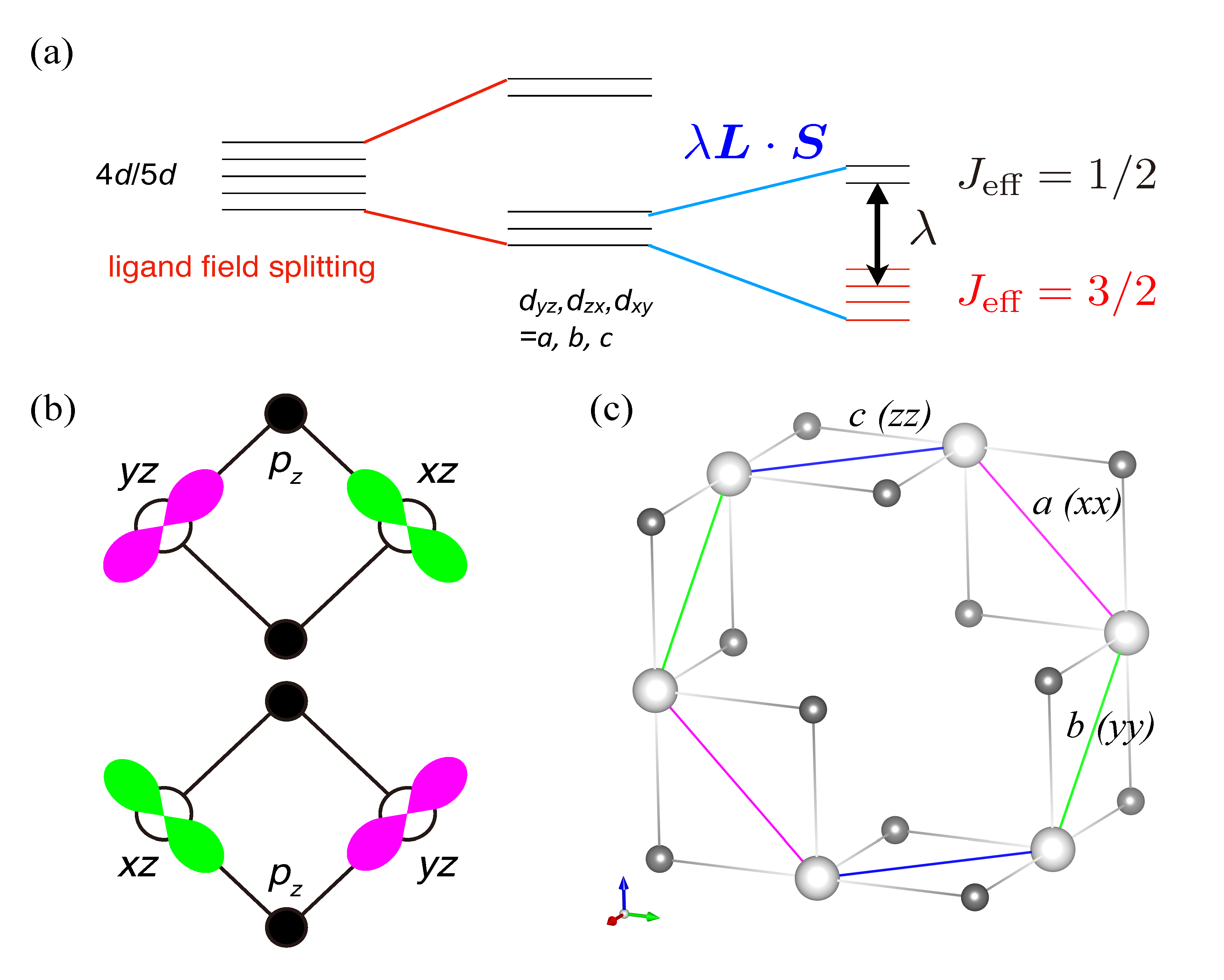}
\caption{Jackeli-Khaliullin mechanism.
(a) Energy splitting of the $d$-orbitals in the octahedral ligand field.
(b) Superexchange pathways for $c$-bonds between
the two adjacent Ru$^{3+}$ or Ir$^{4+}$ ions.  There are two possible pathways
between $yz$ and $xz,$ and between $xz$ and $yz.$
(c) Structure of $\alpha$-RuCl$_3$ or
$\alpha$-$A_2$IrO$_3$-type iridates ($A$ = Li, Na). If it is seen from the (111)
direction, the structure is basically the layered honeycomb lattice, and the bond
direction becomes ideal to realize the Kitaev interaction.
Magenta, light green, and blue bonds represent $a$-, $b$-, and $c$-bonds on the
$yz$-, $zx$-, and $xy$-planes, respectively.
We note that we used magenta instead of red used in the following part for $a$-bonds
to show them more clearly.}
\label{jackeli}
\end{figure}

The Kitaev model on the honeycomb lattice is defined as follows.
\begin{align}
        H_\textrm{Kitaev} &=K_x\sum_{\langle jk\rangle \in a} S_j^x S_k^x+K_y\sum_{\langle jk\rangle \in b} S_j^y S_k^y+K_z\sum_{\langle jk\rangle \in c} S_j^z S_k^z
\end{align}
where $\langle jk\rangle \in \alpha$ means that a nearest neighbor bond $\langle jk\rangle$
belongs to the $\alpha$-bond, and $K_x,\,K_y,$ and $K_z$ are real parameters.
$a$-, $b$-, and $c$-bonds are defined as bonds on the $yz$-, $zx$-, and $xy$-planes,
respectively, as shown in Fig.~\ref{jackeli}(c). This $abc$-notation will be used
in the main text.\footnote{Some may think that it is better to use $K_a,\,K_b,$ and $K_c$
for Kitaev parameters, but here we followed the standard notation.}
This model actually has a gapless or gapped spin liquid ground state depending on its
parameter, as discussed in Appendix~\ref{psg}.  Here we would only discuss how such a bond-dependent
exchange frustration arises in real materials. In fact,
Jackeli and Khaliullin~\cite{Jackeli2009} discovered that the onsite spin-orbit coupling of the Ir$^{4+}$ (or Ru$^{3+}$) ion
in the octahedral coordination can indeed produce this type of models in the Mott insulator
limit with a strong Hubbard $U,$ while
there is also a proposal for a topological insulator in the itinerant limit~\cite{Shitade2009}.

We will consider the low-spin (\textit{i.e.} spin-1/2) $d^5$ system of the transition metal.
Following Jackeli and Khaliullin, we assume iridates (\textit{i.e.} Ir-oxides) with the Ir$^{4+}$
ions in the strong (infinite) octahedral ligand field, but the same thing will apply
to other $d^5$ metal ions like Ru$^{3+}$ as long as they show the low-spin configuration.
The strong octahedral ligand field breaks 5-fold degenerate 5$d$-orbitals into 2-fold
degenerate $e_g$-orbitals and 3-fold degenerate $t_{2g}$-orbitals.\footnote{These
$t_{2g}$-orbitals split into a famous $J_\textrm{eff}=1/2$ doublet
and a $J_\textrm{eff}=3/2$ quartet by SOC, as shown in Fig.~\ref{jackeli}(a),
and here we will use the former.}  The $d^5$ electronic
configuration can be regarded as a situation where one hole is put on the closed
$t_{2g}$-shell.  Moving on to the hole picture, the local ground state for this
hole has 6-fold degeneracy, 3 coming from the orbital degrees of freedom and
2 coming from the spin degrees of freedom.

In this hole picture, the onsite SOC can be treated as follows.  The effective orbital
angular momentum operator $\bm{l}_{\textrm{eff},\,j}$ can be defined for each $t_{2g}$-manifold
of Ir$^{4+}$ because which orbital the hole belongs to among the $yz$-, $xz$- and $xy$-orbitals
(we use a basis set $^t( \ket{yz},\,\ket{xz},\,\ket{xy})$ for these orbitals,
respectively, and represent the transformation/rotation of this triplet by a $3\times 3$
matrix) can be regarded as the vector representation of the (cubic) rotational symmetry.
Clearly, it is a triplet with $l_{\textrm{eff}}=1$ with
$\ket{l^z=0}=\ket{xy},\,\ket{l^z=\pm 1}=-(i\ket{xz}\pm \ket{yz})/\sqrt{2},$ and
onsite SOC conserves the local total angular momentum
$\bm{J}_{\textrm{eff},\,j}=\bm{l}_{\textrm{eff},\,j}+\bm{S}_j$ for each $j.$
The onsite SOC has the form of antiferromagnetic interaction between the effective orbital
angular momentum and the spin angular momentum for each $j$ in the hole picture, as follows.
\begin{equation}
        H_\textrm{SOC}=\sum_j [\lambda \bm{l}_{\textrm{eff},\,j}\cdot \bm{S}_j +\Delta_z (l_{\textrm{eff},\,j}^z)^2],
\end{equation}
where $\lambda > 0$ is the strength of the onsite effective SOC and
$\Delta_z$ is the tetragonal distortion of the IrO$_6$ octahedra along the $z$-direction.
When $\Delta_z=0$ the ground state Kramers doublet is clearly a $J_\textrm{eff}=1/2$ doublet,
and in the general case, the ground state doublet (pseudospin) can be written as
\begin{align}
        \ket{\tilde{\uparrow}} &=\sin \theta \ket{0}\otimes\ket{\uparrow}-\cos\theta \ket{+1}\otimes\ket{\downarrow}, \\
        \ket{\tilde{\downarrow}} &=\sin \theta \ket{0}\otimes\ket{\downarrow}-\cos\theta \ket{-1}\otimes\ket{\uparrow},
\end{align}
where the left-hand side of $\otimes$ means the value of $l_{\textrm{eff}}^z,$ and
the right-hand side means $S^z,$ while $\theta$ parametrizes
the tetragonal distortion as $\tan (2\theta) =2\sqrt{2} \lambda / (\lambda-2\Delta_z).$
In the following, we assume $\Delta_z=0,\,\sin\theta =1/\sqrt{3},$ \textit{i.e.} the completely cubic case, for simplicity.

As shown in Fig.~\ref{jackeli}(b), the superexchange hopping pathways between the two adjacent
Ir$^{4+}$ ions $j$ and $k$ in the case of $c$-bonds
via the oxygen $p$-orbitals with 90-degree configuration
can be written as the following matrix.
\begin{equation}
        H_\textrm{hop}^{k\gets j}=
        \begin{pmatrix}
                0 & -t & 0 \\
                -t & 0 & 0 \\
                0 & 0 & 0
        \end{pmatrix}
        \otimes I_2,
\end{equation}
where $I_m$ is the $m \times m$ identity matrix, acting on the spin space in this case,
and $t$ is the real hopping parameter between $xz$ and
$yz$-orbitals via the oxygen $p$-orbitals as shown in Fig.~\ref{jackeli}(b).
The most important observation is that this matrix has no matrix elements
between the adjacent $J_\textrm{eff}=1/2$ doublets and,
therefore, there is no contribution to the antiferromagnetic exchange interaction
due to the Pauli principle
if we project this superexchange interaction onto the $J_\textrm{eff}=1/2$ pseudospin model.
Thus, if we make the $J_\textrm{eff}=1/2$ pseudospin model from this hopping, the strongest
interaction we have to consider is the contribution from the second-order perturbation
involving the onsite ferromagnetic Hund interaction between the $J_\textrm{eff}=3/2$
and $J_\textrm{eff}=1/2$ orbitals on the $k$th site.  The second-order contribution
is just a hole going from $J_\textrm{eff}=1/2$ on $j$ to $J_\textrm{eff}=3/2$ on $k$ and
then coming back from $J_\textrm{eff}=3/2$ on $k$ to $J_\textrm{eff}=1/2$ on $j,$ while,
on the other side, the other hole remains sitting in the $J_\textrm{eff}=1/2$ manifold on $k.$
This perturbative contribution can roughly be estimated as the following effective Hamiltonian.
\begin{align}
    H_\textrm{eff}^{t_{2g}} &= \frac{H_\textrm{hop}^{j\gets k}H_\textrm{hop}^{k\gets j}}{-(\Lambda-J_H \bm{S}_j \cdot \bm{S}_k)} \\
        &\sim \frac{1}{\Lambda^2}H_\textrm{hop}^{j\gets k}H_\textrm{hop}^{k\gets j}\cdot (-J_H \bm{S}_j \cdot \bm{S}_k) \nonumber \\
        &=-\frac{J_H t^2}{\Lambda^2} P_{l^z=\pm1}(\bm{l}_{\textrm{eff},\,j})\bm{S}_j \cdot \bm{S}_k,
\end{align}
where $J_H>0$ is the ferromagnetic Hund interaction inside the same ion, $\Lambda$
is the potential energy for the excited $J_\textrm{eff}=3/2$ state which is
almost proportional to $\lambda,$ and $P_{l^z=\pm1}$
is a projection operator onto the manifold with a condition $l^z=\pm1.$
By projecting this effective Hamiltonian onto the pseudospin system with only
$J_\textrm{eff}=1/2$ degrees of freedom, we finally get the ferromagnetic Ising
interaction with anisotropy along the $z$-direction  because the
operator $P_{l^z=\pm1}(\bm{l}_{\textrm{eff},\,j})$ and the projection onto $J_\textrm{eff}=1/2$
will completely kill the terms $S_j^+ S_k^-$ and $S_j^- S_k^+.$
The final form of the interaction between two adjacent spins becomes
\begin{equation}
        H_\textrm{eff}^{J_\textrm{eff}=1/2}\sim -\frac{J_Ht^2}{\lambda^2}\cos\theta (\sin\theta+\cos\theta)J_{\textrm{eff},\,j}^z J_{\textrm{eff},\,k}^z,
\end{equation}
assuming $\Lambda \propto \lambda.$
This $z$-directional anisotropy comes from the oxygen configuration in the $xy$-plane.
Thus, in the honeycomb geometry of a (111) thin film of iridates as shown in Fig.~\ref{jackeli}(c),
the whole interactions between $J_\textrm{eff}=1/2$ pseudospins become the Kitaev model
assuming the perfect cubic coordination and the 90-degree oxygen configuration.

Because the superexchange Kitaev interaction coming from this mechanism
is ferromagnetic, it is advantageous to realize the Kitaev spin liquid phase in the Kitaev-Heisenberg model
(\textit{i.e.} the sum of the Kitaev model and the nearest-neighbor Heisenberg model)
on the honeycomb lattice, which is known to be more stable in the ferromagnetic case than
in the antiferromagnetic case~\cite{Chaloupka2010,Chaloupka2013}.

Additionally, Kitaev~\cite{Kitaev2006} discussed the classification of
symmetry-enriched topological (SET) phases of the Kitaev model and the toric code~\cite{Kitaev2003}
based on a modern theory of weak symmetry breaking, but we will follow the PSG
theory~\cite{Wen2002} in this thesis for simplicity.
Thermodynamic properties of the Kitaev model were also examined by QMC
due to the accidental absence of a sign problem on this model~\cite{Nasu2014}.
The sign problem of QMC is a major theoretical difficulty to study spin liquids,
but we will not discuss this point in this thesis.\footnote{The origin of the sign problem
of the $\mathrm{SU}(4)$ Heisenberg models on 2D bipartite lattices is quite complicated.}

\section{$\mathrm{SU}(N)$ spin systems and large-$N$ limits}\label{largen}

Here we review the physics of one-dimensional (1D) and 2D $\mathrm{SU}(N)$ systems.
Increasing the number of flavors to a large $N$ actually leads to the third
type of quantum frustration. Even in the 1D case there is macroscopic
degeneracy of classical ground states in 1D $\mathrm{SU}(N)$ antiferromagnets when
$N>2,$ suggesting a possibility that increasing $N$ results in a large quantum (zero-point)
oscillation in any dimensions.  When $N=2,$ the corresponding classical model is
the Ising model with two states $\sigma =$ $\uparrow,\,\downarrow$ per site.
Thus, antiferromagnetic ground states are just twofold degenerate in the 1D Ising model.
When $N>2,$ (even with $N=3$) if we label an onsite degree of freedom by A, B, and C,
then the ground states of this classical model are already macroscopically degenerate even
in one dimension, including \textit{e.g.} ABABABABABAB, ABCABCABCABC, ACBCACBCACBC, etc.
This degenerate classical ground state manifold allows us to construct a highly-entangled
quantum ground state by a macroscopic superposition of such states, leading to
a possibility of realizing a \textit{long-range entangled} state in higher dimensions.
Although a real $N \to \infty$ limit is classical, for an intermediate $N$
quantum fluctuation gets stronger than either $N=2$ or $N \to \infty$
and it possibly leads to a new QSL state.

First, let us discuss how to generalize the $\mathrm{SU}(2)$ Heisenberg interaction
to general cases.  As we already discussed, the $\mathrm{SU}(2)$ Heisenberg interaction
can be rewritten in terms of a swapping operator $P_{jk}$ between the $j$th and $k$th
sites.  Thus, a natural generalization of the $\mathrm{SU}(N)$ Heisenberg interaction
with $N>2$ is also written by this swapping operator for $\mathrm{SU}(N)$ fundamental
representations.

\begin{align}
    H_{\mathrm{SU}(N)} = \frac{J}{N} \sum_{\langle jk \rangle} P_{jk}, \label{sun1}
\end{align}
where $J$ is the Heisenberg term.  The Hilbert space is defined by putting
a fundamental representation spin on each site of the lattice.  We note that
any representations can be used to define a similar model, but we only consider
a fundamental representation in this thesis.  Let us simply check that this is a natural
generalization for the $N=2$ and spin-1/2 case by looking at a two-body model.
The energy splitting according to the representation can be described by the following
Young tableaux.  Each box represents fundamental representation of $\mathrm{SU}(N),$
and change in the dimension of the representation is shown below.

\begin{align}
{\Yvcentermath1 \yng(1) \otimes \yng(1)} &= {\Yvcentermath1 \yng(1,1) \oplus \yng(2)} \\
N \times N &= \frac{N(N-1)}{2} + \frac{N(N+1)}{2}.
\end{align}
As easily seen from the above diagrams, a two-body model $P_{jk}$ simply separates
the antisymmetrized state from the symmetrized state and $N(N-1)/2$ states
have a lower energy out of the original $N^2$ states.
We note that in the $N=2$ case only a singlet state has the lower energy, but
in the $N>2$ case the two-body solution still has (macroscopic) degeneracy and
finding a quantum ground state out of degenerate ``valence bond'' coverings
is already a nontrivial problem when the system is constructed over some periodic lattice.

Next, let us discuss the 1D chain of the $\mathrm{SU}(N)$ Heisenberg model~\cite{Sutherland1975}.
An accurate description requires non-Abelian bosonization~\cite{Affleck1986SUN}, but we will qualitatively
explain it using a na\"ive Tomonaga-Luttinger liquid theory~\cite{Bloch1933,Tomonaga1950,Luttinger1963,Affleck1988mua}.
The $\mathrm{SU}(N)$ Heisenberg model can always be derived from the $\mathrm{SU}(N)$
Hubbard model at $1/N$ filling:
\begin{align}
    H_\textrm{Hubbard} = -t \sum_{\langle jk \rangle} (c_{j\alpha}^\dagger c_{k\alpha} +h.c.) + \frac{U}{2}\sum_j n_j(n_j -1),
\end{align}
where a fermion $c_{j\alpha}$ has a flavor (index) $\alpha = 1,\,2,\dots,\,N,$ and
a number operator $n_j = \sum_\alpha c_{j\alpha}^\dagger c_{j\alpha}.$
At $1/N$ filling with a large $U/|t|$ a metal-insulator
transition to a Mott insulator always happens.  In the $U/|t| \to \infty$ limit, the Hilbert
space is spanned by states with exactly one fermion per site, and the (degenerate)
second-order perturbation inside this Hilbert space is always reduced to Eq.~\eqref{sun1}.

A metal-insulator transition into a Mott insulator usually accompanies some magnetic order,
but in one dimension the enhanced quantum fluctuation is known to suppress any long-range
ordering, leading to a gapless liquid state with spinon excitations, as is well-known
in the case $N=2.$  Every correlation decays algebraically, resulting in a ``solvable''
liquid state.  This is the famous Tomonaga-Luttinger liquid theory, where the charge
and ``spin'' degrees of freedom are separated, and only the charge sector is gapped
in the Mott-insulating phase.  Thus, the only thing left in the $U/|t| \to \infty$ limit,
\textit{i.e.} Eq.~\eqref{sun1}, is a gapless ``spin'' liquid with $N-1$ fermionic
spinon excitations.  They are described by the theory of bosonization.
We note that a 1D $\mathrm{SU}(4)$ Heisenberg chain is also free of a sign problem
and its thermodynamic property has been investigated very well~\cite{Frischmuth1999}.
The results almost agree with the previous studies based on bosonization~\cite{Affleck1986SUN},
so we have unbiased reproduction of the effective theory in the $N=4$ case.
If we separate $N=4$ degrees of freedom into the spin sector and the orbital sector,
we can name fractionalized orbital excitations orbitalons.

From now on we would like to discuss an $\mathrm{SU}(N)$ Heisenberg on the 2D square lattice.
In this case a large $N$ limit is actually useful for an intermediate $N$ region,
though for $N \leq 4$ the ground state is known to be ordered in most numerical simulations.
In a real $N \to \infty$ limit, the mean-field theory suggests the ground state to be
a chiral spin liquid (CSL) state.  CSL is not a QSL in a strict sense because it spontaneously
breaks the time-reversal symmetry with an effective magnetic field acting on quasiparticles.
However, apart from that, CSL states are usually regarded as a variant of QSLs without
an explicit long-range correlation (in two-body operators) and with fractionalized
topological excitations assuming the existence of a gap.

In the same spirit as we used the Hubbard model to describe spinons, we can
represent spin operators by fermions in the fundamental representation of $\mathrm{SU}(N).$
Since swapping operators can be decomposed into a product of two hopping terms
$f_\alpha^\dagger f_\beta$ and $f_\beta^\dagger f_\alpha,$ assuming $\mathrm{SU}(N)$
spin operators are represented as
$S_\alpha^\beta=f_\alpha^\dagger f_\beta$ for each site,\footnote{$S_\alpha^\beta$
operators are redundant because only $N^2 -1$ components are independent} so
\begin{align}
    H_{\mathrm{SU}(N)} &\propto \sum_{\langle jk \rangle,\alpha,\beta} f_{j\alpha}^\dagger f_{j\beta} f_{k\beta}^\dagger f_{k\alpha},
\end{align}
where $f_{j\alpha}$ is a spinon annihilation operator with $\alpha = 1,\dots,N$ on the
$j$th site.  By imposing a constraint $\sum_\alpha f_{j\alpha}^\dagger f_{j\alpha}=1$ for
each $j,$ which is the same as the $U \to \infty$ limit of the Hubbard model, the model
is exactly mapped to the original $\mathrm{SU}(N)$ Heisenberg model.
The four-fermion terms can be decomposed by a mean-field approximation~\cite{Affleck1988am,Marston1989}.

After introducing this (Schwinger-Wigner) fermionic representation,
taking an $N \to \infty$ limit is the same as considering a classical solution
at a saddle point,\footnote{There is a subtlety when taking this limit~\cite{Hermele2009}.}
and the problem results in finding a solution of the following self-consistent equations.
\begin{align}
    H_f &= -\sum_{\langle jk \rangle,\alpha} \left(\chi_{jk} f_{j\alpha}^\dagger f_{k\alpha} + h.c. \right), \\
    \chi_{jk} &= \left\langle \sum_\beta f_{k\beta}^\dagger f_{j\beta} \right\rangle,
\end{align}
where the expectation value is taken for a free-fermionic model $H_f.$

\begin{figure}
\centering
\includegraphics[width=8cm]{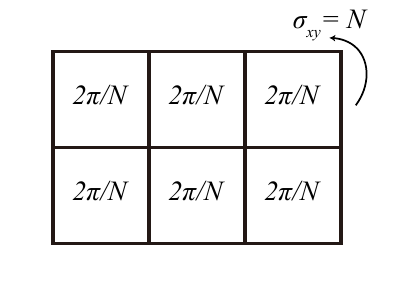}
\caption{Chiral spin liquid state proposed in Ref.~\citenum{Hermele2009}.
The fractional flux inside each plaquette breaks the time-reversal symmetry
giving a quantum Hall conductance of $\sigma_{xy} = N$ in a dimensionless form.}
\label{csl}
\end{figure}

Though it is still approximate to find out a solution in the sense that
there is no way to impose a local number constraint in classical calculations,
for $5 \leq N \leq 10$ (or even $5 \leq N$), the following CSL solution
is known to be a large-$N$ solution~\cite{Hermele2009}: $|\chi_{jk}| = \chi_0$ is a
constant and the phase of $\chi_{jk}$ is determined for each square plaquette to carry
a $2\pi/N$ flux, \textit{i.e.}
$\prod_{\langle jk \rangle \in C} \chi_{jk} = \chi_0^4 e^\frac{2\pi i}{N}$ for
each plaquette $C.$  We will omit a fluxoid quantum to make it dimensionless.

This solution is very similar to the lattice quantum Hall state with a fractional
flux quantum $2\pi/N$ per plaquette and it is expected to host anyonic excitations,
useful for universal quantum computation, after the Gutzwiller projection.
This ansatz is the same as Affleck-Marston's
when $N=2,$ and only in this case it does not break the time-reversal
symmetry.\footnote{This is because $\pi = -\pi$ (mod $2\pi$).  Such a magnetic field
is only available on neutron stars~\cite{Lieb1994} and exists only \textit{emergently}
on earth.}  When $N>2,$ it is a CSL with
a fractional magnetic flux, and only if $N>4$ it can possibly be a ground state
of the $\mathrm{SU}(N)$ Heisenberg model on the square lattice.  Though a large-$N$ limit
is always a classical saddle-point solution, it is a good starting point even for an
intermediate $N,$ and it is expected that the $1/N$ correction includes a quantum
fluctuation which is neglected in the mean-field model without a gauge degree of
freedom.  A real QSL/CSL must have a gauge fluctuation, which is exemplified
in the Kitaev model for example [see Appendix~\ref{psg}], so we should always confirm
that the quantum fluctuation does not destroy the classical state, as is the case
with a small $N.$

\section{Lieb-Schultz-Mattis theorem and its extension}

In order to clarify the relation between the absence of spontaneous symmetry breaking
or long-range ordering and the existence of fractionalized excitations, we would like
to discuss an important theorem called Lieb-Schultz-Mattis (LSM) theorem~\cite{LSM1961} and
its extension by Oshikawa~\cite{Oshikawa2000,Oshikawa2003} and Hastings~\cite{Hastings2005}.
The discussion here follows Ref.~\citenum{MT} and some proofs are included in Appendix~\ref{lsma}.

Let us begin with the 1D case, where the quantum fluctuation is strong enough to destroy
any kinds of magnetic ordering.  For a 1D $\mathrm{SU}(2)$ Heisenberg (or more generally XXZ)
chain, the Bethe ansatz solution indicates the existence of gapless spinon excitations
above the ground state.  Such spinon excitations are sometimes regarded as fractionalized
because they are spin-1/2 instead of spin-1 for magnons.
The gapless nature of the $S=1/2$ Heisenberg chain is protected by the LSM
theorem~\cite{LSM1961} and there cannot exist a gapped ground
state with no ground state degeneracy (GSD) for the spin-1/2 (or half-odd-integer spin) chain~\cite{Affleck1986}.
We assume the lattice translation symmetry, the $S_\textrm{tot}^z$ conservation, and the bond-centered space-inversion (or
time-reversal) symmetry of the Hamiltonian, all of which are natural for quantum spin liquids,
and also assume that the ground state is unique in a finite system.  Then,
it is straightforward to prove the LSM theorem by introducing the following twist operator $U.$
\begin{equation}
        U:=\exp\bigl[i\sum_{j=1}^L \frac{2\pi j}{L} S_j^z\bigr],
\end{equation}
where $\bm{S}_j$ is a spin operator in the usual definition (assuming $\hbar=1$) and $\bm{S}_j= \bm{\sigma}_j/2$ for the spin-1/2 case.
This $U$ can be regarded as a creation operator for the lowest-energy ``spin wave'' excitation, so
$\ket{\Psi_t}:=U\ket{\Psi_0}$ would be the first excited state.  This is orthogonal to
the ground state, \textit{i.e.} $\braket{\Psi_0 | \Psi_t}=0$ as follows.
\begin{align}
        \braket{\Psi_0 | \Psi_t} &=\braket{\Psi_0 |U| \Psi_0}=\braket{\Psi_0 |T^{-1}UT| \Psi_0} \nonumber \\
        &=\braket{\Psi_0 |U \exp\bigl[-i\frac{2\pi}{L} S_\textrm{tot}^z \bigr] \exp(2\pi i S_1^z)| \Psi_0}=-\braket{\Psi_0 |U| \Psi_0},
\end{align}
using $S_\textrm{tot}^z=0$ for the ground state, and an operator identity $\exp(2\pi i S_1^z)=-1$ for half-odd spins, where $T$ is a translation operator.
With a 1D local Hamiltonian $H,$ the energy difference $\braket{\Psi_t|H|\Psi_t}-\braket{\Psi_0|H|\Psi_0}=\mathcal{O}(L^{-1})$
becomes zero due to the inversion (or time-reversal) symmetry
of the ground state.  Here we illustrate the proof
in the specific case of the XXZ model.  The XXZ model is defined as,
\begin{align}
        H &= \sum_j H_j, \\
        H_j &=S_j^x S_{j+1}^x + S_j^y S_{j+1}^y + \Delta S_j^z S_{j+1}^z.
\end{align}
Therefore, $\braket{\Psi_t|H_j|\Psi_t}-\braket{\Psi_0|H_j|\Psi_0}\propto\braket{\Psi_0|\frac{i}{L}[S_j^+ S_{j+1}^- - S_j^- S_{j+1}^+] +\mathcal{O}(L^{-2}) |\Psi_0}$ and the order $1/L$ term vanishes due to the inversion (or time-reversal) symmetry.  Thus, $\ket{\Psi_t}$ becomes the degenerate ground state in the thermodynamic limit.
This proves that either the system is gapless or the ground state
is not unique (\textit{i.e.} spontaneous symmetry breaking in 1D)
in the thermodynamic limit of half-odd spin chains.
This is consistent with the gapless nature of the spin-1/2 (and half-odd spin) Heisenberg chain(s).
Nevertheless, as for the spin chain with an integer spin quantum number,
the situation is different because $\exp(2\pi i S_1^z)=1.$

LSM-type theorems are more important in higher dimensions because GSD
suggests the existence of a so-called topological order.
The generalization of
the LSM theorem was done by Oshikawa~\cite{Oshikawa2000,Oshikawa2003} and more rigorously
by Hastings and others~\cite{Hastings2005,Nachtergaele2007}.  The assertion for
$\mathrm{SU}(2)$-symmetric quantum spin models on the lattice from the
Hastings-Oshikawa-Lieb-Schultz-Mattis (HOLSM) theorem~\cite{LSM1961,Hastings2005,Oshikawa2000}
is the following.

\begin{thm}
For the spin system in 2D or higher dimensions, assuming
the translation symmetry for the Hamiltonian and there are an
odd number of total spin quantum numbers in the unit cell,
the ground state of the lattice spin system
must either be gapless, break the spin-space or translation symmetry, or have
multiple GSD.
\end{thm}

This theorem is intuitively understandable by the following arguments~\cite{Zaletel2015}.
For simplicity, we here only consider the case with a spin-1/2 degree of freedom per unit cell.
If we map a spin-1/2 lattice model into hard-core bosons, where spin up
is an empty site and spin down is a site occupied by a boson.
Then, a ground state with no GSD must have a half-odd filling of bosons.
To get a featureless insulator\footnote{A featureless insulator is usually defined as a symmetric gapped phase with a unique ground state.}~\cite{Kimchi2013}
from this bosonic system with a translation
symmetry, the bosons must be fractionalized into half-charged entities, which
is distributed uniformly in the lattice.  Translated back to the spin language,
this implies that to obtain a symmetric ground state, we need a spin-1/2 excitation
in the bulk, but there is no local excitation carrying a spin-1/2 degree of freedom,
and therefore it must be nonlocal (\textit{i.e.} topological).

In two or higher dimensions, GSD always implies the existence of topological order in gapped systems.
We do not discuss the direct relationship between GSD and the nature of the topological
order, but we quickly review an easy example of topological order.
A 2D $Z_2$ topological order is the simplest Abelian topological order in closed
gapped systems.  This topological order is realized in the ground state of Kitaev's toric code~\cite{Kitaev2003}
or the Kitaev model~\cite{Kitaev2006} in the gapped phase [see Appendix~\ref{psg}].
These models have physically proven that the ground states with a topological order
always carry fractionalized excitations above the energy gap.
The $Z_2$ topological order is known to possess GSD depending on the genus
if it is defined on the closed surface.  If the genus of the surface is $g,$
then GSD is $4^g.$  The dependence of the ground states property on the global topology
suggests the existence of a \textit{long-range entanglement} in the system.

Many numerical results of a spin-1/2 antiferromagnetic Heisenberg model
on the kagome lattice actually suggests the absence of magnetic ordering
at very low temperature~\cite{Leung1992,Yan2011}.  We can simply conclude from the HOLSM theorem that,
assuming the absence of spontaneous symmetry breaking, this model has
either a gapless ground state or a gapped ground state with multiple GSD
because it has odd number of spin-1/2 degrees of freedom in the unit cell.
In either case, we can conclude that the ground state of
the spin-1/2 antiferromagnetic Heisenberg model on the kagome lattice should
be exotic with a fractionalized excitation beyond the Ginzburg-Landau theory,\footnote{In the gapless case,
this point is subtle but we can say that the excitations are exotic in the sense that
it is still fractionalized even if we gap out these excitations without breaking the symmetry.}
and we will
refer to this ground state as kagome spin liquid~\cite{MT,He2017}.
Whether this kagome spin liquid
is gapped or gapless is still under debate among both theorists and experimentalists.
A gapped spin liquid with a $Z_2$ topological order is sometimes called
$Z_2$ spin liquid, which is one of the most important candidates of the kagome spin liquids~\cite{Yan2011}.
Another candidate is a Dirac spin liquid~\cite{He2017}, which is similar to the Affleck-Marston state.

Though the meaning of a long-range entanglement in gapped systems is clear
based on this topological order/GSD, it is subtle in gapless systems.
There are many measures for it, such as entanglement entropy and entanglement spectrum,
but we will not seek this direction deeply.
We note that in the case of Ref.~\citenum{Corboz2012} the bond dimension $D$ for
iPEPS calculations is used as a measure for quantum entanglement.  More generally,
in tensor network calculations including density matrix renormalization group (DMRG)
this bond dimension is known to be a good measure to detect quantum entanglement of
the ground state.

In relation to the main focus of this thesis, the extension of the theorem by
Affleck and Lieb for $\mathrm{SU}(N)$ spin systems~\cite{Affleck1986} is more important.
This is called Lieb-Schultz-Mattis-Affleck (LSMA) theorem and
will be discussed in detail in Appendix~\ref{lsma}.

\section{Dirac spin liquids in the $\mathrm{SU}(4)$ Heisenberg model}

\begin{figure}
\centering
\includegraphics[width=6cm]{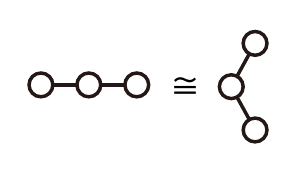}
\caption{Accidental isomorphism between $\mathfrak{su}(4)$ and $\mathfrak{so}(6).$}
\label{dynkin}
\end{figure}

So far we explained previous candidates for QSLs with geometric/exchange frustrations.
A fairly new approach was the $\mathrm{SU}(N)$ magnetism and it has many advantages.
Gapless excitations can be guaranteed by the LSMA theorem [see Appendix~\ref{lsma}],
and QSLs with an $\mathrm{SU}(N)$ symmetry without a symmetry breaking (not CSL)
would be a great playground to study fractionalization of excitations.  We already
presented a theoretical background, but here we would like to introduce one concrete
example of $\mathrm{SU}(N)$ spin liquids, which is a main target of this thesis.
We introduce parton mean-field theories for the $\mathrm{SU}(4)$ Heisenberg
model.  There are mainly two types of representations, a Schwinger-Wigner fermionic
representation~\cite{Wen2002} and a Wang-Vishwanath Majorana representation~\cite{Wang2009}.
The former is the same as Affleck-Marston's theory or the one used in Sec.~\ref{largen}, so
let us first review an $\mathrm{SO}(6)$ Majorana representation for $\mathrm{SU}(4)$ spins.
Though we will not use this representation in the main text, we quickly review it
because it is beautiful and useful for $N=4.$

There is a mathematical accidental isomorphism between Lie algebras $\mathfrak{so}(6)$
and $\mathfrak{su}(4),$ which is clearly reflected in their Dynkin diagrams [see Fig.~\ref{dynkin}].
An accidental isomorphism is always for Lie algebras, but
we abuse terminology like $\mathrm{SO}(6) \cong \mathrm{SU}(4),$ for simplicity,
to mention this fact.\footnote{Here, $\cong$ means local isomorphism.}
Since $\mathrm{SU}(4) \cong \mathrm{SO}(6),$ there is also an isomorphism between
an antisymmetric tensor representation of $\mathrm{SU}(4)$ and a vector representation
of $\mathrm{SO}(6).$  Though we will not explicitly show this isomorphism,
it is the reason behind the fact that we can construct an $\mathrm{SO}(6)$ Majorana representation.

The representation is similar to the one used by Kitaev for the
$\mathrm{SU}(2)$ spin~\cite{Kitaev2006} except for the number of physical subspaces.
First, similarly to the previous section, we divide the $\mathrm{SU}(4)$ fundamental
representation into spin and orbital degrees of freedom.  Then, a spin $\bm{S}_j$
and an orbital $\bm{T}_j$ can be decomposed into a cross product of two $\mathrm{SO}(3)$
Majorana fermions.
\begin{align}
    S_j^{\gamma} &= -\frac{i}{4} \varepsilon^{\alpha \beta \gamma} \eta_j^\alpha \eta_j^\beta, \\
    T_j^{\gamma} &= -\frac{i}{4} \varepsilon^{\alpha \beta \gamma} \theta_j^\alpha \theta_j^\beta,
\end{align}
where $\varepsilon^{\alpha \beta \gamma}$ is a Levi-Civita symbol, and $\bm{\eta}$ and $\bm{\theta}$ are $\mathrm{SO}(3)$ Majorana fermions with $\{\eta_j^\alpha, \eta_k^\beta\}=\{\theta_j^\alpha, \theta_k^\beta\} = 2 \delta_{jk}\delta^{\alpha \beta},$ and $\{\eta_j^\alpha, \theta_k^\beta\}=0.$
These 6 Majorana fermions per site have an $\mathrm{SU}(4) \cong \mathrm{SO}(6)$ symmetry.
This representation is redundant and for each site an extended Hilbert space for
Majorana fermions has a dimension $(\sqrt{2})^6= 8.$  Thus, we have to halve the dimension
and project them onto the physical subspace in an $\mathrm{SO}(6)$-symmetric way.

The simplest and most useful constraint for the projection is
\begin{align}
    i\eta_j^x \eta_j^y \eta_j^z \theta_j^x \theta_j^y \theta_j^z = 1 \quad \textrm{for}\,\forall j, \label{case1}
\end{align}
or
\begin{align}
    i\eta_j^x \eta_j^y \eta_j^z \theta_j^x \theta_j^y \theta_j^z = -1 \quad \textrm{for}\,\forall j. \label{case2}
\end{align}
Differently from Kitaev's representation, both Eq.~\eqref{case1} and Eq.~\eqref{case2} can
simplify the original Hamiltonian and result in the same Majorana Hamiltonian.
In either case, all higher order terms in the original $\mathrm{SU}(4)$ Heisenberg model
are reduced to quartic terms:
\begin{align}
    H_\textrm{Majorana} \propto -\frac{1}{8}\sum_{\langle jk \rangle} \left( i\bm{\eta}_j \cdot \bm{\eta}_k + i\bm{\theta}_j \cdot \bm{\theta}_k \right)^2.
\end{align}
Thus, at a saddle point we can simply define a real mean field to solve a self-consistent
equation by $\chi_{jk}^R = \langle i\bm{\eta}_j \cdot \bm{\eta}_k + i\bm{\theta}_j \cdot \bm{\theta}_k \rangle,$ and
\begin{align}
    H_\textrm{MF}^R = \sum_{\langle jk \rangle} \left[ -\frac{\chi_{jk}^R}{4} \left( i\bm{\eta}_j \cdot \bm{\eta}_k + i\bm{\theta}_j \cdot \bm{\theta}_k \right) + \frac{(\chi_{jk}^R)^2}{8} \right].
\end{align}
We note that the mean field $\chi_{jk}^R = -\chi_{kj}^R$ is always real, and Majorana
fermions cannot feel a complex magnetic field like in a quantum Hall state.  This is one
important difference between a Majorana $\chi_{jk}^R$ mean field and a complex
$\chi_{jk}$ mean field.

We note that there is no conservation of the fermion number except for the $Z_2$ parity, so
usually we make a mean-field ansatz wavefunction by filling a Fermi sea until half filling,
and do a Gutzwiller projection to the physical subspace, which is an approach similar to
the Kitaev model [see Appendix~\ref{psg}].  Two different fermionic approaches are
\textit{a priori} describing symmetric $\mathrm{SU}(4)$ spin liquids equally well
with a symmetric flux ansatz which does not break any symmetry of the Hamiltonian.
Since Lieb's theorem~\cite{Lieb1994} is not applicable to the quarter-filling case,
there is no \textit{a priori} guess for the lowest-energy mean
field.\footnote{Lieb's theorem may be applicable to the Majorana representation,
but it works only within this representation.}  In order to
systematically compare energies for different mean-field assumptions, a variational
Monte Carlo (VMC) method~\cite{Gubernatis2016} is the most powerful numerical tool.
Although we will not review the technical details for this method, as well as tensor
network methods, we trust the results of VMC and infinite projected entangled-pair state
(iPEPS) calculations~\cite{Corboz2011,Corboz2012SUN}, and will not argue about
the appropriateness of their methods.

From combined VMC and iPEPS calculations, the $\mathrm{SU}(4)$ Heisenberg model
on the honeycomb lattice is expected to host a QSOL~\cite{Corboz2012}.
The state is roughly described
by a $\pi$-flux Schwinger-Wigner ansatz with an algebraic decay in correlation.
They compared 0-flux and $\pi$-flux states for both Schwinger-Wigner and Wang-Vishwanath
representations, and found the $\pi$-flux Schwinger-Wigner state has the lowest
energy, very close to the ground state energy.  Since the spectrum of this $\pi$-flux state
is described by a Dirac fermion (spinon) in the mean-field theory, the ground
state must be a Dirac spin liquid with doubly degenerate Dirac cones.  The gauge
structure is unknown in the previous study.
The Dirac cone spectrum is discussed in detail in Appendix~\ref{psg}.
If we use the language of spin-orbital systems, the unbroken $\mathrm{SU}(4)$ symmetry
makes two types of fractionalized excitations, spinons and orbitalons, equivalent.
This point would be discussed again in Chapter~\ref{sum}.

We note that this $\pi$-flux ansatz is consistent with the famous Affleck-Marston
argument~\cite{Affleck1988am},
though there is no reason to assume such guiding principles to find out the correct
flux sector. Indeed, a similar numerical analysis has been done for the hyperhoneycomb
lattice~\cite{Natori2018su4}, but it does not obey the Affleck-Marston rule.

\section{Cold atomic realization}

Motivated by theoretical interests, experiments to realize the $\mathrm{SU}(N)$ magnetism
in reality are also ongoing.  Usual spin systems only have the $\mathrm{SU}(2)$ symmetry
at most, so we have to seek for unusual experimental tools to increase the symmetry.
Approaching $\mathrm{SU}(3)$ quantum chromodynamics (QCD) requires a very high energy, so
we would need a low-energy effective $\mathrm{SU}(N)$ symmetry in table-top systems.
First, we would like to review the realization of $\mathrm{SU}(N)$ systems in ultracold atoms.
Although the main topic of this thesis is magnetic materials, atomic systems can also
be regarded as some quantum simulator of spin models.  Especially, the breakthrough
in optical technology enables us to make an optical lattice, and inside this optical
lattice we can simulate a periodic model Hamiltonian where atoms are hopping between
modulated effective potentials induced by light.  In order to realize ``a Mott insulator''
of atoms, we mainly focus on fermionic atom gases to realize spin models, where
correlated electrons are replaced by interacting atoms themselves.
This section follows a review paper~\cite{Cazalilla2014}.

Fermionic condensate can be realized in alkaline-earth atoms.  We also include
atoms like Yb into alkaline-earth atoms, though Yb is rare-earth.
Those atoms (Sr, Yb, etc.) are often used as Fermi gases, and we here focus on
alkaline-earth-atomic Fermi gases.
For an alkaline-earth atom in the symmetric ground state ($^1S_0$), there are no degree
of freedom with spin or orbital angular momentum, so
nuclear spin ($F > 0$) is decoupled from the electronic state due
to the absence of hyperfine interactions.
Because of the electronic-nuclear spin decoupling in the fermionic isotopes,
the scattering parameters involving the $^1S_0$
and $^3P_0$ states have to be independent of its nuclear spin.
Thus, in the so-called clock states, all of the scattering lengths become equal.
Under these conditions, the interaction and kinetic parts of the Hamiltonian are
\textit{emergently} $\mathrm{SU}(N)$-symmetric, where $N = 2F + 1.$  Especially,
$^{173}$Yb gases have the $\mathrm{SU}(6)$ symmetry and $N$ up to 10 is
likely to be feasible~\cite{Cazalilla2009,Gorshkov2010}.

In addition to isolated gases, condensed matter systems like
the $\mathrm{SU}(N)$ Hubbard (or Heisenberg) model can be implemented in optical lattices.
In order to simulate the periodic (Bloch) potential experienced by electrons
in crystalline systems, we can use ultracold atomic gases
by confining them in periodic arrays of light potentials~\cite{Giorgini2008}.
Thus, $\mathrm{SU}(N)$ physics discussed in previous sections can be realized
in cold atoms.

Though most Fermi gases on the optical lattice are treated by the $\mathrm{SU}(N)$ Hubbard
model, in reality there exists a symmetry-breaking term, even in an ideal setup.
In the case of $F=3/2$ and $N=4,$ the symmetry is reduced to
$\mathrm{SO}(5) \subset \mathrm{SU}(4)$ by additional interactions~\cite{Congjun2003}.
This is because the coupling of spin-3/2 and spin-3/2 results in two
independent interaction terms with a total spin-0 and spin-2.  We note
that spin-1 is impossible because of the statistics.  These terms
in the form of a 4-component spinor no longer have an $\mathrm{SU}(4)$ symmetry,
while they still have a (hidden) $\mathrm{SO}(5)$ symmetry.
This mathematical structure will be discussed again in Sec.~\ref{hund}.
It was proposed that $^{135}$Ba and $^{137}$Ba are close to an ideal
$\mathrm{SU}(4)$-symmetric line~\cite{Congjun2003}.  Thus, these atoms
are the most important candidates for $\mathrm{SU}(4)$ magnetism
in ultracold systems.

Though $\mathrm{SU}(N)$ Heisenberg models with an even $N$ may essentially be
realized in the optical lattice,\footnote{The realization of $\mathrm{SU}(N)$ with
an odd $N$ might still be difficult.} the realization in magnetic materials also
has many advantages because every technology accumulated for many decades in condensed
matter physics is directly applicable.  Magnetic materials can be investigated
in moderate environment and requires no extreme technology of cooling or a laser control.
From a theoretical perspective, a question ``what is a realistic spin-(orbital)
model feasible in real magnetic materials'' is an important unresolved problem,
though such problems are reduced to a technological one in cold atomic systems.
This perspective in condensed matter theory has long been neglected, and,
until Jackeli and Khaliullin~\cite{Jackeli2009} discovered iridates as candidate Kitaev spin liquids,
the importance of discussing the material realization of some ``designer'' Hamiltonian~\cite{Kaul2013}
was underestimated.  From now on, we will concentrate on such open questions
especially for $\mathrm{SU}(N)$ spin models.

\section{Spin-orbital systems and quantum spin-orbital liquids}

A spin-orbital system is another important candidate for $\mathrm{SU}(N)$ magnetism,
especially in the case of $N=4,$ as will be discussed in Chapter~\ref{main}.
Both spin and orbital degrees of freedom are angular momenta, so it is
a ``magnetic material'' in a usual sense.
Before going on to the realization of the $\mathrm{SU}(N)$
symmetry, we will review the previous studies on orbital physics.

A quantum orbital liquid (QOL) itself has been discussed in some literature~\cite{Keimer2000,Khaliullin2000}.
This notion is defined for a system where orbital degeneracy survives on some metal ion.
LaTiO$_3$ is an original candidate for this orbital liquid, an extension of the RVB theory
to the orbital sector active in the $d^1$ electronic configuration~\cite{Khaliullin2000}.
In the same spirit as QSL is a state without a magnetic transition,
if the Jahn-Teller (JT) transition does not break an effective symmetry
between multiple degenerate orbitals even at low temperature, the state
is usually called orbital liquid, especially QOL if this is due to the
quantum fluctuation/entanglement of orbital degrees of freedom.\footnote{We are not
sure whether a ``pure'' QOL is a well-defined notion because SOC in real materials
always mixes two degrees of freedom.}
There is a nice review paper for orbital physics in general~\cite{Tokura2000}.

A possibility that the orbital fluctuation enhances the spin fluctuation,
leading to a QSOL (quantum spin-orbital liquid), has been discussed for a long
time in the Kugel-Khomskii-type models~\cite{Feiner1997}, but finding a real
material candidate is not an easy task.  Though the coupling between spin and orbital
sectors is strong especially in the $d^9$ system, we need to confine a $d^9$
ion in a rigid octahedral cage to protect the orbital degeneracy.
As already discussed, BCSO (Ba$_3$CuSb$_2$O$_9$) is a prominent candidate for
a QSOL~\cite{Zhou2011,Nakatsuji2012,Corboz2012}, where both spin and orbital degrees of
freedom are fluctuating at the lowest temperature.  Based on the crystallographic structure
presented in Ref.~\citenum{Katayama2015}, both Cu and Sb ions are in a good octahedral
coordination.  Especially, Cu is in the $2+$ state with an orbital degeneracy between
the $d_{x^2-y^2}$ and $d_{z^2}$ orbitals which is as active as a spin degeneracy,
forming a decorated honeycomb lattice.  Thus, in this structure both spin and orbital
degrees of freedom can be unfrozen.  A QSOL realized in BCSO is a combination of a QSL and
a QOL.  Though there is a possibility that disorder plays an important role
in this material~\cite{Smerald2015}, experiments clearly show surviving quantum fluctuations
for both spin and orbital degrees of freedom.  In the case of BCSO,
finite-frequency electron spin resonance (ESR)~\cite{Han2015} and
extended X-ray absorption fine structure (EXAFS)~\cite{Nakatsuji2012} are used
to observe quantum orbital fluctuations dynamically.  They should still be
important tools, so we will discuss this experimental approach later again.

Previously, such orbital liquid states are thought to be impossible because
the fluctuation between two wavefunctions of different orbitals always couples
to the lattice motion (Jahn-Teller coupling).  Especially, an orbital liquid without
a cooperative JT order may abandon an energy gain $O(1000)$ times larger than that of QSLs.
This is because of the energy discrepancy between electronic and phononic
(lattice) degrees of freedom. 
However, as we shall see, in the case that a QOL stabilizes a symmetric
coordination of ligands (\textit{e.g.} octahedral coordination) and the lattice
(phonon) energy is still minimized at this symmetric coordination even with an
electronic fluctuation, the energy scale difference does not matter.

Though BCSO was a good candidate for QSOLs, the estimated parameters for BCSO are
rather far from the model with an exact $\mathrm{SU}(4)$ symmetry~\cite{Smerald2014}.
Moreover, SOC and the directional dependence of the orbital hopping usually
break both the spin-space and orbital-space $\mathrm{SU}(2)$ symmetries.
It would seem even more
difficult to realize an $\mathrm{SU}(N)$-symmetric system in real magnets with SOC,
and thus it is very challenging to find an $\mathrm{SU}(N)$ symmetry in materials
with a strong SOC.

The organization of this thesis is as follows.
In Chapter~\ref{main}, we first propose a honeycomb magnetic material with an emergent
$\mathrm{SU}(4)$ symmetry, derive its effective Hamiltonian, extend the discussion
to 3D systems, and give a new perspective on the protection of topological properties
by crystalline symmetries.  This part follows the organization of Ref.~\citenum{Yamada2018}.
In the latter half of Chapter~\ref{main}, we discuss the triangular lattice case,
boundary condition effects, a Hund coupling effect, and flux variables determination
for 3D tricoordinated lattices.
In Chapter~\ref{sum}, we first summarize the main contents, and then discuss another
candidate system for an $\mathrm{SU}(4)$ symmetry, called twisted bilayer
graphene/dichalcogenide.  Finally, Appendix~\ref{lsma} is discussing one extension
of the LSM theorem, and Appendix~\ref{psg} supplements the definition of a crystalline
(Kitaev) spin liquid with a concrete example.

\chapter{Emergent $\mathrm{SU}(4)$ symmetry and its realization}\label{main}

As we saw in the Introduction, $\mathrm{SU}(N)$ systems are new important candidates for QSLs,
but are restricted to some artificial systems like cold atoms.  Thus, we would
like to discuss a possible realization in magnetic materials.
In this chapter, we mostly focus on $\alpha$-ZrCl$_3$ and its low-energy effective
model. We also discuss how to generalize the result to other materials or lattices.
In addition to what were discussed in the Introduction, metal-organic frameworks (MOFs) are
another playground for
$\mathrm{SU}(4)$ magnetism, and a variety of candidate materials will enable us to seek
many unknown spin-orbital liquids beyond a honeycomb Dirac spin liquid.

\section{Honeycomb materials}

\begin{figure}
\centering
\includegraphics[width=10cm]{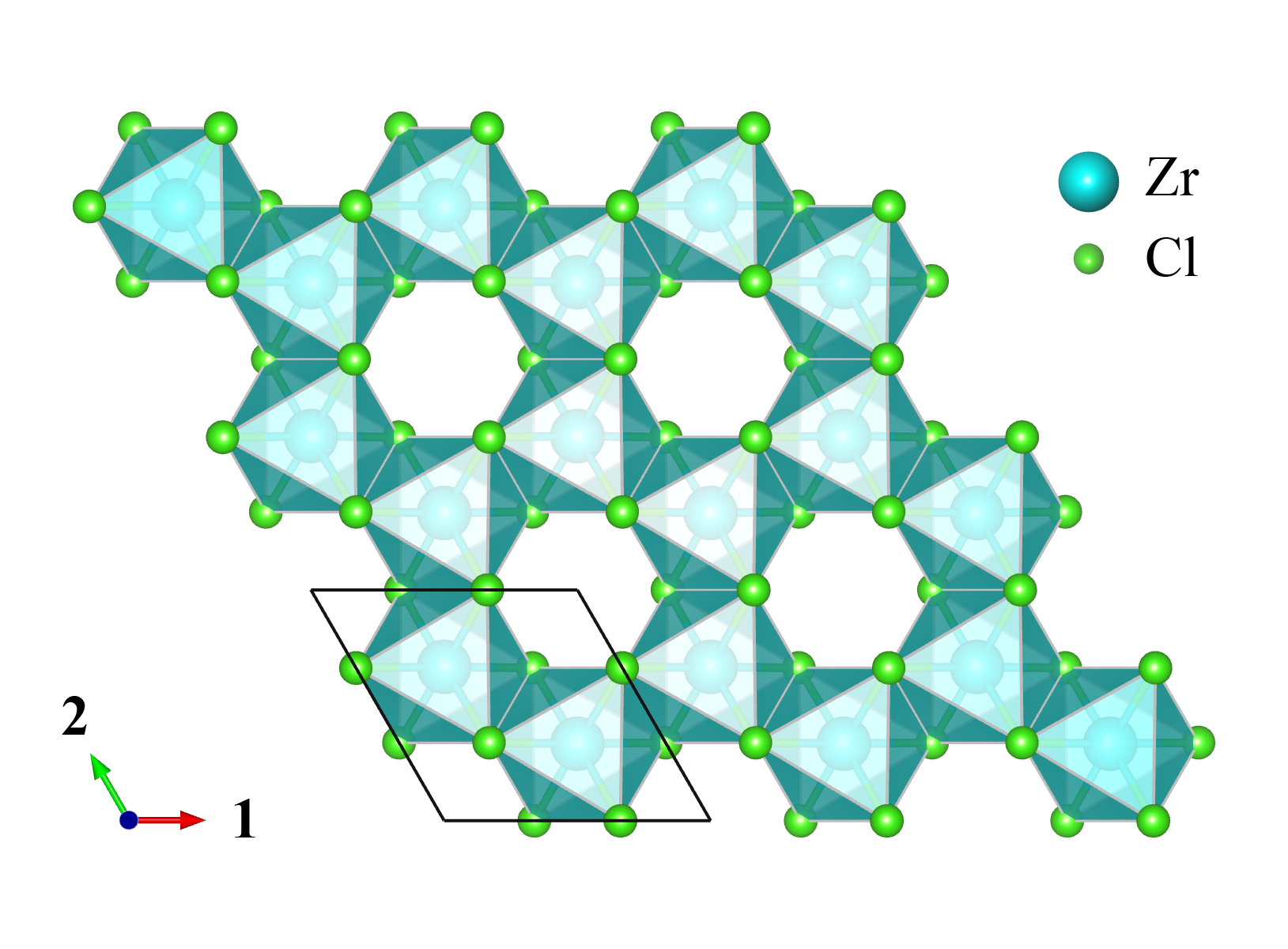}
\caption{Geometric structure of honeycomb $\alpha$-ZrCl$_3.$
Cyan and light green spheres represent Zr and Cl, respectively.
The crystallographic axes are shown and labelled as the 1- and 2-directions.
Reprinted figure with permission from \cite{Yamada2018} Copyright 2017 by the American Physical Society.}
\label{zrcl}
\end{figure}

As we mentioned in Chapter~\ref{intro}, we propose
$\alpha$-ZrCl$_3$ with a honeycomb geometry as the first candidate for
the $d^1$ honeycomb system, as shown in Fig.~\ref{zrcl}.
More generally, we consider the class of materials
$\alpha$-$MX_3$, with $M=$ Ti, Zr, Hf, etc., $X=$ F, Cl, Br, etc.
Their crystal structure is almost the same
as that of $\alpha$-RuCl$_3$ shown in Fig.~\ref{jackeli}(c),
which is known to be an approximate realization of the Kitaev
honeycomb model~\cite{Kitaev2006,Plumb2014}.
However, the electronic structure of $\alpha$-$MX_3$
is different from $\alpha$-RuCl$_3$: 
here, $M$ is in the $3+$ state with a $d^1$ electronic configuration in the
octahedral ligand field.
While in the $d^5$ configuration of Ru$^{3+}$ the $J_\textrm{eff}=1/2$ doublet is the
ground state, a $d^1$ electronic configuration which is a particle-hole inversion counterpart
will potentially realize the $J_\textrm{eff}=3/2$ ground state, as shown in
Fig.~\ref{jackeli}(a).
Our strategy for realizing $\mathrm{SU}(4)$ spin models starts with a low-energy
quartet of electronic states with the
effective angular momentum $J_\textrm{eff}=3/2$ on each $M.$

For this description to be valid, the SOC has to be strong enough.  As
the atomic number increases from Ti to Hf, SOC gets stronger and the
description by the effective angular momentum becomes more accurate.
The compounds $\alpha$-$M$Cl$_3$ with $M=$ Ti, Zr and related Na$_2$VO$_3$
have been already reported experimentally.
For $\alpha$-TiCl$_3,$ a structural transition and opening
of the spin gap at $T=217$ K have been reported~\cite{Ogawa1960}.
This implies a small SOC, as it is consistent with
a massively degenerate manifold of spin-singlets
expected in the limit of a vanishing SOC~\cite{Jackeli2007}. 
In compounds with heavier elements, the strong SOC can convert
this extensively degenerate manifold of product states into
a resonating quantum state. 
Thus, we expect realization of the $\mathrm{SU}(4)$ QSOL due to strong
SOC with metal ions heavier than Ti.  In the following, we pick up
$\alpha$-ZrCl$_3$ as an example, although the same analysis should apply
to $\alpha$-HfCl$_3,$ and $A_2M^\prime$O$_3$ ($A=$ Na, Li, etc.,
$M^\prime=$ Nb, Ta, etc.) as well.

\begin{figure}
\centering
\includegraphics[width=12cm]{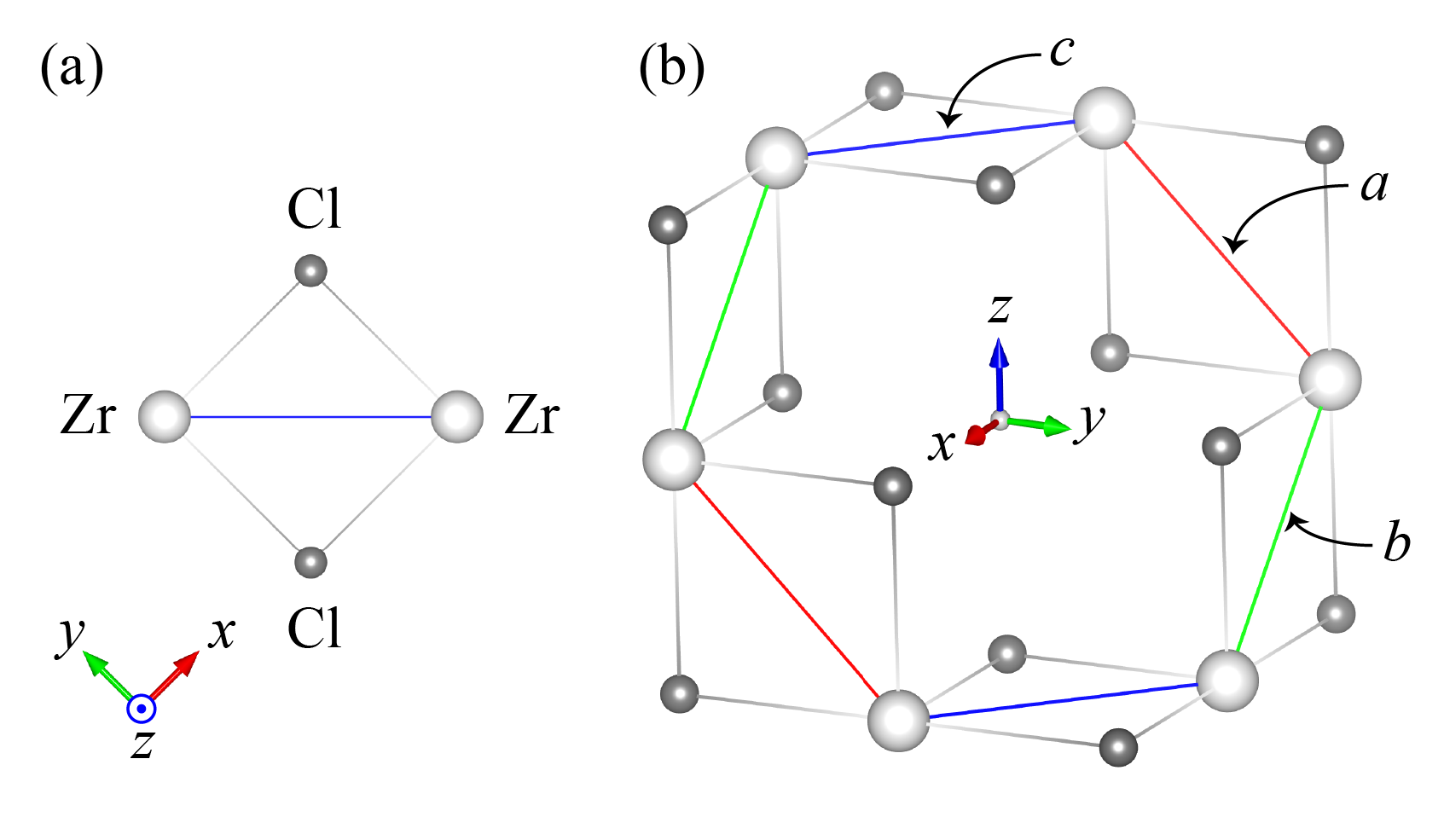}
\caption{(a) Superexchange pathways between two Zr ions connected by
a $c$-bond (blue) in $\alpha$-ZrCl$_3.$
White and grey spheres represent Zr and Cl atoms, respectively.
(b) Three different types of bonds in $\alpha$-ZrCl$_3.$
Red, light green, and blue bonds represent $a$-, $b$-, and $c$-bonds on the
$yz$-, $zx$-, and $xy$-planes, respectively.
Reprinted figure with permission from \cite{Yamada2018} Copyright 2017 by the American Physical Society.}
\label{honeycomb}
\end{figure}

\section{Effective Hamiltonian}\label{uij}

In the strong-ligand-field limit, the description with one electron
in the threefold degenerate $t_{2g}$-shell for $\alpha$-ZrCl$_3$ becomes accurate.
We denote these $d_{yz}$, $d_{zx}$, and $d_{xy}$-orbitals
by $a,$ $b,$ $c,$ respectively.
Let $a_{j\sigma},$ $b_{j\sigma}$ and $c_{j\sigma}$ represent annihilation operators on these orbitals on the $j$th site of Zr$^{3+}$ with spin-$\sigma$, and
$n_{\xi\sigma j}$ with $\xi \in \{a,b,c\}$ be the corresponding number operators.
We also use this $(a,\,b,\,c)=(yz,\,zx,\,xy)$ notation to label bonds:
each Zr --- Zr bond is called $\xi$-bond ($\xi=a,$ $b,$ $c$)
when the superexchange pathway is on the $\xi$-plane,\footnote{The Cartesian $xyz$ axes are defined as
in Fig.~\ref{honeycomb}(b).} as illustrated in Fig.~\ref{honeycomb}.

We define a $J_\textrm{eff}=3/2$ spinor as \mbox{$\psi = (\psi_{\uparrow \uparrow},\psi_{\uparrow \downarrow},\psi_{\downarrow \uparrow},\psi_{\downarrow \downarrow})^t = (\psi_{3/2},\psi_{-3/2},\psi_{1/2},\psi_{-1/2})^t ,$} where $\psi_{J^z}$ is the annihilation operator for the $\ket{J=3/2, J^z}$ state.
Assuming the SOC is the largest electronic energy scale, except for the ligand field splitting,
fermionic operators can be rewritten by the quartet
$\psi_{j\tau\sigma}$ as follows.
\begin{align}
        a_{j\sigma}^\dagger &\to \frac{\sigma}{\sqrt{6}} (\psi_{j\uparrow \bar{\sigma}}^\dagger-\sqrt{3}\psi_{j\downarrow \sigma}^\dagger), \label{Eq.a} \\
        b_{j\sigma}^\dagger &\to \frac{i}{\sqrt{6}} (\psi_{j\uparrow \bar{\sigma}}^\dagger+\sqrt{3}\psi_{j\downarrow \sigma}^\dagger), \label{Eq.b} \\
        c_{j\sigma}^\dagger &\to \sqrt{\frac{2}{3}}\psi_{j\uparrow\sigma}^\dagger, \label{Eq.c}
\end{align}
where the indices $\tau$ and $\sigma$ of $\psi_{j\tau\sigma}$ label the pseudoorbital
and pseudospin indices, respectively.  Here $\bar{\sigma}$ is an opposite spin to $\sigma.$
We begin from the following Hubbard Hamiltonian for
$\alpha$-ZrCl$_3.$
\begin{align}
        H =& -t \sum_{\sigma, \langle ij \rangle \in \alpha} (\beta_{i\sigma}^\dagger \gamma_{j\sigma}+\gamma_{i\sigma}^\dagger \beta_{j\sigma})+ h.c.
+ \frac{U}{2} \sum_{j, (\delta,\sigma) \neq (\delta^\prime,\sigma^\prime)} n_{\delta\sigma j}n_{\delta^\prime \sigma^\prime j}, \label{Eq.original}
\end{align}
where $t$ is a real-valued hopping parameter through the hopping shown in Fig.~\ref{honeycomb}(a), $U>0$ is the Hubbard interaction, $\langle ij \rangle \in \alpha$ means that the bond $\langle ij\rangle$ is
an $\alpha$-bond,
$\langle \alpha,\beta,\gamma \rangle$ runs over every cyclic permutation of $\langle a,b,c \rangle,$ and $\delta,\delta^\prime \in \{a,b,c\}.$
By inserting Eqs.~\eqref{Eq.a}-\eqref{Eq.c}, we get
\begin{equation}
        H= -\frac{t}{\sqrt{3}} \sum_{\langle ij \rangle} \psi_i^\dagger U_{ij} \psi_j +h.c.
        + \frac{U}{2} \sum_{j} \psi_j^\dagger \psi_j (\psi_j^\dagger \psi_j-1), \label{Eq.Hub1}
\end{equation}
where $\psi_j$ is the $J_\textrm{eff}=3/2$ spinor on the $j$th site,
and $U_{ij}=U_{ji}$ is a $4\times 4$ matrix
\begin{equation}
        U_{ij} = \begin{cases}
    U^a = \tau^y \otimes I_2 & (\langle ij \rangle \in a) \\
    U^b = -\tau^x \otimes \sigma^z & (\langle ij \rangle \in b) \\
    U^c = -\tau^x \otimes \sigma^y & (\langle ij \rangle \in c)
  \end{cases},
\end{equation}
where $\bm{\tau}$ and $\bm{\sigma}$ are Pauli matrices
acting on the $\tau$ and $\sigma$ indices
of $\psi_{j\tau\sigma},$ respectively.
We note that $U^{a,b,c}$ are unitary and Hermitian, and thus $U_{ji}={U_{ij}}^\dagger
= U_{ij}$.

Now we consider a (local) $\mathrm{SU}(4)$ gauge transformation,
\begin{equation}
        \psi_j \to g_j\cdot \psi_j, \qquad
        U_{ij} \to g_i U_{ij} g_j^\dagger,
\end{equation}
where $g_j$ is an element of $\mathrm{SU}(4)$ defined for each site $j$.
For every loop $C$ on the lattice, the $\mathrm{SU}(4)$ flux defined by the product $\prod_{\langle ij \rangle \in C} U_{ij}$ is invariant under the gauge transformation.

Remarkably, 
for each elementary hexagonal loop (which we call plaquette) $p$
in the honeycomb lattice with the coloring illustrated
in Fig.~\ref{honeycomb}(b),
\begin{equation}
        \prod_{\langle ij \rangle \in \hexagon_p} U_{ij}=U^a U^b U^c U^a U^b U^c=(U^a U^b U^c)^2 =-I_4,
\end{equation}
which corresponds to just an Abelian phase $\pi$.
Since all the flux matrices on the honeycomb lattice can be made of some product of
these plaquettes,
there is an $\mathrm{SU}(4)$ gauge transformation to reduce
the model~\eqref{Eq.Hub1} to the $\pi$-flux
Hubbard model $H$ with a global $\mathrm{SU}(4)$ symmetry, as proven in Sec.~\ref{boundary}.
\begin{equation}
        H = -\frac{t}{\sqrt{3}} \sum_{\langle ij \rangle} \eta_{ij} \psi_i^{\dagger} \psi_j + h.c.
        + \frac{U}{2} \sum_{j} \psi_j^{\dagger} \psi_j (\psi_j^{\dagger} \psi_j -1), \label{Eq.piflux1}
\end{equation}
where the definition of $\eta_{ij}=\pm 1$, arranged
to insert a $\pi$ flux inside each plaquette, is included in Sec.~\ref{boundary}.
At quarter filling, \textit{i.e.} one electron per site,
which is the case in $\alpha$-ZrCl$_3,$ the system
becomes a Mott insulator for a sufficiently large $U/|t|.$
We note that in this Mott regime other contributions, such as a Hund coupling, can be ignored
as discussed in Sec.~\ref{hund}.
The low-energy effective Hamiltonian for the spin and orbital degrees of
freedom, obtained by the second-order perturbation theory in $t/U,$
is the Kugel-Khomskii model exactly at the $\mathrm{SU}(4)$ point~\eqref{Eq.KK_SU4},
with $\bm{S}=\bm{\sigma}/2,$ $\bm{T}=\bm{\tau}/2,$ and $J=8t^2/(3U)$
in the transformed basis set.
We note that the effective Hamiltonian does not depend on the phase
factor $\eta_{ij}$, as it cancels out in the second-order perturbation
in $t/U$.
Corboz \textit{et al.} argued that this $\mathrm{SU}(4)$ 
Heisenberg model on the honeycomb lattice hosts a gapless
QSOL~\cite{Corboz2012}.
Therefore, we have found a possible realization of gapless QSOL 
in $\alpha$-ZrCl$_3$ with an \textit{emergent} $\mathrm{SU}(4)$ symmetry.

The nontrivial nature of this model
may be understood in terms of the LSMA theorem for 
the $\mathrm{SU}(N)$ spin systems~\cite{LSM1961,Affleck1986,Lajko2017,YHO2018},
generalized to higher dimensions~\cite{LSM1961,Affleck1988,Oshikawa2000,Hastings2005,Totsuka2017}.
As a result,
under the $\mathrm{SU}(N)$ symmetry and the translation symmetry,
the ground state of the $\mathrm{SU}(N)$ spin system with $n$ spins
of the fundamental representation  per unit cell cannot be unique,
if there is a non-vanishing excitation gap and $n/N$ is not an integer [see Appendix~\ref{lsma}].
This rules out a featureless Mott insulator phase,
which is defined as a gapped phase with a unique ground state,
namely without any spontaneous symmetry breaking or topological order.

For the honeycomb lattice $(n=2)$ there is no LSMA constraint
for an $\mathrm{SU}(2)$ spin system~\cite{Jian2016}.
Nevertheless, for the $\mathrm{SU}(4)$ spin system
we discuss in this thesis, a two-fold ground-state degeneracy
is required to open a gap.
This suggests the stability of a gapless QSOL phase of
the $\mathrm{SU}(4)$ Heisenberg model on the honeycomb lattice.
Especially, assuming the $\pi$-flux Dirac spin-orbital liquid ansatz
proposed in Ref.~\citenum{Corboz2012} is correct, a trivial mass gap for the Dirac spectrum
is forbidden unless the $\mathrm{SU}(4)$ or translation symmetry is broken.

\begin{figure}
\centering
\includegraphics[width=8.6cm]{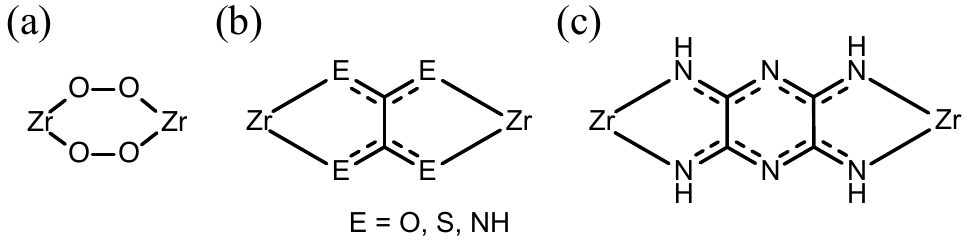}
\caption{Other possible superexchange pathways between two metal ions.
(a) Zr --- O --- O --- Zr.
(b) Oxalate-based metal-organic motif. ($E=$ O, S, NH.)
(c) Tetraaminopyrazine-bridged metal-organic motif.
Reprinted figure with permission from \cite{Yamada2018} Copyright 2017 by the American Physical Society.}
\label{super}
\end{figure}

\section{Other possible structures}

In addition to 3D inorganic polymorphs, MOFs with motifs listed in Fig.~\ref{super}
are an interesting playground to explore a variety of $\mathrm{SU}(4)$ QSOLs.
Actually, Kitaev spin liquids can be realized
in MOFs by a mechanism similar to the one in iridates~\cite{Yamada2017MOF,Jackeli2009}.
Since the present derivation of an emergent $\mathrm{SU}(4)$ symmetry shares
the same $t_{2g}$ hopping model as in Ref.~\citenum{Jackeli2009}, it is also expected
to apply to Zr- or Hf-based MOFs.
While Fig.~\ref{super}(a) is a longer superexchange pathway expected
in oxides similar to triangular iridates~\cite{Catuneanu2015},
Fig.~\ref{super}(b) and (c) show the superexchange pathways possible in
Zr- or Hf-based MOFs.  With these oxalate- or tetraaminopyrazine-based
ligands, we can expect the two independent superexchange pathways
similar to $\alpha$-ZrCl$_3$ as discussed in
Ref.~\citenum{Yamada2017MOF}.

Following the case of the honeycomb lattice, we can repeat the same analysis
to derive the effective spin-orbital model for each 3D tricoordinated lattice.
Recently, the classification of spin liquids on various tricoordinated lattices
attracts much attention, so it is worth investigating~\cite{Hermanns2015Weyl,Hermanns2015BCS,Obrien2016}.
All the tricoordinated lattices considered in this thesis are listed in
Table~\ref{lattice}.  This table is based on the classification of tricoordinated
nets by Wells~\cite{Wells1977}.  We use a Schl\"afli symbol $(p,c)$ to label a lattice,
where $p$ is the shortest elementary loop length of the lattice,
and $c=3$ means the tricoordination of the vertices.
For example, (6,3) is the 2D honeycomb lattice, and
all the other
lattices are 3D tricoordinated lattices, distinguished by additional letters
following Wells~\cite{Wells1977}.  $8^2.10$-$a$ is a nonuniform lattice and, thus,
the notation is different from the other lattices.

Generalizing the discussion on the honeycomb lattice, if the $\mathrm{SU}(4)$ flux for
any loop $C$ is reduced to an Abelian phase $\zeta_C = \pm 1$ as
$    \prod_{\langle ij \rangle \in C} U_{ij} =
\zeta_C I_4 \quad (\textrm{for}\,^\forall C)$, 
the Hubbard model acquires the $\mathrm{SU}(4)$ symmetry.
We have examined this for each lattice in Table~\ref{lattice},
where a checkmark is put on the $\mathrm{SU}(4)$ column
if the above condition holds. Details are included in Sec.~\ref{trico}.
Moreover, in order to form a stable structure with the present mechanism,
the bonds from each site must form 120 degrees and an octahedral coordination.
This condition is again checked for each lattice, and indicated
in the 120\degree~bond column~\cite{Obrien2016} of Table~\ref{lattice}.
We also put a checkmark on the LSMA column,
when the LSMA theorem implies a ground state degeneracy or
gapless excitations for the $\mathrm{SU}(4)$-symmetric Hubbard model. 
For example, the LSMA constraint applies to
the (8,3)-$b$ lattice, since $n/N=6/4$ is fractional.

\begin{table}[H]
        \centering
        \caption{\label{lattice}Tricoordinated lattices discussed in this thesis.
Space groups are shown in number indices. Nonsymmorphic ones are underlined.
$n$ is the number of sites per unit cell.}
        \begin{tabular}{ccccccc}
                Wells' notation & Lattice name & $\mathrm{SU}(4)$ & \mbox{120\degree} bond & $n$ & Space group & LSMA \\
                \hline
                (10,3)-$a$ & hyperoctagon & \checkmark\footnotemark[2] & \checkmark & 4 & \underline{\textbf{214}} & \checkmark\footnotemark[3] \\
                (10,3)-$b$ & hyperhoneycomb & \checkmark\footnotemark[2] & \checkmark & 4 & \underline{\textbf{70}} & \checkmark\footnotemark[3] \\
                (10,3)-$c$ & & $-$ & $-$ & 6 & \underline{\textbf{151}} & \checkmark \\
                (10,3)-$d$ & $-$ & \checkmark\footnotemark[2] & $-$ & 8 & \underline{\textbf{52}} & \checkmark\footnotemark[3] \\
                (9,3)-$a$ & hypernonagon & $-$ & $-$ & 12 & \textbf{166} & $-$ \\
                $8^2.10$-$a$ & $-$ & \checkmark & \checkmark & 8 & \underline{\textbf{141}} & $-$ \\
				(8,3)-$b$ & hyperhexagon & \checkmark & \checkmark & 6 & \textbf{166} & \checkmark\footnotemark[4] \\
                $-$ & stripyhoneycomb & \checkmark & \checkmark & 8 & \underline{\textbf{66}} & $-$ \\
                (6,3) & 2D honeycomb & \checkmark & \checkmark & 2 & & \checkmark\footnotemark[5]
        \end{tabular}
\end{table}
\footnotetext[2]{The product of hopping matrices along every elementary loop is unity, 
resulting in the $\mathrm{SU}(4)$ Hubbard model with zero flux.}
\footnotetext[3]{Nonsymmorphic symmetries of the lattice are enough to protect a QSOL state,
hosting an XSOL state.}
\footnotetext[4]{Although the model has a $\pi$ flux, with an appropriate gauge choice the unit cell is not enlarged.  Therefore, the LSMA theorem straightforwardly applies to the $\pi$-flux $\mathrm{SU}(4)$ Hubbard model.}
\footnotetext[5]{While the standard LSMA theorem is not effective for
the $\pi$-flux $\mathrm{SU}(4)$ Hubbard model here,
the magnetic translation symmetry works to protect a QSOL state~\cite{LRO2017}.}

\section{Crystalline spin liquids and crystalline spin-orbital liquids}

Crystalline spin liquids (XSL)~\cite{Yamada2017XSL} are defined
originally for Kitaev models and the discussion is in Ref.~\citenum{Yamada2017XSL}.
We would quickly review the definition and generalize this notion to
$\mathrm{SU}(4)$-symmetric models based on the LSMA theorem.

In the context of gapless Kitaev spin liquids as originally proposed in
Ref.~\citenum{Yamada2017XSL}, a crystalline spin liquid is
defined as a spin liquid state where a gapless point (or a gapped
topological phase) is protected not just by the unbroken time-reversal
or translation symmetry, but by the space group symmetry of the lattice.
This is a simple analogy with a topological crystalline insulator, where
a symmetry-protected topological order is protected by some space group symmetry.

Differently from topological crystalline insulators, the classification or identification
of crystalline spin liquids is not easy.  This is because a symmetry could be
implemented \textit{projectively} in spin liquids and the representation of the symmetry (action)
becomes a projective (fractionalized) one.  The classification depends not only on
its original symmetry of the lattice but also on its PSG, so there are a macroscopic
number of possible crystalline spin liquids.  The only thing we can do is to identify
the mechanism of the symmetry protection for each specific case.
In Ref.~\citenum{Yamada2017XSL}, two Kitaev spin liquids are identified, one with
3D Dirac cones, and the other with a nodal line protected by the lattice symmetry,
not by the time-reversal symmetry.  The former is discussed in Appendix~\ref{psg},
and the latter is a nodal-line spin liquid robust under the time-reversal breaking,
both of which are beyond the classification of Kitaev spin liquids based on the internal
symmetries~\cite{Obrien2016}.

Sometimes, however, extended LSM-type theorems can prove the existence of a gapless point
or a topological state in the gapped case.  Thus, the LSM theorem can potentially
prove that some spin liquid is XSL without a microscopic investigation,
if we ignore whether it is gapped or gapless~\cite{WPVZ2015}.
This is a subtle point, but LSM-type theorems
extended to include a nonsymmorphic symmetry is very powerful to discuss the
property of spin liquids abstractly [see also Appendix~\ref{lsma}].
We note that this type of spin liquids are called filling-enforced QSLs in Ref.~\citenum{MT}

\begin{figure}
\centering
\includegraphics[width=\textwidth]{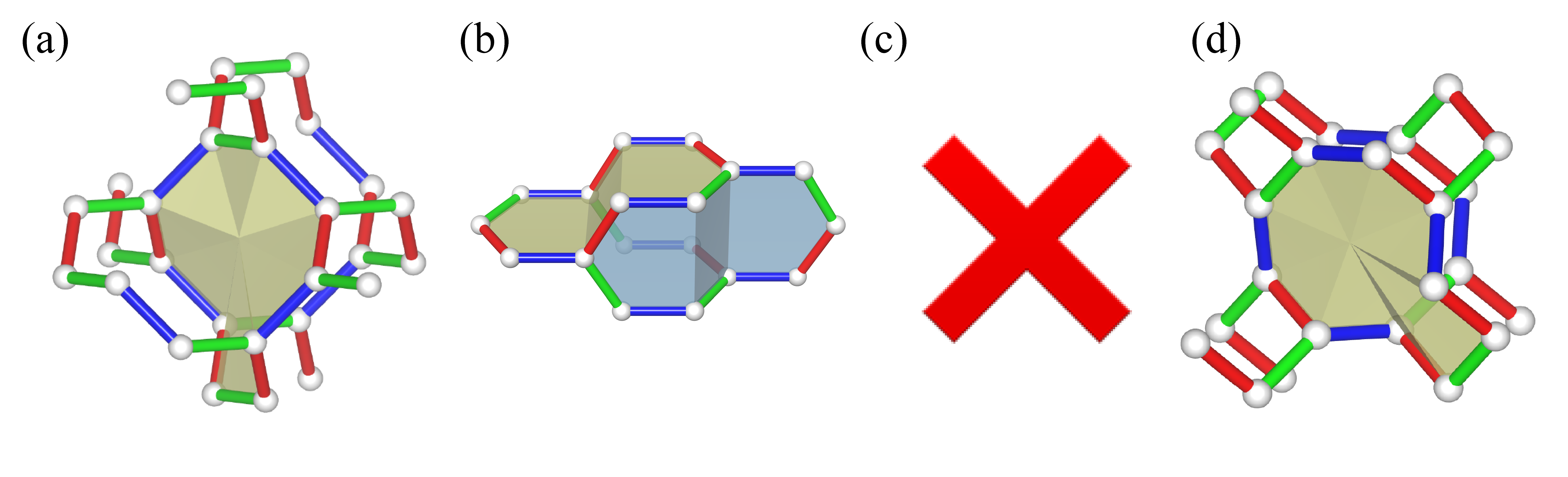}
\caption{(10,3) lattices.
(a) (10,3)-$a$ hyperoctagon lattice.
(b) (10,3)-$b$ hyperhoneycomb lattice.
(c) (10,3)-$c$ does not support the $\mathrm{SU}(4)$ symmetry.
(d) (10,3)-$d$ lattice.}
\label{xsol}
\end{figure}

Next, we would like to discuss the generalization of the concept of
XSL to $\mathrm{SU}(4)$-symmetric models.
In the (10,3) lattices [see Fig.~\ref{xsol}] listed in Table~\ref{lattice}, the unit cell
consists of a multiple of 4 sites, and thus the generalized LSMA theorem
seems to allow a featureless insulator if we only consider the translation.
Following Refs.~\citenum{PTAV2013,WPVZ2015,PWJZ2017}, however,
we can effectively reduce the size
of the unit cell by dividing the unit cell by the nonsymmorphic
symmetry, and thus the filling constraint becomes tighter with a
nonsymmorphic space group.  Even in the (10,3) lattices, the gapless
QSOL state can be protected by the further extension of the LSMA
theorem. We call them crystalline spin-orbital liquids
(XSOLs) in the sense that these exotic phases are protected
in the presence of both the $\mathrm{SU}(4)$
symmetry and (nonsymmorphic) space group symmetries.  We put a checkmark
on the LSMA column of Table~\ref{lattice} if either the standard or
extended LSMA theorem applies.

\section{Triangular $d^1$ system}

\begin{figure}
\centering
\includegraphics[width=6cm]{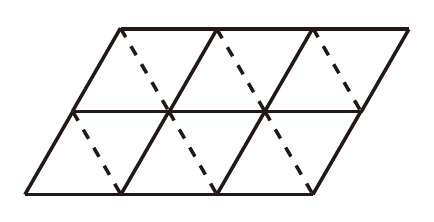}
\caption{Triangluar $d^1$ model.  Solid bonds have the $\mathrm{SU}(4)$ Heisenberg
interaction, but dashed bonds have an exotic interaction Eq.~\eqref{Eq.exotic}.
If we ignore dashed bonds, it becomes the $\mathrm{SU}(4)$
Heisenberg model on the square lattice~\cite{Corboz2011}.
}
\label{tri}
\end{figure}

It would be interesting to investigate $\mathrm{SU}(4)$
Heisenberg models on nontricoordinated lattices.  Especially, on the
lattice with 1 or 3 sites per unit cell, the LSMA theorem can
exclude the possibility of a simply gapped $\mathbb{Z}_2$ spin liquid
and suggests a $\mathbb{Z}_4$ QSOL or new SET phases instead. This can be understood by applying
the proof of the LSMA theorem to a cylinder boundary condition
because the fourfold GSD on a cylinder suggests the existence of a
gapless edge mode, or a topological order beyond $\mathbb{Z}_2$ topological order,
for example.
The case of the triangular lattice is also mentioned in Ref.~\citenum{Natori2018su4}.

From now on, we only consider a triangular lattice case for simplicity
because it may be relevant to some accumulated
graphene/transition metal dichalcogenide (TMDC) systems~\cite{Constantin2019}.
We can easily expect the existence of an unknown spin liquid state even for
the $\mathrm{SU}(4)$ Heisenberg model on the triangular lattice.  However, unfortunately
real triangluar $d^1$ systems cannot host an exact $\mathrm{SU}(4)$ Heisenberg model.
Instead, we found a new ``$\Gamma^5$'' flux inside each triangluar plaquette
and the resulting spin-orbital model becomes exotic, reflecting this additional
(non-Abelian) flux.

Similarly to Ba$_3$IrTi$_2$O$_9$~\cite{Catuneanu2015}, we can imagine
a triangular $d^1$ system as a starting point.  In this case, each triangular
plaquette binds the following flux:
\begin{align}
    \prod_{\langle ij \rangle \in \triangle} U_{ij}=U^a U^b U^c =: i\Gamma^5.
\end{align}
We note that the representation of $\Gamma^5$ here is different from Sec.~\ref{hund}.
For simplicity, we use a chiral representation as follows:
\begin{align}
    \Gamma^5=-\tau^z \otimes I_2 =
    \begin{pmatrix}
        -I_2 & 0 \\
        0 & I_2
    \end{pmatrix}.
\end{align}

A gauge transformation can always concentrate a flux matrix to only one bond
for each triangular plaquette, so it is enough to focus on one bond $\langle ij \rangle$ with
$U_{ij} = i\Gamma^5$ in order to derive an effective spin-orbital model by the second-order
perturbation in $t/U.$  The rest of the bonds are all $\mathrm{SU}(4)$-symmetric, in which case
the discussion is completely parallel to the honeycomb case.  As for a bond with
$U_{ij} = i\Gamma^5,$ the second-order perturbation leads to the following spin-orbital
model:
\begin{equation}
    H_{ij} = J\Bigl(\bm{S}_i \cdot \bm{S}_j+\frac{1}{4}\Bigr)\Bigl(T_i^z T_j^z-T_i^x T_j^x-T_i^y T_j^y+\frac{1}{4}\Bigr), \label{Eq.exotic}
\end{equation}
if $\langle ij \rangle$ is a dashed bond shown in Fig~\ref{tri}.
This term breaks the $\mathrm{SU}(4)$ symmetry, but still has a high symmetry, $\mathrm{SU}(2)
\times \mathrm{SU}(2).$
We can expect an exotic frustration, which is absent even in the $\mathrm{SU}(N)$ Heisenberg
model.  This is a new Hamiltonian which we first derived, and there is no previous study
for this model, so it is worthwhile to study it in the future.

Discussions here are essentially relevant to 1T-TaS$_2$~\cite{Law2017,Yu2017,Murayama2020}
in a symmetric phase without a structural distortion.  However, it is usually regarded as a
spin-1/2 system after the charge density wave transition.  If the symmetric
phase survives at very low temperature, 1T-TaS$_2$ should also be an
important playground for the quasi-$\mathrm{SU}(4)$ magnetism.

NaZrO$_2$ is also a candidate for the same triangular $d^1$ state,
though the density functional theory (DFT) claims that it is
a nonmagnetic metallic state~\cite{Assadi2018}.  It could possibly
lead to the above model after the Mott transition.
A DFT study for LiZrO$_2$ was also found~\cite{Singh2004}.

\section{Boundary condition effects on the $\mathrm{SU}(N)$ gauge transformation}\label{boundary}

Until here we concentrate on the physical realization and implication, but from now
on we will discuss more about the mathematical structure of our theory.
In this section, we would like to discuss the mathematical construction of the gauge transformation.
First, we begin from the 1D Hubbard model with an open boundary condition (OBC).
\begin{equation}
		H_\textrm{1DOBC}=-t\sum_{j=1}^{L-1} \psi_j^\dagger U_{j,j+1} \psi_{j+1} + h.c. + \frac{U}{2} \sum_{j=1}^L \psi_j^\dagger \psi_j (\psi_j^\dagger \psi_j-1),
\end{equation}
where $L$ is a system size, $\psi_j$ is a $N$-component spinor, $U_{j,j+1}$ is an $N \times N$
unitary matrix defined on the $j$th site, and $t$ and $U$ are real-valued hopping and Hubbard
terms, respectively.
The (local) gauge transformation is simply given by the following string operator $g_j.$
\begin{align}
		g_j &= \prod_{k=1}^{j-1} U_{k,k+1}, \\
        \psi_j^\prime &= g_j\cdot \psi_j, \\
		U_{j,j+1}^\prime &= g_j U_{j,j+1} g_{j+1}^\dagger = I_N.
\end{align}
Thus, 1D Hubbard model with OBC is
a trivial case where we can always make it $\mathrm{SU}(N)$-symmetric.
\begin{equation}
		H_\textrm{1DOBC}=-t\sum_{j=1}^{L-1} \psi_j^{\prime\dagger} \psi_{j+1}^\prime + h.c. + \frac{U}{2} \sum_{j=1}^L \psi_j^{\prime\dagger} \psi_j^\prime (\psi_j^{\prime\dagger} \psi_j^\prime-1),
\end{equation}

Therefore, in 1D electronic systems on a linear chain with
nearest-neighbor hoppings only,
if the $N\times N$ hopping matrices are all unitary,
the tight-binding Hubbard model is trivially gauge-equivalent to
the 1D $\mathrm{SU}(N)$ Hubbard model~\cite{Arovas1995,Pati1998,Itoi2000,Kugel2015}.
Such emergence of the $\mathrm{SU}(N)$ symmetry by the gauge transformation
becomes more nontrivial in higher dimensions because there is a
topological obstruction coming from the lattice geometry
and also a possibility to realize topological ground state
degeneracy, which is impossible in 1D systems~\cite{Chen2011}.

Before going to higher dimensions, it is
instructive to consider the 1D Hubbard model with a periodic boundary condition (PBC).
\begin{equation}
		H_\textrm{1DPBC}=-t\sum_{j=1}^{L} \psi_j^\dagger U_{j,j+1} \psi_{j+1} + h.c. + \frac{U}{2} \sum_{j=1}^L \psi_j^\dagger \psi_j (\psi_j^\dagger \psi_j-1),
\end{equation}
where $\psi_{L+1}$ is identified as $\psi_1.$  Clearly the gauge transformation
does not change the flux inside the loop, so there is a necessary condition
to have a gauge transformation which makes the Hamiltonian $\mathrm{SU}(N)$-symmetric,
\begin{equation}
        \prod_{j=1}^L U_{j,j+1} = \zeta I_N,
\end{equation}
with some $|\zeta|=1.$
This is also a sufficient condition.  If we apply the same gauge transformation
$g_j = \prod_{k=1}^{j-1} U_{k,k+1}$ as the OBC case for $j=1,\dots,L,$
the transformed matrices become
\begin{equation}
		U_{j,j+1}^\prime = \begin{cases}
				\prod_{k=1}^L U_{k,k+1} = \zeta I_N & (j=L) \\
				I_N & (\textrm{otherwise})
\end{cases}.
\end{equation}
Thus, the resulting Hamiltonian is completely $\mathrm{SU}(N)$-symmetric with a factor $\zeta,$
\begin{equation}
		H_\textrm{1DPBC}=-t\Bigl( \sum_{j=1}^{L-1} \psi_j^{\prime\dagger} \psi_{j+1}^\prime + \zeta \psi_L^{\prime\dagger} \psi_{1}^\prime \Bigr) + h.c. + \frac{U}{2} \sum_{j=1}^L \psi_j^{\prime\dagger} \psi_j^\prime (\psi_j^{\prime\dagger} \psi_j^\prime-1).
\end{equation}
It must be noted that $\zeta$ cannot be eliminated by any gauge transformation and
thus it is physical and called (magnetic) flux.

\begin{figure}
\centering
\includegraphics[width=12cm]{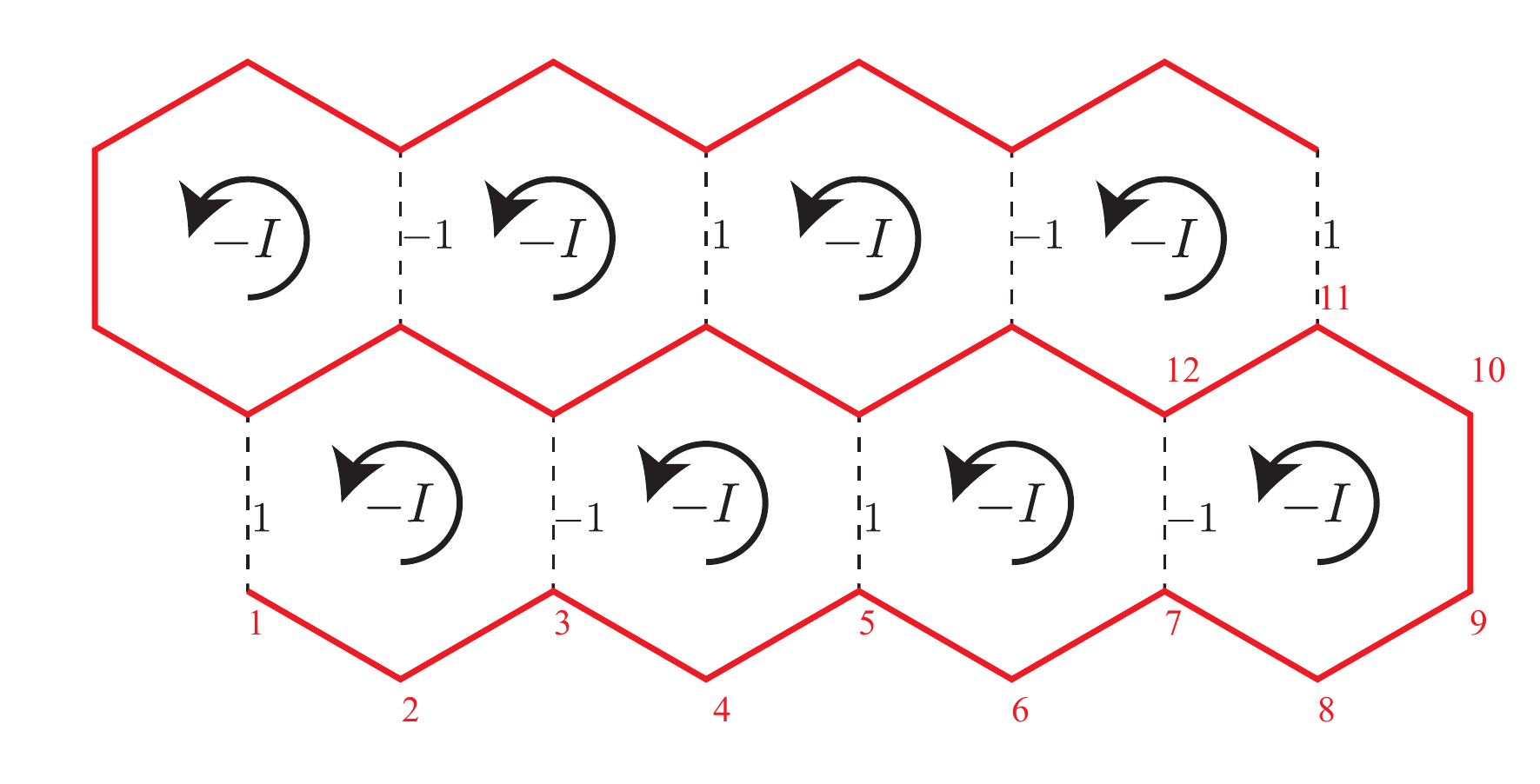}
\caption{Flake of the honeycomb lattice to show how the gauge transformation
works for OBC.  Along the red solid line, we used 1D gauge transformation and
the flux constraints automatically determines the transformed hopping matrices for
the rest of the bonds shown in black dashed lines.}
\label{gauge}
\end{figure}

As for OBC, it is almost trivial to extend the proof of the existence of the
gauge transformation to higher dimensions.  This can be achieved by drawing the lattice with
a single stroke of the brush.  For simplicity, we use the finite-size 2D
honeycomb lattice
with OBC.  We begin from the following Hamiltonian.
\begin{equation}
		H_\textrm{2D}= -\frac{t}{\sqrt{3}} \sum_{\langle ij \rangle} \psi_i^\dagger U_{ij} \psi_j +h.c.
        + \frac{U}{2} \sum_{j} \psi_j^\dagger \psi_j (\psi_j^\dagger \psi_j-1), \label{Eq.Hub}
\end{equation}
where $U_{ij}$ is again an $N\times N$ unitary matrix defined for each bond,
and $\psi_j$ is the $N$-component spinor on the $j$th site.
Assuming each site is numbered in order
for some nearest-neighbor site to have the subsequent number,
we can do the same gauge transformation as the 1D OBC case.
Again, this gauge transformation does not change the flux value for any loops,
so there is a necessary condition to get an $\mathrm{SU}(N)$-symmetric model for each hexagonal
plaquette (elementary loop) $p.$
\begin{equation}
		\prod_{\langle ij \rangle \in p} U_{ij} = \zeta_p I_N \qquad (\textrm{for}\,^\forall p). \label{Eq.flux}
\end{equation}
This condition is actually sufficient for OBC (assuming the existence of a single stroke
path).  We take a flake of the honeycomb lattice shown in Fig.~\ref{gauge}.
For simplicity, we use $\zeta_p = -1$ for $\alpha$-ZrCl$_3,$
but $\zeta_p$ can generally depend on each plaquette $p.$

If we draw a single stroke path shown as the red solid line in Fig.~\ref{gauge},
all the unitary matrices on the red bonds become identity by the gauge transformation
for the 1D red line.  Remaining are black dashed bonds, but their hopping matrices
are fixed by the flux condition (Eq.~\eqref{Eq.flux}).  In the case of Fig.~\ref{gauge},
around the bottom plaquettes the hopping matrices are determined from right to left
because five of the surrounding matrices are made identity one by one for each plaquette.
By continuing this, all the unitary matrices are transformed into some $\eta_{ij}$ times
identity with $|\eta_{ij}|=1,$ and thus the Hamiltonian becomes completely $\mathrm{SU}(N)$-symmetric. We call the following transformed gauge ``theorists' gauge''.
\begin{equation}
		H_\textrm{2D} = -\frac{t}{\sqrt{3}} \sum_{\langle ij \rangle} \eta_{ij} \psi_i^{\prime \dagger} \psi_j^\prime + h.c.
        + \frac{U}{2} \sum_{j} \psi_j^{\prime\dagger} \psi_j^\prime (\psi_j^{\prime\dagger} \psi_j^\prime-1), \label{Eq.piflux}
\end{equation}
where $\eta_{ij}=1$ for red bonds, while the sign of $\eta_{ij}=\pm 1$ depends on each bond
for black dashed bonds as indicated in Fig.~\ref{gauge} by the number near each black dashed bond.
This is nothing but the model called a $\pi$-flux Hubbard model on the honeycomb lattice
and the model can be constructed by changing the sign of the $c$-bonds alternately
along the perpendicular direction.  This gauge transformation effectively doubles
the size of the unit cell.

Finally, we would like to discuss the 2D PBC case.  In this case, we cannot find a
gauge transformation, even if we assume the flux condition (Eq.~\eqref{Eq.flux}) for
every hexagonal plaquette.  The final obstructions
to be considered are global (or topological) ones, which are two types of noncontractible
loops on the 2D torus.  The noncontractible loops in the same homotopy class are related by
the flux conditions, so it is enough to consider only two noncontractible loops $C_1$ and
$C_2$ along the $1$- and $2$-directions, respectively.  Assuming the size of the torus
to be $L_1 \times L_2$ original unit cells, the lengths of $C_1$ and $C_2$ become
multiples of $L_1$ and $L_2,$ respectively.  The necessary and sufficient conditions
to find a gauge transformation in addition to Eq.~\eqref{Eq.flux} are two new
flux conditions for $C_1$ and $C_2,$
\begin{equation}
		\prod_{\langle ij \rangle \in C_1} U_{ij} = \zeta_{C_1} I_N, \qquad \prod_{\langle ij \rangle \in C_2} U_{ij} = \zeta_{C_2} I_N.
\end{equation}

In general these fluxes cannot be Abelian for any sets of unitary matrices $U_{ij}.$
Thus, we specifically consider the model of $\alpha$-ZrCl$_3$ discussed previously.
In this model, all the hopping matrices are accidentally written by Pauli matrices, and
their products only take some Pauli matrices times a complex number, which actually
only takes $1,i,-1,-i.$  In other words, their products are included in the Pauli group
on 2 qubits.  In this group, any element to the power of 4 becomes identity, so
the flux inside the two noncontractible loops become trivial if both $L_1$ and $L_2$ are
multiples of 4.  This is a condition to find a gauge transformation to make the model
explicitly $\mathrm{SU}(N)$-symmetric with a symmetric boundary condition, \textit{i.e.}
a boundary
condition where both $C_1$ and $C_2$ have a zero flux.  If we allow a more general
boundary condition with a $\pi$ flux inside $C_1$ or $C_2,$ then the conditions
for $L_1$ or $L_2$ become milder.

Our effective model for the honeycomb $\alpha$-ZrCl$_3$ was derived based on
the superexchange interactions between the Zr$^{3+}$ ions constructed
from its geometry.  However, similar superexchange interactions can also arise
in the other structures listed in Fig.~\ref{super}, or in face-shared systems.
We note that ZrCl$_3$ has some polymorphs and a chain compound $\beta$-ZrCl$_3$
with face-shared Cl octahedra~\cite{Watts1966} can also host
a 1D $\mathrm{SU}(4)$ Heisenberg model~\cite{Kugel2015}.

Since a nonlayered structure of Na$_2$VO$_3$ has already been reported~\cite{Rudorff1956},
we can expect various 3D polymorphs
of ZrCl$_3$ or $A_2M^\prime$O$_3$ with $A=$ Na, Li and $M^\prime=$ Nb, Ta, similarly
to 3D $\beta$-Li$_2$IrO$_3$~\cite{Takayama2015} and $\gamma$-Li$_2$IrO$_3$~\cite{Modic2014}.

The generalization from the 2D case to the 3D case is straightforward.
The difference
is that in 3 dimensions not all the fluxes of the plaquettes (or elementary loops in Sec.~\ref{trico})
can be determined independently.  This is called volume constraint and will
be discussed in Sec.~\ref{trico}.

\section{Hidden $\mathrm{SO}(4)$ symmetry in the Hund coupling}\label{hund}

In reality, the multiorbital Hubbard model is not as simple as that with a Hubbard interaction
which has been discussed in previous sections.  The multiorbital Hubbard model usually
includes four interaction terms $U,$ $U^\prime,$ $J_H,$ and $J_H^\prime.$  As discussed
by Kanamori~\cite{Kanamori1963}, $U$ and $U^\prime$ have a similar magnitude, while
$J_H$ and $J_H^\prime (\sim J_H)$ are much smaller because they are from the exchange
integral between different $d$-orbitals.  Thus, it is natural to begin by assuming
$U = U^\prime$ and $J_H = J_H^\prime = 0$ as the first approximation as was done
so far, though we must consider $U^\prime - U \sim 2J_H$ and Hund couplings to be
perturbations of an order $J_H/U \sim \mathcal{O}(0.1).$  We assume $J_H/U \sim 0.1$
in $\alpha$-ZrCl$_3.$  At least from the stability condition $J_H/U$ has to be smaller
than $1 / 3$ in any case.

We here only consider an onsite Hund coupling $J_H=J_H^\prime,$ for simplicity.
There are other possible perturbations like further-neighbor
interactions, but we can expect that such effects are smaller than that of the Hund coupling
similarly to $\alpha$-RuCl$_3.$  Actually, in Kitaev materials like
$\alpha$-RuCl$_3$ the nearest-neighbor Kitaev interaction and the third-neighbor
Heisenberg interaction are expected to be comparable~\cite{Winter2016},
but this is probably due to fine tuning
happening in the $J_\textrm{eff}=1/2$ manifold and the Kiteav interaction has to be
smaller than the na\"ive superexchange interaction expected in the whole $t_{2g}$ orbitals
because of the destructive interference which cancels out the direct hopping between
the $J_\textrm{eff}=1/2$ manifold~\cite{Jackeli2009}.
In our $J_\textrm{eff}=3/2$ models realized \textit{e.g.} in $\alpha$-ZrCl$_3,$ such an accidental
reduction of the highest-order contribution
does not occur even in the nearest-neighbor interactions, so we expect the magnetic interaction
in $\alpha$-ZrCl$_3$ is much larger than the dominant Kiteav interaction in $\alpha$-RuCl$_3,$
and thus one- or two-order larger than the third-neighbor Heisenberg interactions in the
case of $\alpha$-ZrCl$_3.$

Next, in order to evaluate the effect of the Hund coupling,
we will change the ordering of the $J_\textrm{eff}=3/2$ bases to compare the model
with a so-called $\mathrm{SO}(5)$-symmetric Hubbard model discussed in the literature
on $S=3/2$ cold atomic systems~\cite{Congjun2003,Congjun2005,Congjun2006},
\begin{equation}
		\psi = (\psi_{3/2},\psi_{1/2},\psi_{-1/2},\psi_{-3/2})^t = (\psi_{\uparrow \uparrow},\psi_{\downarrow \uparrow},\psi_{\downarrow \downarrow},\psi_{\uparrow \downarrow})^t.
\end{equation}
In this basis it is easy to see a hidden $\mathrm{SO}(4)$ symmetry,
which is a subgroup of $\mathrm{SO}(5) \simeq \mathrm{Sp}(4) \subset \mathrm{SU}(4)$
in the original model.

We will now show that the Hund coupling in $\alpha$-ZrCl$_3$ actually possesses
the $\mathrm{SO}(5) \simeq \mathrm{Sp}(4)$ symmetry, although the hopping matrices
break a part of this symmetry. If we add a Hund coupling for the hopping model inside
the $t_{2g}$ orbitals~\cite{Georges2013}, the Hamiltonian becomes
\begin{align}
        H =& -t \sum_{\sigma, \langle ij \rangle \in \alpha} (\beta_{i\sigma}^\dagger \gamma_{j\sigma}+\gamma_{i\sigma}^\dagger \beta_{j\sigma})+ h.c. \nonumber \\
		&+ \sum_{j} \left[ \frac{U-3J_H}{2} N_j(N_j-1) -2J_H \bm{s}_j^2 -\frac{J_H}{2} \bm{L}_j^2 +\frac{5}{2}J_H N_j \right],
\end{align}
where $\langle ij \rangle \in \alpha$ means that the bond $\langle ij\rangle$ is
an $\alpha$-bond,
$\langle \alpha,\beta,\gamma \rangle$ runs over every cyclic permutation of $\langle a,b,c \rangle,$ $N_j$ is a number operator, $\bm{s}_j$ is a total spin, and $\bm{L}_j$ is a total
effective angular momentum.  In this form the stability condition $J_H/U<1/3$ is apparent.
Assuming a strong spin-orbit coupling limit $\lambda \gg |t|,\,J_H,$
we project the Hilbert space onto the $J_\textrm{eff}=3/2$ manifold.
We note that we will ignore doublon/holon excitations with higher energies in the following discussions.
In the original gauge before the gauge transformation,
which we call ``lab gauge'', the projected Hamiltonian becomes
\begin{equation}
        H= -\frac{t}{\sqrt{3}} \sum_{\langle ij \rangle} \psi_i^\dagger V_{ij} \psi_j +h.c.
		+ \sum_{j} \left[ \frac{U-3J_H}{2} \psi_j^\dagger \psi_j (\psi_j^\dagger \psi_j-1)-
		\frac{4}{9}J_H \bm{J}_j^2 +\frac{5}{2}J_H \psi_j^\dagger \psi_j \right],
\end{equation}
where $\bm{J}_j = \bm{s}_j + \bm{L}_j$ is a total effective angular momentum operator
with a condition $J=3/2$ after the projection, and
\begin{equation}
V_{ij} = \begin{cases}
    V^a = \tau^z \otimes \sigma^y = \Gamma^3 & (\langle ij \rangle \in a) \\
    V^b = -\tau^z \otimes \sigma^x = -\Gamma^2 & (\langle ij \rangle \in b) \\
    V^c = -\tau^y \otimes I_2 = \Gamma^1 & (\langle ij \rangle \in c)
  \end{cases}.
\end{equation}
We used $\bm{s}_j=\bm{J}_j/3$ and $\bm{L}_j=2\bm{J}_j/3$ inside the $J_\textrm{eff}=3/2$ manifold
derived from the Wigner-Eckart theorem. Thus, ignoring the hopping terms,
the Hubbard and Hund couplings possess a hidden $\mathrm{SO}(5) \simeq \mathrm{Sp}(4)$ symmetry
in the same way as the $S=3/2$ cold atomic systems with a spin-preserving interaction.

The hopping term partially breaks this $\mathrm{SO}(5)$ symmetry. To see this
we use anticommuting Dirac gamma matrices in Ref.~\cite{Congjun2003} defined as
\begin{equation}
(\Gamma^1,\Gamma^2,\Gamma^3,\Gamma^4,\Gamma^5) = (-\tau^y \otimes I_2, \tau^z \otimes \sigma^x, \tau^z \otimes \sigma^y, \tau^z \otimes \sigma^z, -\tau^x \otimes I_2).
\end{equation}
Gamma matrices $\Gamma^p$ ($p=1,\dots,5$) are forming an $\mathrm{SO}(5)$ vector, which
transforms as a vector in the same rotation for the hidden $\mathrm{SO}(5)$ symmetry
of the Hund coupling. There is no way to eliminate the non-Abelian hopping just by the
$\mathrm{SO}(5) \simeq \mathrm{Sp}(4)$ gauge transformation, but we can rotate
$\mathrm{SO}(5)$ vectors locally to eliminate the bond dependence of the hopping.

For example, we can rotate all $V_{ij}$s to $\Gamma^5$ and then the Hamiltonian
becomes almost uniform up to the same factors $\eta_{ij}=\pm 1$ as discussed in Sec.~\ref{boundary}:
\begin{equation}
		H = -\frac{t}{\sqrt{3}} \sum_{\langle ij \rangle} \eta_{ij} \psi_i^{\prime \dagger} \Gamma^5 \psi_j^\prime + h.c.
		+ \sum_{j} \left[ \frac{U-3J_H}{2} \psi_j^{\prime \dagger} \psi_j^\prime (\psi_j^{\prime \dagger} \psi_j^\prime-1)-
		\frac{4}{9}J_H \bm{J}_j^{\prime 2} +\frac{5}{2}J_H \psi_j^{\prime \dagger} \psi_j^\prime \right]. \label{eq.so5}
\end{equation}
This model explicitly has a hidden $\mathrm{SO}(4)$ symmetry because $\Gamma^5$ is
invariant under the $\mathrm{SO}(4)$ subgroup of the $\mathrm{SO}(5)$ rotation
which keeps a vector $(0,0,0,0,1)$ invariant.
The last term is constant in the large $(U-3J_H)$ limit at quarter filling,
so the first meaningful contribution of an order $J_H/U \sim \mathcal{O}(0.1)$ would
be the $\mathrm{SO}(4)$-invariant perturbation coming from the term
$(4J_H/9) \bm{J}_j^{\prime 2},$ which separates the degeneracy of the virtual state
with two electrons per site into $J=0$ and $J=2.$ However, this effect is again
$\mathcal{O}(0.1)$ and, thus, we can expect this $\mathrm{SU}(4)$ breaking perturbation
to be negligible.

We note that the $\mathrm{SO}(5) \simeq \mathrm{Sp}(4)$ gauge transformation is just
a subgroup of the $\mathrm{SU}(4)$ gauge transformation, and it is not enough
to go to ``theorists' gauge'' without any non-Abelian hopping matrices.
In fact, Dirac gamma matrices are not included in the generator of the $\mathrm{Sp}(4)$
rotation for $\psi$ and the rotation is generated by $\Gamma^{pq} = -(i/2)[\Gamma^p,\Gamma^q] = -i\Gamma^p \Gamma^q$
($1\leq p,q \leq 5$)~\cite{Congjun2003}. Since the number of gamma matrices is
conserved mod 2 by the $\mathrm{SO}(5) \simeq \mathrm{Sp}(4)$ rotation,
the hopping matrices written by one gamma matrix cannot be rotated to an $\mathrm{SO}(5)$
scalar by the $\mathrm{SO}(5)$ gauge transformation, and this is why $\Gamma^5$ cannot
be eliminated in Eq.~\eqref{eq.so5}.

In this analysis, we only considered the extreme limit $\lambda \gg J_H$ for simplicity
to prove that the $\mathrm{SU}(4)$-breaking term comes from the order of
$\mathcal{O}(0.1)$ by employing the $\mathrm{SO}(5)$ gauge transformation intensively.
While we no longer expect the existence of a hidden $\mathrm{SO}(4)$ symmetry in
a general case, it is not difficult to show that in the second-order perturbation the
contribution breaking the original $\mathrm{SU}(4)$ symmetry always involves an
virtual state with an energy higher than the lowest order by $\lambda$ or $J_H.$
Anyway, we can conclude that, as long as we ignore higher order contributions of
$\mathcal{O}(0.1),$ the emergent $\mathrm{SU}(4)$ symmetry would be robust.

We note that recently it was argued that $\mathcal{O}(0.1)$ perturbation
of $J_H$ and $U^\prime - U$ would not destabilize the $\mathrm{SU}(4)$ spin liquid
in the case of BCSO~\cite{Natori2019}.  Although it is not clear this result is applicable to
$\alpha$-ZrCl$_3,$ we can expect that the stability region of a size $\mathcal{O}(0.1)$
will be reproduced for $\alpha$-ZrCl$_3,$ too, by similar mean-field and variational
calculations.  While this is a preliminary discussion, further studies will disclose
the effects of $J_H$ and $U^\prime - U$ in the future.

\section{Flux configurations for various tricoordinated lattices}\label{trico}

\begin{figure}
\centering
\includegraphics[width=10cm]{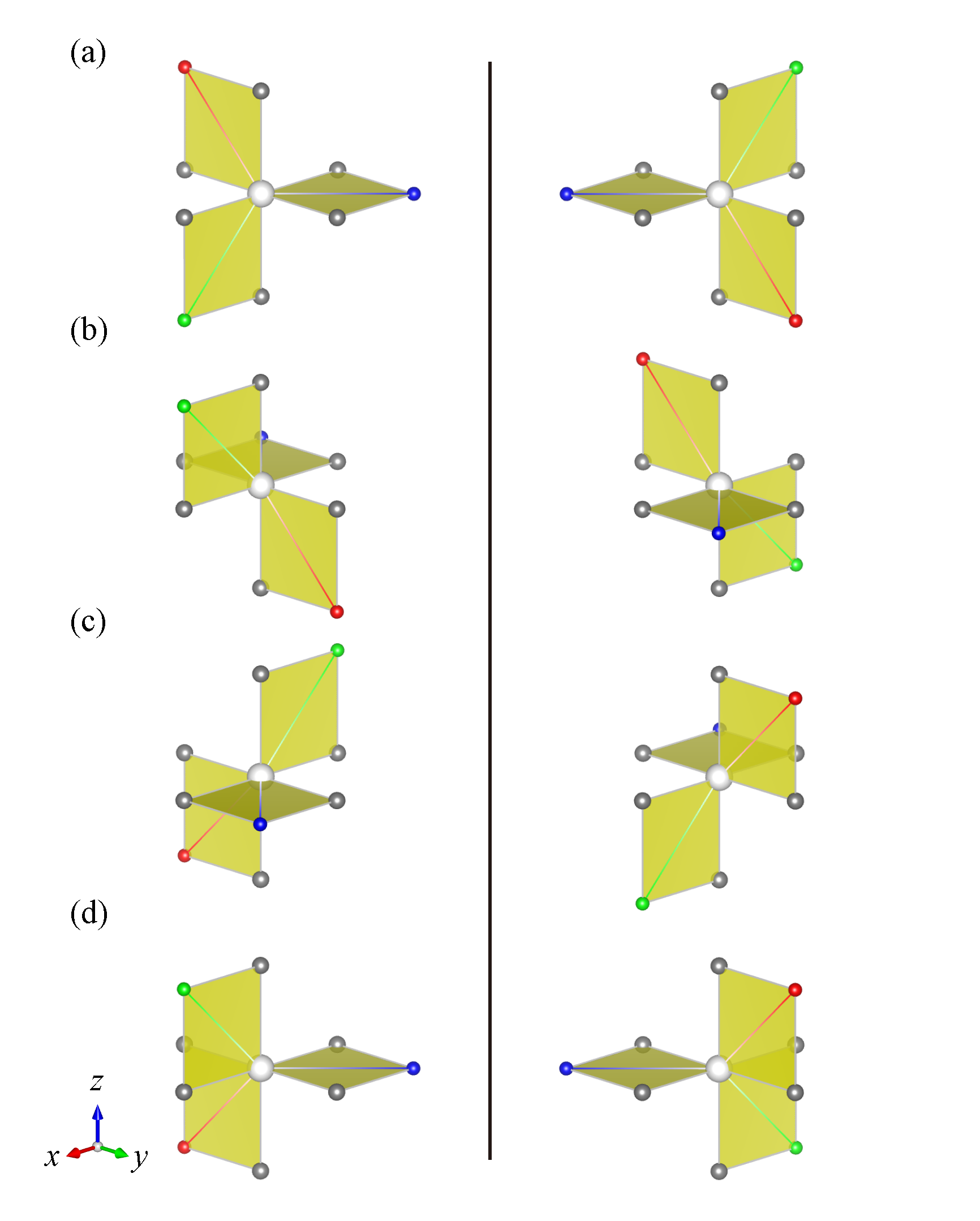}
\caption{All possible ways to connect three bonds in the 3D tricoordinated lattices.
(a) is the same one as that in the 2D honeycomb lattice, while (b), (c), and (d)
are produced by rotating (a) by $180\degree$ around the $x,$ $y,$ and $z$-axes, respectively.
The left-hand side and the right-hand side are related by the inversion for
each figure.}
\label{phase}
\end{figure}

The flux configurations for the tricoordinated lattices listed in Table~\ref{lattice}
can be treated similarly to the Kitaev models on tricoordinated lattices~\cite{Kitaev2006,Obrien2016}
except for the difference in the gauge group.
Following Kitaev~\cite{Kitaev2006}, we use terminology of the lattice gauge theory.
The link variables $U_{ij}$ are Hermitian and unitary (in this case)
$4\times 4$ matrices defined for each bond (link) $\langle ij \rangle$ of the lattice. 
Each link variable depends on its type (color) of the bond as
\begin{equation}
        U_{ij} = \begin{cases}
    U^a = \tau^y \otimes I_2 & (\langle ij \rangle \in a) \\
    U^b = -\tau^x \otimes \sigma^z & (\langle ij \rangle \in b) \\
    U^c = -\tau^x \otimes \sigma^y & (\langle ij \rangle \in c)
  \end{cases}, \label{eq.lab}
\end{equation}
where $\bm{\tau}$ and $\bm{\sigma}$ are independent
Pauli matrices, following the original gauge (basis) used in Sec.~\ref{uij}
(not the one used in the previous section). The bond type $abc$ is determined from
which plane this bond belongs to in the same way as $\alpha$-ZrCl$_3.$
We note that in the 3D case we actually have six types of bonds with additional
$\pm 1$ factors, so $U_{ij} = \pm U^a,\,\pm U^b,\,\pm U^c$ depending on a detailed
structure of the bond $\langle ij \rangle.$ This comes from the spatial dependence of
the sign of the wavefunctions of the $d$-orbitals.

These additional $\pm 1$ factors can simply be gauged out in the following way.
In the 2D honeycomb lattice, there is no sign difference in the same bond type
because all of them are related by the translation symmetry. In some 3D lattices,
even if the two bonds belong to the same type, the hopping matrices can
differ because they are related not by the translation symmetry, but
by the screw or glide symmetry. Accompanied by the reflection or rotation, this symmetry can
actually change the sign of the hopping matrix by $-1$ according to the shape of the
$t_{2g}$-orbitals. When seen from the metal site, it is a $180\degree$ rotation around
the $x,$ $y,$ or $z$-axis. If we consider the signs of the $t_{2g}$-orbitals,
it is clear that $180\degree$ rotation changes the signs of some orbitals,
while the inversion does not change the signs of the $d$-orbitals.
As shown in Fig.~\ref{phase}, there are 8 types of metal sites, and
all of them are related by some $180\degree$ rotation, which causes the sign difference,
up to inversion. Fortunately, however, this additional sign
can be eliminated by some gauge transformation, \textit{i.e.}
local rotations of the definition of the
effective angular momentum $l=1$ of the $t_{2g}$-orbitals.
For example, if the metal site is rotated around the $x$-axis by $180\degree,$
the configuration of the surrounding ligands changes
from Fig.~\ref{phase}(a) to Fig.~\ref{phase}(b). Then, according to the rotation,
we rotate the definition of the angular momentum $l=1$ around the $x$-axis by $180\degree,$
which can be done just by flipping the sign of the $yz$-orbital.
Similarly, for the ones shown in Fig.~\ref{phase}(c), we just flip the
sign of $zx$-orbital. Then, if we connect these two, Fig.~\ref{phase}(b) and (c), along
the $xy$-plane, we obtain an additional $-1$ phase from this gauge transformation, and
it completely cancels out the sign in question.
If we do a similar local rotation in the fictitious orbital space for each metal site
according to the physical $180\degree$ rotation, all the hopping matrices will be returned
to the original ones in Eq.~\eqref{eq.lab}, and after all we do not have to care about
the subtle difference among the same bond type. Thus, Eq.~\eqref{eq.lab} is still valid
after this ``$\mathbb{Z}_2$'' gauge transformation.

In order to find a gauge transformation
to get an $\mathrm{SU}(4)$ Hubbard model, we have to check that every Wilson loop operator
is Abelian.  In an abuse of language, each Wilson loop will be called flux inside the loop.
We regard a Wilson loop operator $I_4$ as a zero flux, and $-I_4$ as a $\pi$ flux.
In order to get a desired gauge transformation,
it is enough to show that the flux inside every elementary loop $C$ is Abelian:
\begin{equation}
        \prod_{\langle ij \rangle \in C} U_{ij} = \zeta_C I_4, \label{Eq.fluxfree}
\end{equation}
with some phase factors $|\zeta_C|=1.$

Since $U_{ij}^2=I_4,$ not all the fluxes are independent. In the case of a $\mathbb{Z}_2$ gauge
field, the constraints between multiple fluxes are called volume constraints~\cite{Obrien2016}.
However, due to the non-Abelian nature of the flux structure, it is subtle
whether they apply.  Fortunately, the above $U^\alpha$ ($\alpha=a,\,b,\,c$) obeys
the following anticommutation relations.
\begin{equation}
        \{U^\alpha,U^\beta\} = 2\delta^{\alpha\beta}I_4.
\end{equation}
This algebraic relation proves the product of the fluxes of the loops surrounding some
volume must vanish (volume constraints).  Moreover, we can easily show that, if every
bond color is used even times in each loop, which is a natural consequence for the lattices
admitting materials realization, the flux inside should always be Abelian with
$\zeta_C=\pm 1.$  Actually, every lattice included in Table~\ref{flux} obeys this
condition, so we have already proven all of them have an Abelian flux value.

The remaining subtle problem is which flux these elementary loops have, a zero flux,
or a $\pi$ flux. To check this, we need to investigate every loop one by one.
To calculate every flux value systematically, we often use space group symmetries to relate
two elementary loops, even though the system is in the strong spin-orbit
coupling limit. We note that the threefold rotation symmetry of the $xyz$-axes
of the Cartesian coordinate is not clear in the original gauge in Sec.~\ref{uij}.
This symmetry is important for some 3D models, although the spin quantization
axis along the (111) direction will make this symmetry explicit.
We have checked all the elementary loops in the tricoordinated lattices listed
here. In most cases, elementary loops of the same length
have the same flux due to some symmetry.
Only the flux value for the shortest elementary loops is shown in Table~\ref{flux}.

\begin{landscape}
\begin{table}
        \centering
        \caption{\label{flux}Flux value of tricoordinated lattices.  Only the flux value for the shortest elementary loops is shown here.  Nonsymmorphic space group numbers are underlined.  NS means that nonsymmorphic symmetries of the lattice are enough to protect a quantum spin-orbital liquid state.  In addition to the contents of Table~\ref{lattice}, we also include O'Keeffe's three-letter codes~\cite{OKeeffe2003regular,OKeeffe2003semiregular}.}
        \begin{tabular}{cccccccccc}
                        Wells' & Lattice & O'Keeffe's & Minimal & Flux & 120-degree & Number & \multicolumn{2}{c}{Space group} & LSMA \\
                        notation & name & code & loop length & value & bond & of sites & symbol & No. & constraints \\
                \hline
				(10,3)-$a$ & hyperoctagon & \textbf{srs} & 10 & 0-flux & \checkmark & 4 & $I4_1 32$ &\underline{\textbf{214}} & NS \checkmark \\
                (10,3)-$b$ & hyperhoneycomb & \textbf{ths} & 10 & 0-flux & \checkmark & 4 & $Fddd$\footnotemark[6] & \underline{\textbf{70}} & NS \checkmark \\
                (10,3)-$d$ & & \textbf{utp} & 10 & 0-flux & $-$ & 8 & $Pnna$\footnotemark[7] & \underline{\textbf{52}} & NS \checkmark \\
				nonuniform & $8^2.10$-$a$ & \textbf{lig} & 8 & $\pi$-flux & \checkmark & 8 & $I4_1 /amd$ & \underline{\textbf{141}} & $-$ \\
				(8,3)-$b$ & hyperhexagon & \textbf{etb} & 8 & $\pi$-flux & \checkmark & 6 & $R\bar{3}m$ & \textbf{166} & \checkmark \\
                nonuniform & stripyhoneycomb & \textbf{clh} & 6 & $\pi$-flux & \checkmark & 8 & $Cccm$\footnotemark[8] & \underline{\textbf{66}} & $-$ \\
				(6,3) & 2D honeycomb & \textbf{hcb} & 6 & $\pi$-flux & \checkmark & 2 & & & \checkmark
        \end{tabular}
\end{table}

\footnotetext[6]{The most symmetric case should be $I4_1 /amd,$ including $Fddd$.  Actually, $Fddd$ is enough for the filling constraint.}
\footnotetext[7]{There exists another phase with a $Pbcn$ symmetry.  Both symmetries are enough for the filling constraint.}
\footnotetext[8]{There exists a more symmetric phase with a $P4_2/mmc$ symmetry, but it is not enough for the filling constraint.}

\end{landscape}

\subsection{(10,3)-$a$}

\begin{figure}
\centering
\includegraphics[width=0.8\textwidth]{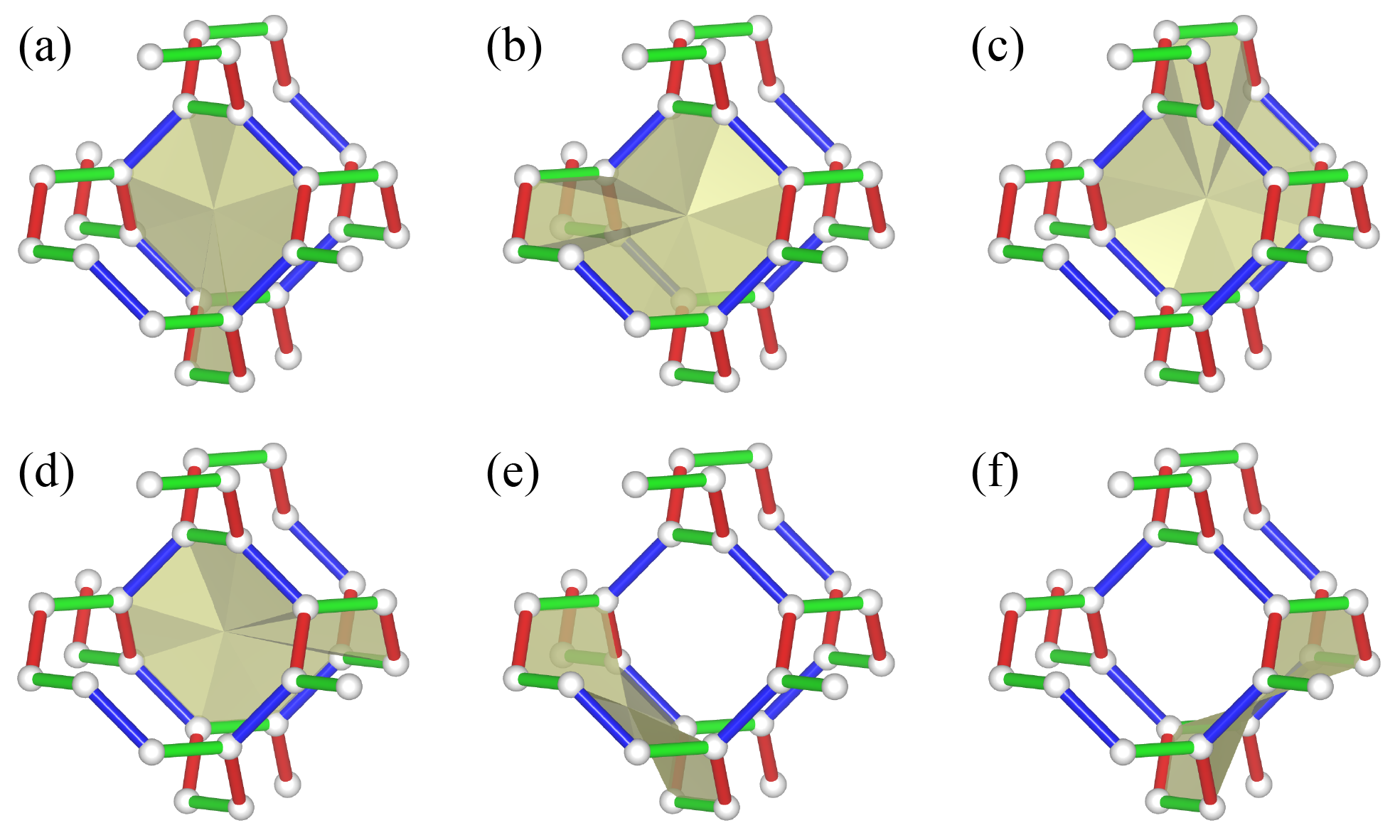}
\caption{Part of (10,3)-$a.$ All the six elementary loops~\cite{Hermanns2014} are highlighted by
yellow surfaces.  Loops (a)-(d) are related by the fourfold screw rotation,
and loops (e) and (f) are again related by the same symmetry.}
\label{10a}
\end{figure}

First of all, nonsymmorphic symmetries are useful to determine the flux value
because nonsymmorphic transformations often do not change the bond coloring
and effectively reduce the number of elementary loops.  As a concrete example,
we take the hyperoctagon lattice (10,3)-$a$ to show its usefulness.  (10,3)-$a$ has
six elementary loops of length 10~\cite{Hermanns2014}, and 4 of them are related by
the fourfold screw rotation symmetry [see Fig.~\ref{10a}(a)-(d)].
This fourfold screw exchanges the $b$-bonds for
the $c$-bonds, but this will not affect the flux value if the flux is Abelian
because the choice of the $xyz$-axes and its chirality is arbitrary.
The rest two elementary loops [see Fig.~\ref{10a}(e)-(f)] accidentally have the same coloring as
they are related by the screw symmetry.
Therefore, it is enough to check only two elementary loops, (a) and (e).
\begin{align}
U^cU^aU^cU^aU^bU^aU^cU^aU^cU^b &=(U^cU^a)^2U^b(U^aU^c)^2U^b=I_4, \\
U^bU^aU^bU^aU^cU^aU^bU^aU^bU^c &=(U^bU^a)^2U^c(U^aU^b)^2U^c=I_4.
\end{align}
From the above symmetry arguments, or from volume constraints,
we can conclude that all the six elementary loops
(of length 10) in (10,3)-$a$ have a zero flux.  This result agrees with the fact
that this zero-flux configuration is the unique $Z_2$ flux configuration
that obeys all the lattice symmetries of (10,3)-$a$~\cite{Obrien2016}.

\subsection{(10,3)-$b$}

\begin{figure}
\centering
\includegraphics[width=0.5\textwidth]{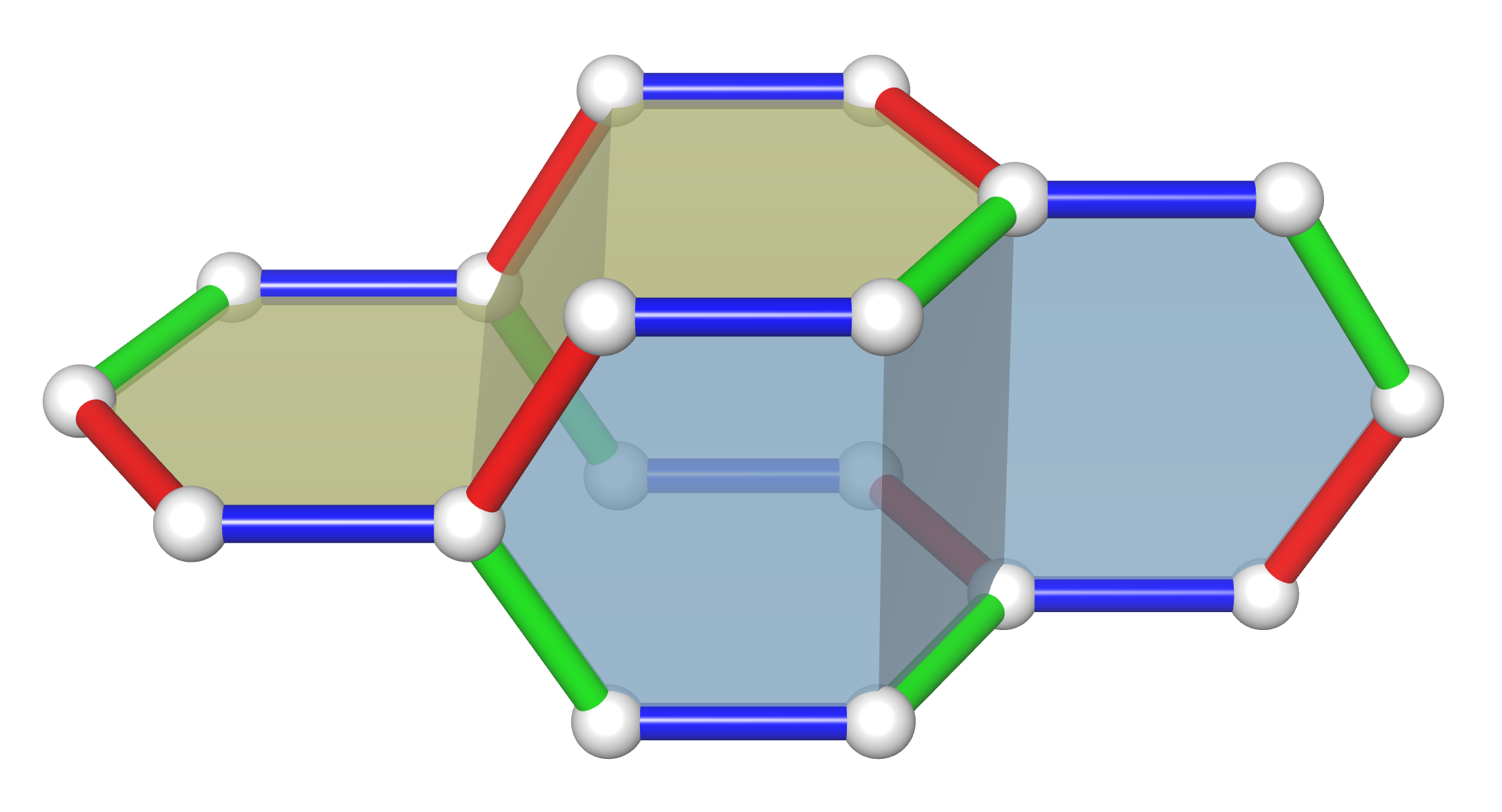}
\caption{Part of (10,3)-$b$ including four loops forming a volume constraint.
Two elementary loops with different coloring patterns are highlighted by
yellow and cyan surfaces, respectively.}
\label{10b}
\end{figure}

Among various point group symmetries, the inversion symmetry of the lattice
is the most useful.  As is the case in the honeycomb lattice, if an elementary
loop has an inversion center, then the flux inside this loop becomes the
square of some Pauli matrices times a complex number, which actually
only takes $1,i,-1,-i.$  Therefore, the existence of an inversion center
automatically proves that the flux is Abelian and should be $0$ or $\pi.$
This is another proof that a non-Abelian flux vanishes on some lattices.
This applies, for example, to the hyperhoneycomb lattice (10,3)-$b.$
All the four elementary loops of length 10 (10-loops) have an inversion center,
making the direct calculation easier.  We can classify these four 10-loops
into two pairs, where two loops are related by the glide mirror symmetry
with the same coloring pattern for each pair.  Therefore, it is enough to check
two loops, shown in the yellow and cyan surfaces, respectively, in Fig.~\ref{10b}.
\begin{align}
U^b U^c U^a U^c U^a U^b U^c U^a U^c U^a &= [U^b (U^c U^a)^2]^2 = I_4. \\
U^a U^c U^b U^c U^b U^a U^c U^b U^c U^b &= [U^a (U^c U^b)^2]^2 = I_4.
\end{align}
Therefore, all the four elementary loops in (10,3)-$b$ have a zero flux.

\subsection{(10,3)-$d$}

\begin{figure}
\centering
\includegraphics[width=0.5\textwidth]{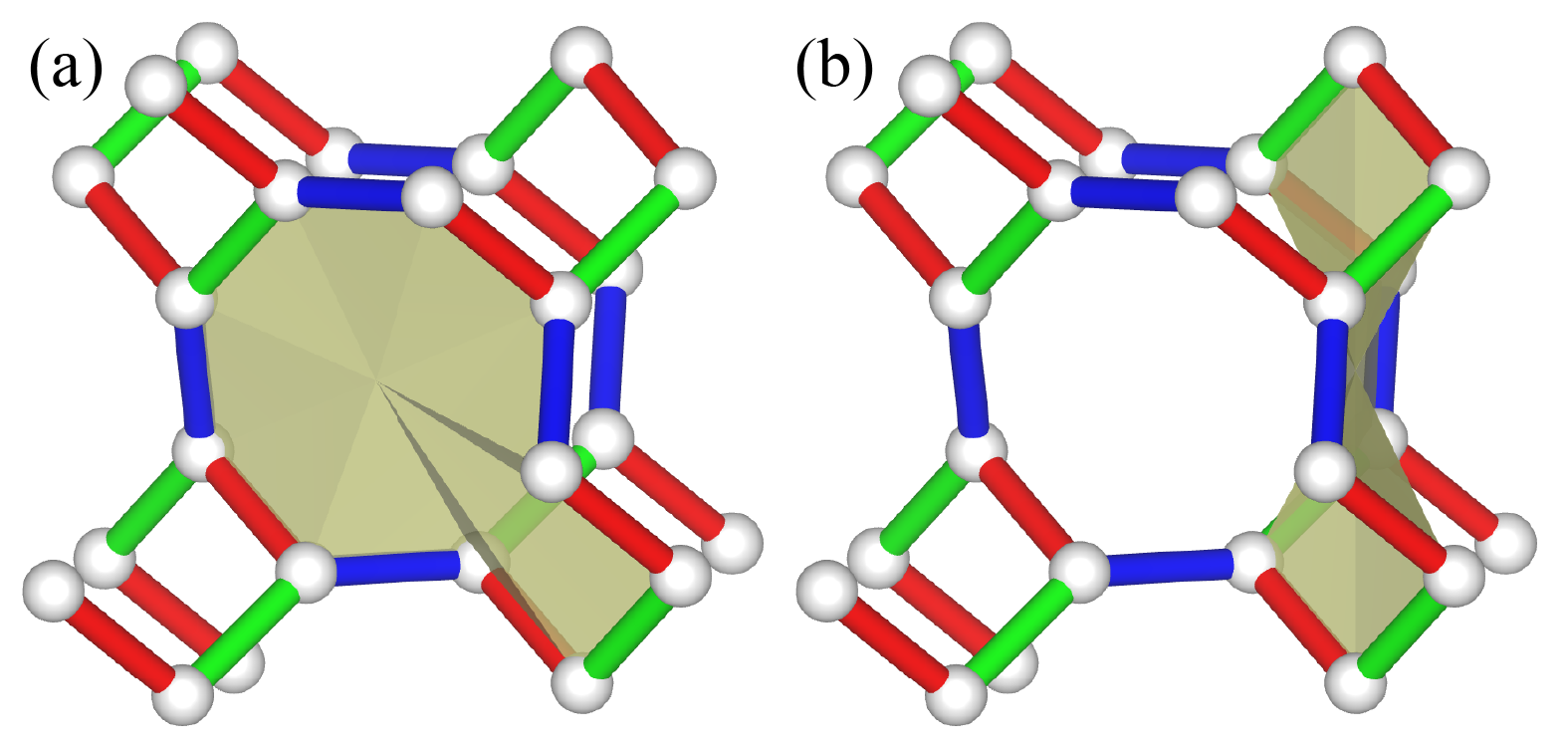}
\caption{Part of (10,3)-$d.$ (a) One of the type-A loops highlighted by the yellow surface.
(b) One of the type-B loops highlighted by the yellow surface.}
\label{10d}
\end{figure}

The structure of (10,3)-$d$ is related to (10,3)-$a$ because they share the same projection
onto the (001) plane, the 2D squareoctagon lattice.  Due to the difference in the chiralities
of the square spirals, the unit cell is enlarged in (10,3)-$d$ and possess 8 elementary loops
(of length 10) per unit cell.

Since this lattice does not allow any 120-degree configuration, we cannot simply
decide the bond coloring.  If we take the most symmetric bond coloring discussed in~\cite{Yamada2017XSL},
then the calculation becomes simple.  We can divide 8 elementary loops of length 10 into
two types.  Four type-A loops are spiraling up the octagon spiral
and then spiraling down the square spiral [see Fig.~\ref{10d}(a)].
All the four type-A loops are related by
the inversion symmetry or the twofold screw rotation symmetry
(the combination of them is the glide mirror symmetry), and thus have the same flux.
Four type-B loops are spiraling up the square spiral and then
spiraling down the nearest-neighbor square spiral [see Fig.~\ref{10d}(b)].
Four type-B loops are
related by the twofold screw rotation symmetry or by the glide mirror symmetry,
and have the same flux.  Thus, it is enough to check one for each type.
\begin{align}
U^b U^c U^a U^c U^a U^b U^a U^c U^a U^c &= U^b (U^c U^a)^2 U^b (U^a U^c)^2 = I_4. \\
U^b U^a U^b U^a U^c U^b U^a U^b U^a U^c &= [(U^b U^a)^2 U^c]^2 = I_4.
\end{align}
The direct calculation tells us that the hopping model is in a zero-flux configuration.

\subsection{$8^2.10$-$a$}\label{810acolor}

\begin{figure}
\centering
\includegraphics[width=0.8\textwidth]{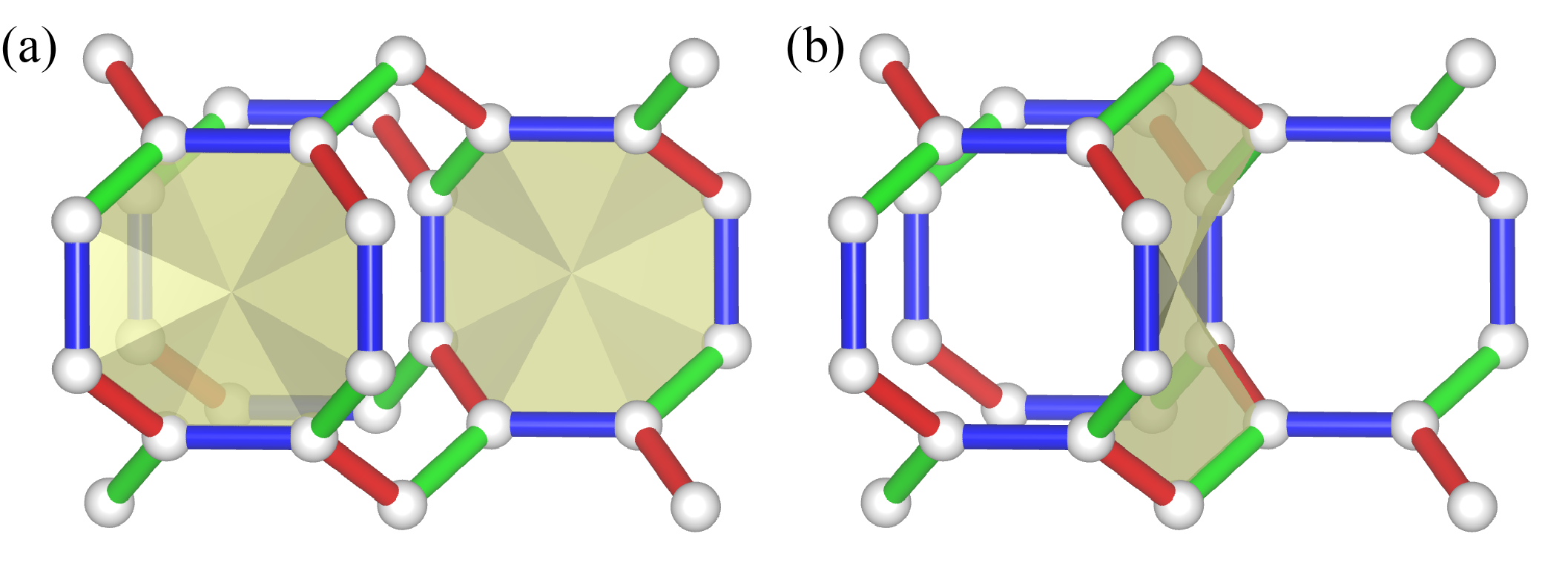}
\caption{Part of $8^2.10$-$a.$ (a) All the two 8-loops are shown by yellow
surfaces.  They are related by the fourfold screw rotation symmetry.
(b) One of the four 10-loops is shown by the yellow surface.
The rest are produced by applying the fourfold screw rotation around the
square spiral.}
\label{810a}
\end{figure}

$8^2.10$-$a$ is nonuniform, but Archimedean.  Therefore, each site is included in the
two types of elementary loops, some of length 8 and others of length 10.
The unit cell includes two elementary loops of length 8 (8-loops) [see Fig.~\ref{810a}(a)]
and four elementary loops of length 10 (10-loops) [see Fig.~\ref{810a}(b)].
It is enough to check one of the 8-loops and one of the 10-loops
because all the elementary loops of the same length are related by the fourfold screw
rotation symmetry.
\begin{align}
U^a U^c U^b U^c U^a U^c U^b U^c &= [U^a U^c U^b U^c]^2 = -I_4. \\
U^c U^a U^b U^a U^b U^c U^a U^b U^a U^b &= [U^c (U^a U^b)^2]^2 = I_4.
\end{align}
Therefore, all the 8-loops have a $\pi$ flux and all the 10-loops have a zero flux.
We note that the hopping
model in this $\pi$-flux configuration does not break the original translation symmetry~\cite{Yamada2017XSL}.

\subsection{(8,3)-$b$}

\begin{figure}
\centering
\includegraphics[width=0.5\textwidth]{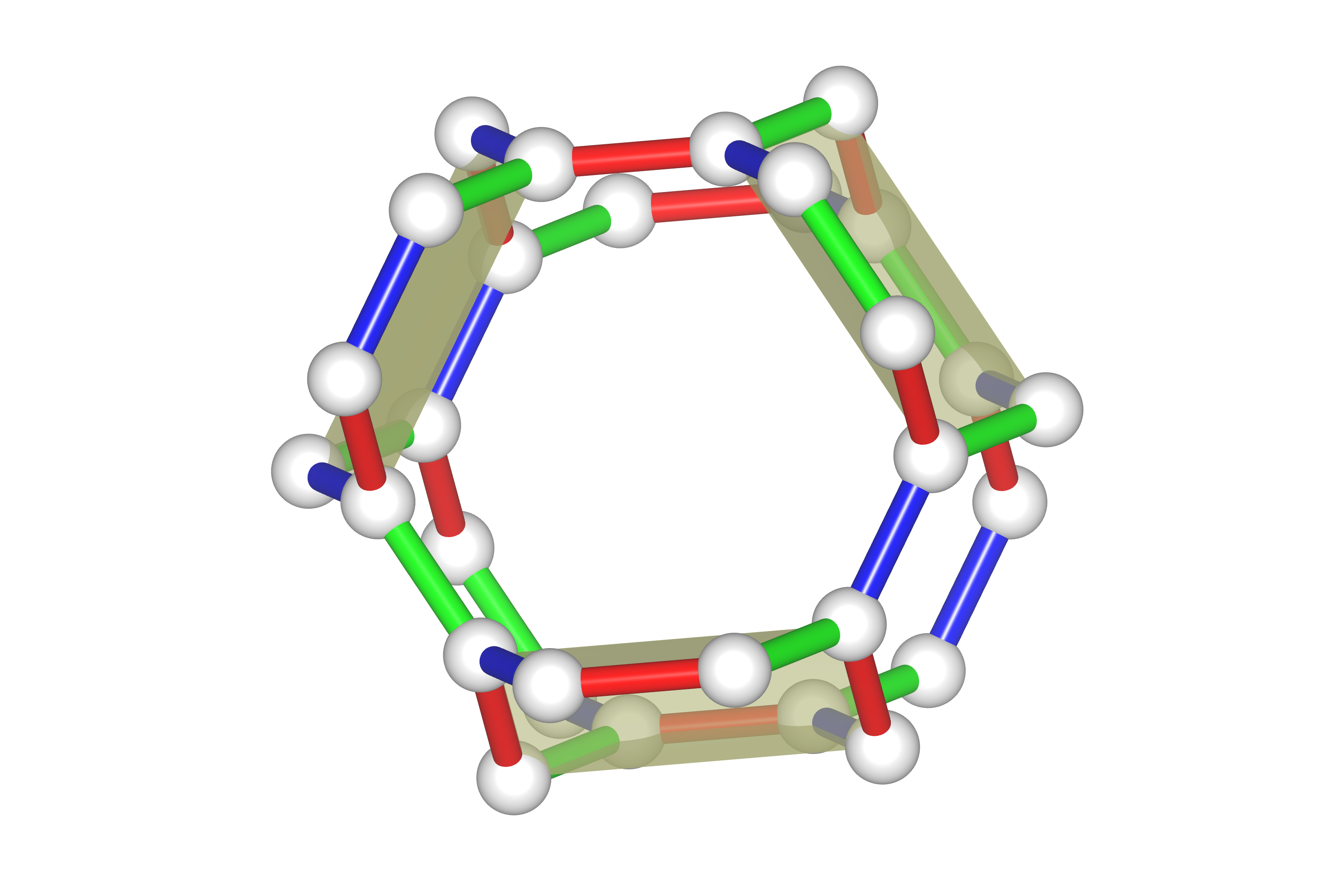}
\caption{Part of (8,3)-$b.$ All the three elementary loops of length 8 are
highlighted by yellow surfaces.  They are related by the threefold rotation symmetry.}
\label{8b}
\end{figure}

The hyperhexagon lattice (8,3)-$b$ has three elementary loops of length 8, and they
are related by the threefold rotation symmetry changing the $xyz$-axes, as shown in
Fig.~\ref{8b}.  Therefore, it is
enough to check only one of them.  The direct calculation tells us that it has a $\pi$ flux.
\begin{equation}
U^a U^c U^b U^c U^a U^c U^b U^c = [U^a U^c U^b U^c]^2 = -I_4.
\end{equation}

Therefore, (8,3)-$b$ is in the $\pi$-flux configuration. We note that there is another
elementary loop of length 12, but the flux value is immediately determined to
be zero due to the accidental fourfold symmetry of the coloring.
It is worth mentioning the hopping
model in this $\pi$-flux configuration does not break the original translation symmetry,
and thus the LSMA theorem applies as it is to the $\pi$-flux $\mathrm{SU}(4)$ Hubbard model, as well
as the $\mathrm{SU}(4)$ Heisenberg model.

\subsection{Stripyhoneycomb lattice}

\begin{figure}
\centering
\includegraphics[width=\textwidth]{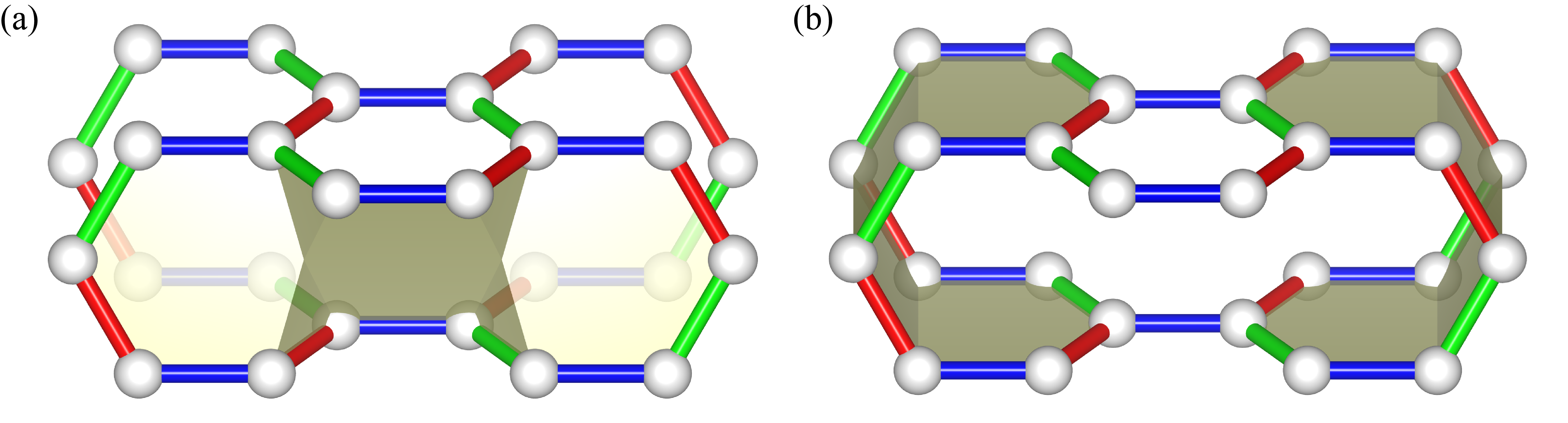}
\caption{Part of the stripyhoneycomb lattice. (a) A loop of length 14 is highlighted.
(b) A pair of loops of length 12 are highlighted. They are related by the inversion
symmetry (or the volume constraint) and thus have the same flux.}
\label{stripy}
\end{figure}

The stripyhoneycomb lattice is nonuniform, so the length of the shortest elementary loops
differs in space.  Every elementary loop of length 6 is the same as the honeycomb, and thus
has a $\pi$ flux.  The structure includes two types of the $\pi$-flux hexagons aligning
in different planes~\cite{Kimchi2014}.
In addition, there exist a long loop of length 14 (14-loop) and a twisted
loop of length 12 (12-loop) [see Fig.~\ref{stripy}].
These four types of elementary loops are enough to determine the flux values.

One 14-loop shown in Fig.~\ref{stripy}(a) has a zero flux because
\begin{equation}
		U^a U^c U^a U^b U^c U^b U^c U^a U^c U^a U^b U^c U^b U^c = [U^a U^c U^a (U^b U^c)^2]^2=I_4.
\end{equation}

One 12-loop shown on the right-hand side of Fig.~\ref{stripy}(b) also has a zero flux because
\begin{equation}
U^a U^b U^c U^a U^b U^c U^b U^a U^c U^b U^a U^c = (U^a U^b U^c)^2 (U^b U^a U^c)^2= I_4.
\end{equation}

There are many other tricoordinated lattices not discussed in this thesis,
so it is future work to determine the flux values for all the possible tricoordinated
lattices.

\chapter{Summary and Discussions}\label{sum}

\section{Summary}

As discussed in the Introduction, the $\mathrm{SU}(N)$ magnetism has a distinct feature
with additional degrees of freedom to realize new QSLs beyond geometric/exchange frustration.
Especially, a stable Dirac spin liquid is expected in the $\mathrm{SU}(4)$ Heisenberg
model on the honeycomb lattice, but no material candidates were found for this exotic
model, and even the realization in cold atoms has not been achieved.

In summary, we newly found that, as a consequence of the combination of the octahedral ligand
field and SOC, an $\mathrm{SU}(4)$ symmetry \emph{emerges} in $\alpha$-ZrCl$_3.$
This is contrary to the ordinary expectation that SOC reduces the symmetry of spins.
The derivation is similar to Ref.~\citenum{Jackeli2009}, but we employed the language
of a lattice gauge theory to simplify the discussions.  This would pave a new way
to realize the $\mathrm{SU}(4)$ magnetism in real materials, not restricted
to cold atomic systems.

In addition to the $\alpha$-ZrCl$_3$ (or $A_2M^\prime$O$_3$) family we have discussed, 
Zr- or Hf-based MOFs could also realize $\mathrm{SU}(4)$ Heisenberg models
on various tricoordinated lattices.
Especially, 3D (10,3)-$a$~\cite{Coronado2001}, (10,3)-$b$~\cite{Zhang2012hyper},
and $8^2.10$-$a$~\cite{ClementeLeon2013,Yamada2017XSL} lattices, as well as the 2D honeycomb
lattice~\cite{Zhang2014}, were already
realized in some MOFs with an oxalate ligand.
Thus we can expect that microscopic
models defined by Eq.~\eqref{Eq.original} on various tricoordinated lattices will apply
in the same way as the honeycomb $\alpha$-ZrCl$_3$ if we replace the metal ions of
these MOFs with Zr$^{3+},$ Hf$^{3+},$ Nb$^{4+},$ or Ta$^{4+}$~\cite{Yamada2017MOF}.

Such orbital physics can be sought in other systems like $f$-electron systems.
For example, ErCl$_3$ may have twofold orbital degeneracy at low temperature~\cite{Kramer1999,Kramer2000}.
In most cases, orbitals have twofold degeneracy, so the highest achievable symmetry
of QSOLs in spin-orbital materials is $\mathrm{SU}(4).$
Whether it is possible to realize $\mathrm{SU}(6)$ spin systems in spin-orbital
systems is an interesting open question. So far a cold atomic system is the only
candidate for $\mathrm{SU}(6).$

The JT term which couples the orbital to the lattice has been ignored so far.
Usually, this term breaks a symmetry of the lattice, resulting a JT transition
to the low-symmetry phase~\cite{Tokura2000}.  In order for the symmetric phase to survive,
the itinerant quantum fluctuation which can tunnel between classical ground states
may be necessary.
Thus, the competition between QSOLs and JT phases (orbital order) can be understood
by the spinon/orbitalon band width $W \sim J = 8t^2/(3U)$~\cite{Khaliullin2000}.
If $J$ is large enough to stabilize
the (orbital) symmetric state, then the kinetic energy gain of orbitalons may destabilize
the JT order.  Thus, such energy gain may be maximized around the Mott transition, and
thus the 4$d$- or 5$d$-materials with a smaller $U$ may be beneficial.
In the Dirac spin-orbital liquid phase, the Dirac dispersion of mobile spinons and
orbitalons result in characteristic specific heat and thermal conductivity.
The specific heat $C$ behaves as $C \propto T^2$ as the temperature $T$ goes to $0,$
and with a magnetic field it should behave $C \propto T$ in Dirac spin liquids
within the mean-field approximation~\cite{Ran2007}.
In reality, the gauge field also contributes to $C.$  The correction from the gauge field
is a future problem, but there is a possibility that a characteristic correction exists
in the gauge sector if the low-energy gauge theory is $\mathrm{SU}(4)$ QCD.

Experimentally, muon spin resonance or nuclear magnetic resonance (NMR)
experiments can rule out the existence of long-range magnetic ordering or spin freezing
in the spin sector. In the orbital sector, a possible experimental signature
to observe the absence of orbital ordering or freezing should be
ESR~\cite{Han2015} or EXAFS~\cite{Nakatsuji2012}, similarly to BCSO.
Especially, (finite-frequency) ESR can observe the dynamical JT effect~\cite{Nasu2013,Nasu2015},
where the $g$-factor isotropy directly signals the quantum fluctuation between
different orbitals~\cite{Han2015,Bersuker1975,Bible1970}.
For example, in the case of BCSO~\cite{Han2015}, the orbital ordering of the $d_{x^2-y^2}$ and $d_{z^2}$ orbitals directly couples to the tetragonal distortion of the octahedron.  Thus, the strained direction of the anisotropic $g$-tensor signals the direction of the orbital ``polarization'' between the two $e_g$ orbitals.
This is also applicable to our $t_{2g}$ case because of the shape difference in the $J_\textrm{eff}=3/2$
orbitals~\cite{Romhanyi2017}, and the static JT distortion will result in
the anisotropy in the in-plane $g$-factors~\cite{Iwahara2017}. Here we note that
the trigonal distortion existing \textit{a priori} in real materials only splits the degeneracy
between the out-of-plane and in-plane $g$-factors, and the splitting of the two
in-plane modes clearly indicates some (\textit{e.g.} tetragonal) distortion.
The emergent $\mathrm{SU}(4)$ symmetry would result in changing the
universality class of critical phenomena, or in an accidental coincidence between the
time scales of two different excitations for spins and orbitals observed
by NMR and ESR, respectively.

On the other hand, the direct detection of orbitalons may be challenging.
(Charged) orbitalons carry an orbital angular momentum as well as heat.
Magnetically an orbital angular momentum is indistinguishable and mixed
with a spin by SOC.  However, since the orbital fluctuation is coupled to the lattice,
an electric field, light, or x-rays can directly affect the orbital sector~\cite{Tokura2000}.
Especially, a light beam with an orbital angular momentum has been investigated
recently~\cite{Marrucci2006}, and may be useful for the detection of orbitalons.
It is future work to discover the connection between such technology and fractionalized
orbital excitations.

\section{Discussion and comparison with twisted bilayer graphene}

\begin{figure}
\centering
\includegraphics[width=14cm]{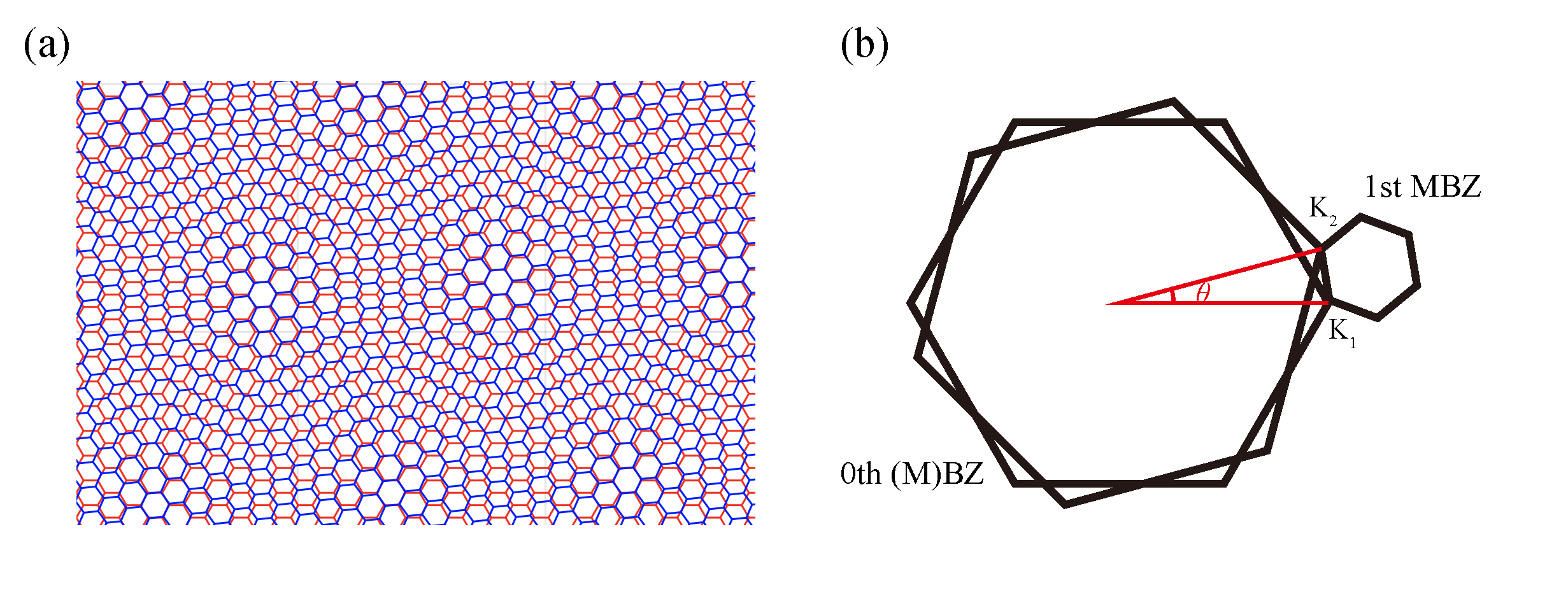}
\caption{(a) Typical moir\'{e} pattern of TBG. The first and second layers are shown
in red and blue, respectively.  (b) Moir\'{e} Brillouin zone (MBZ).
The original (first) Brillouin zone is shown by two large hexagons, each of which
represents a Brillouin zone for each layer.  Since the two layers are twisted by
a (magic) angle $\theta,$ Brillouin zones are also twisted by this angle.
The first MBZ is shown as a smaller hexagon, which connects the two K points
of the original Brillouin zones differed by the angle $\theta.$
When $\theta \sim 1.1\degree$ (a magic angle), flat bands are expected in the effective model.
When $\theta = 0,$ we can regard the original Brillouin zones as the ``zeroth'' MBZ
and due to the degeneracy of two layers and spins the effective model
of TBG has an effective $\mathrm{SU}(4)$ symmetry within the zeroth-order approximation.}
\label{mbz}
\end{figure}

Finally, we would like to mention another candidate material for $\mathrm{SU}(4)$ models.
Specifically, twisted bilayer graphene (TBG) attracted
attention after the discovery of a correlated insulating state and accompanied
superconductivity~\cite{Cao2018,Cao2018ins}.
Graphene is a honeycomb lattice sheet of carbon~\cite{Novoselov2004}.
Bilayer graphene is a van der Waals structure made of two graphene layers.
When two graphene layers are twisted by an angle $\theta,$ the so-called
moir\'{e} pattern appears in the real space [see Fig.~\ref{mbz}(a)].
At some specific $\theta$ called magic angle, the bandwidth approaches zero~\cite{Bistritzer2011},
leading to strong correlation due to a large $U/t$ in the effective Hubbard model.
Though the correct low-energy theory of TBG is complicated
and requires a so-called moir\'{e} Brillouin zone [see Fig.~\ref{mbz}(b)],
the $\mathrm{SU}(4)$ Hubbard model is still a good approximation in the ``zeroth'' order for this system
consisting of spin and valley degrees of freedom with a strong correlation~\cite{Xu2018}.
However, the zeroth-order case $\theta=0$ is weakly correlated and in the strongly
correlated insulating phase ($\theta \sim 1.1\degree$) requires different
maximally localized Wannier functions, and the low-energy model has no $\mathrm{SU}(4)$
symmetry in the first moir\'{e} Brillouin zone~\cite{Nam2017,Yuan2018}.
Thus, $\alpha$-ZrCl$_3$ still has superiority because it has an exact $\mathrm{SU}(4)$
symmetry even in the strongly correlated region $U\to \infty$ with a strong SOC
$\lambda \to \infty.$

After the discovery of TBG, similar 2D heterostructures were also investigated.
Twisted bilayer TMDC is one of them~\cite{Naik2018,Wu2019}.
Those spin-valley systems are also important candidates for $\mathrm{SU}(4)$
magnetism and seeking an ideal $\mathrm{SU}(4)$ system among them would be
important future work.
There is an important DMRG result for the 0-flux $\mathrm{SU}(4)$ Hubbard model on the
honeycomb lattice at quarter filling~\cite{Zhu2019},
though the results are not directly applicable to
$\alpha$-ZrCl$_3$ due to the existence of a $\pi$ flux inside a plaquette.

\appendix

\chapter{Implications from the Lieb-Schultz-Mattis-Affleck theorem}\label{lsma}

The $\mathrm{SU}(N)$ Heisenberg model on the two-dimensional (2D) honeycomb
lattice admits the application of the Lieb-Schultz-Mattis-Affleck (LSMA)
theorem~\cite{LSM1961,Affleck1986,Oshikawa2000,Hastings2005} for $N>2$.
However, the original paper by Affleck and Lieb~\cite{Affleck1986} only
discussed one-dimensional (1D) systems, so we would like to extend the claim to higher dimensions
and systems with a space group symmetry.
Let us first consider a periodic 2D lattice with the
primitive lattice vectors $\bm{a}_{1,2}$,
as defined in Fig.~\ref{zrcl} in the main text.
We define the lattice translation operators $\trans_\mu$
along $\bm{a}_\mu$ for $\mu=1,2$.

Here we consider the case with a fundamental representation on each
site of the honeycomb lattice, which includes the $\mathrm{SU}(4)$ Heisenberg
model discussed in the main text.
We call each basis of the $\mathrm{SU}(N)$ fundamental representation ``flavor''.
The Hamiltonian of the $\mathrm{SU}(N)$ Heisenberg model on the honeycomb lattice in general can
be written as
\begin{equation}
		H_{\mathrm{SU}(N)}=\frac{J_a}{N} \sum_{\langle ij\rangle \in a} P_{ij} +\frac{J_b}{N} \sum_{\langle ij\rangle \in b} P_{ij} +\frac{J_c}{N} \sum_{\langle ij\rangle \in c} P_{ij}, \label{eq.Heisenberg}
\end{equation}
up to constant terms, where $J_\gamma$s are the bond-dependent coupling constants for the
$\gamma$-bonds, as defined in the main text, and $P_{ij}$ is the permutation operator
of the flavors between the $i$th and $j$th sites.  The translation symmetries,
$\trans_1$ and $\trans_2,$ exist independently of the values of $J_\gamma$s, so
the following discussions apply to any positive $J_\gamma$s.
Since the spin-1/2 Heisenberg antiferromagnetic interaction for the
$\mathrm{SU}(2)$ spin can also be written as Eq.~\eqref{eq.Heisenberg} with $N=2$
dimensional Hilbert space at each site.

Now we discuss the generalization of the LSMA theorem to $\mathrm{SU}(N)$ spin
systems~\cite{Affleck1986,Lajko2017,Totsuka2017}
in 2 dimensions following the logic of Ref.~\citenum{Oshikawa2000}.
One of the generators $I^0$ of the $\mathrm{SU}(N)$ in the fundamental
representation is given by the traceless $N\times N$
diagonal matrix:
\begin{equation}
		I^0=\frac{1}{N} \begin{pmatrix}
1 & 0 & \cdots & 0 & 0 \\
0 & 1 &  & 0 & 0\\
\vdots &  & \ddots &  & \vdots \\
0 & 0 &  & 1 & 0 \\
0 & 0 & \cdots & 0 &-(N-1)
\end{pmatrix}.
\end{equation}
We introduce an Abelian gauge field $\bm{\calA} (\bm{r})$,
which couples to the charge $I^0$, where
$\bm{r}$ is the coordinate.

We assume that the (possibly degenerate) ground
states are separated from the continuum of the excited states
by a nonvanishing gap, and that the gap does not collapse during
the flux insertion process discussed below.
We consider the system consisting of $L_1 \times L_2$ unit cells
on a torus, namely with periodic boundary conditions
$\bm{r} \sim \bm{r} + L_1 \bm{a}_1 \sim \bm{r} + L_2 \bm{a}_2$.
A ground state, which is $\mathrm{SU}(N)$-symmetric and 
has a definite crystal momentum
(\textit{i.e.} eigenstate of $\trans_{\mu}$ with $\mu=1,\,2$),
is chosen as the initial state.
We adiabatically increase the gauge field from $\bm{\calA}=0$
to $\bm{\calA} = \bm{k}_1/L_1$, so that
the ``magnetic flux'' contained in the ``hole'' of the torus
increases. When ``magnetic flux'' reaches the unit flux quantum $2\pi,$
the Hamiltonian of the system becomes equivalent to the initial one.
This happens precisely when the Hamiltonian is obtained from
the original Hamiltonian with a large gauge transformation.
The minimal large gauge transformation with respect to the charge
$I^0$ is given by
\begin{equation}
\calU_1 = \exp{\left[
\frac{i}{L_1} \sum_{\bm{r}} \bm{k}_1 \cdot \bm{r} I^0(\bm{r})
\right]},
\end{equation}
where $\bm{k}_\mu$s are primitive reciprocal lattice vectors satisfying
\begin{equation}
 \bm{k}_\mu \cdot \bm{a}_\nu = 2 \pi \delta_{\mu \nu}.
\end{equation}

The large gauge transformation satisfies the commutation relation,
\begin{equation}
 \calU_1 \trans_1   = \trans_1 \calU_1
\exp{\left[
		\frac{2\pi i}{L_1} \Bigl( I^0_T -
\sum_{\bm{r}\cdot\bm{k}_1 = 2\pi (L_1 -1) } L_1 I^0\left( \bm{r} \right) \Bigr)
 \right]} .
\end{equation}
Here $I^0_T = \sum_{\bm{r}} I^0(\bm{r})$.
Since the ground state is assumed to be an $\mathrm{SU}(N)$-singlet when the number of sites
is a multiple of $N,$
it belongs to the eigenstate with $I^0_T = 0.$
Furthermore, because eigenvalues of $I^0(\bm{r})$ are equivalent to $1/N \mod{1},$ we find,
\begin{equation}
 {\trans_1}^{-1} \calU_1 \trans_1  \sim \calU_1 e^{-(2\pi i n L_2/N)},
\end{equation}
where $n$ is the number of sites in the unit cell.

Since the uniform increase in the vector potential does not
change the crystal momentum, this phase factor due to the
large gauge transformation alone gives the
change of the crystal momentum in the flux insertion process.
Choosing $L_2$ to be coprime with $N,$ we find 
a nontrivial phase factor when $n/N$ is not an integer.
This implies that, if $n$ is not an integer multiple of $N$,
the system must be gapless or has degenerate ground states.

For the honeycomb lattice, $n=2,$ and there is no LSMA constraint
for $\mathrm{SU}(2)$ spin systems. In contrast, for the $\mathrm{SU}(4)$ spin system
we discussed in the main text, the ground-state degeneracy (or gapless
excitations) is required even on the honeycomb lattice.
Thus, the resulting quantum spin-orbital liquid (QSOL)~\cite{Corboz2012}
cannot be a ``trivial'' featureless Mott insulator when the symmetry is not
broken spontaneously.

As explained in the above proof, the existence of a nontrivial generator
$I^0$ is important for this theorem.
In the case of $\alpha$-ZrCl$_3$ discussed in the main text, this element is not included
in the generators of the original $\mathrm{SU}(2) \times \mathrm{SU}(2)$ symmetry of the spin-orbital space,
but included in the emergent $\mathrm{SU}(4)$ symmetry in the strong spin-orbit coupling limit.
Thus, we can say that the $\mathrm{SU}(4)$ symmetry actually protects the nontrivial ground state of
the $\mathrm{SU}(4)$ Heisenberg model on the honeycomb lattice.

This proof of the LSMA theorem is not restricted to bosonic systems, and applies to both
bosonic and fermionic systems.  Thus, the generalization to the (zero-flux) $\mathrm{SU}(N)$-symmetric
Hubbard models is straightforward.
With $N$-flavor fermionic degrees of freedom in the $\mathrm{SU}(N)$ fundamental representation
at each site, the necessary condition for the existence of a featureless insulator is
that there exists a multiple of $N$ fundamental representations
per unit cell, which can form an $\mathrm{SU}(N)$ singlet.
We note that the LSMA theorem for $\mathrm{SU}(N)$ spin systems can be derived from the
$U\to \infty$ limit of the $\mathrm{SU}(N)$ Hubbard model at $1/N$ filling.
One can also extend the LSMA theorem to the systems with general representations
 on each site, starting from a Hubbard model.
That is, we include an appropriate onsite ``Hund'' coupling $J_H$ 
in the Hubbard model so that the desired representation have the lowest
energy, and then take the $J_H \to \infty$ limit afterwards.

The generalization to the three-dimensional (3D) case with three translation operators,
$\trans_1,$ $\trans_2,$ and $\trans_3,$ is again straightforward and we will omit the
proof here, but it is useful to extend the
LSMA theorem to the case with a space group symmetry.
Recently, tighter constraints are obtained for nonsymmorphic space
group symmetries~\cite{PTAV2013,WPVZ2015} than what is implied by the
LSMA theorem based on the translation symmetries only.
This is because a nonsymmorphic symmetry behaves as a ``half'' translation, which would
reduce the size of the effective unit cell.

As a demonstration, here we only discuss the constraint given
by one nonsymmorphic (glide mirror or screw rotation) operation $\calG$,
by generalizing the flux insertion argument as in Ref.~\citenum{PTAV2013}.
We note that a tighter condition can be
derived by dividing the torus into the largest flat manifold, which is called
Bieberbach manifold, for some of the nonsymmorphic space groups~\cite{WPVZ2015}.

Among the 157 nonsymmorphic space groups, the 155 except for $I2_1 2_1
2_1$ (No. \textbf{24}) and $I2_1 3$ (No. \textbf{199}) include an
unremovable (essential) glide mirror or screw rotation symmetry
$\calG$~\cite{Konig1999}, so we will concentrate on these 155 to show
how $\calG$ works to impose a stronger constraint on filling.  The
nonsymmorphic operation $\calG$ consists of a point-group operation $G$
followed by a fractional (nonlattice) translation with a vector
$\bm{\alpha}$ in a direction left invariant by $G,$ \textit{i.e.} $\calG: \bm{r}
\mapsto G\bm{r}+\bm{\alpha}$ with $G \bm{\alpha}=\bm{\alpha}.$ We again
assume that the (possibly degenerate) ground states are separated from
the continuum of the excited states by a nonvanishing gap, and that the
gap does not collapse during the flux insertion process discussed below.
A ground state $\ket{\psi},$ which is $\mathrm{SU}(N)$-symmetric and has a
definite eigenvalue of all the crystalline symmetries including $\calG$
(\textit{i.e.} eigenstate of $\calG$), is chosen as the initial state.

We note that, for every nonsymmorphic space group except for $I2_1 2_1
2_1$ (No. \textbf{24}) and its key nonsymmorphic operation $\calG,$ we
can take an appropriate choice of primitive lattice vectors $\bm{a}_1,$
$\bm{a}_2,$ $\bm{a}_3$ with the following properties~\cite{WPVZ2015}:
(i) The associated translation $\bm{\alpha}$ is along the direction of
$\bm{a}_1$, and (ii) The plane spanned by $\bm{a}_2$ and $\bm{a}_3$ is
invariant under $G.$ Assuming this condition, we can show the tightest
condition derived from only one nonsymmorphic operation $\calG.$ For
simplicity, we consider the system consisting of $L_1 \times L_2 \times
L_3$ unit cells on a 3D torus (\textit{i.e.} impose the periodic boundary
conditions $\bm{r} \sim \bm{r} + L_\mu \bm{a}_\mu$
for $\mu=1,2,3$).

We take the smallest reciprocal lattice vector
$\tilde{\bm{k}}_1$ left invariant by $G,$
\textit{i.e.} $G\tilde{\bm{k}}_1=\tilde{\bm{k}}_1$ and
$\tilde{\bm{k}}_1$ generates the invariant sublattice of the
reciprocal lattice along $\tilde{\bm{k}}_1.$
We insert a flux on a torus by introducing a vector potential
$\bm{\calA}=\tilde{\bm{k}}_1/L_1.$
Since the ``magnetic flux'' reaches a multiple of $2\pi$
after this process because $\tilde{\bm{k}}_1$ is a reciprocal lattice vector,
the Hamiltonian of the system becomes equivalent to the initial one.
This happens precisely when the Hamiltonian is obtained from
the original Hamiltonian with a large gauge transformation.
The large gauge transformation to remove the inserted flux is
\begin{equation}
\calU_{\tilde{\bm{k}}_1} = \exp{\left[
\frac{i}{L_1} \sum_{\bm{r}} \tilde{\bm{k}}_1 \cdot \bm{r} I^0(\bm{r})
\right]}.
\end{equation}
Since $\bm{\calA}$ is left invariant under $\calG,$ the inserted flux
does not change the eigenvalues of $\calG.$  Thus, this phase factor due to the
large gauge transformation alone gives the
change of the eigenvalue of $\calG$ for $\ket{\psi}$ in the flux insertion process.
On the other hand,
\begin{equation}
	{\calG}^{-1} \calU_{\tilde{\bm{k}}_1} \calG
 \sim\calU_{\tilde{\bm{k}}_1}
e^{-(2\pi i \Phi_G (\tilde{\bm{k}}_1) n L_2 L_3 /N)},
\end{equation}
where $\Phi_G (\tilde{\bm{k}}_1)=\bm{\alpha} \cdot \tilde{\bm{k}}_1/(2\pi).$
For an unremovable glide or screw
symmetry, this phase factor has to be fractional.\footnote{We can show
that if $\Phi_G (\tilde{\bm{k}}_1)$ is an integer, then this nonsymmorphic
operation is removable, \textit{i.e.} can be reduced to a point-group operation
times a lattice translation by change of origin~\cite{Konig1999}.}
Thus, if we write $\Phi_G (\tilde{\bm{k}}_1)=p/\calS_G$ with $p,\calS_G$
relatively coprime, we can show a tighter bound for the filling
constraint to get a featureless Mott insulator without ground state
degeneracy because $\calS_G>1.$ In fact, to get a featureless Mott
insulator $pn L_2 L_3 /(N\calS_G)$ must at least be integer.  However,
if we choose $L_2$ and $L_3$ relatively prime to $N\calS_G,$ $n$ has to
be a multiple of $N\calS_G.$

If $n$ is not a multiple of $N\calS_G$ for some nonsymmorphic operation $\calG,$
this means the existence of degenerate ground states with a different eigenvalue
of $\calG,$ \textit{i.e.} implies the existence of gapless excitations or a gapped topological
order if the symmetry $\calG$ is not broken.  For example, in the case of the $\mathrm{SU}(4)$
Heisenberg model on the hyperhoneycomb lattice, $n=4,$ and the system can be trivial with
respect to the translation symmetry.  However, the space group of the hyperhoneycomb lattice
includes some nonsymmorphic operations, such as one glide mirror with $\calS_G=2.$
If we assume that nonsymmorphic symmetries are unbroken, the resulting QSOL (a possible
symmetric ground state) cannot be a trivial featureless Mott insulator.  Thus, we can say
this QSOL is protected by the nonsymmorphic space group symmetry of the lattice and
it can be called crystalline spin-orbital liquid (XSOL).

We note that as for the lattice (10,3)-$d,$ it is not enough to consider
only one symmetry operation and one has to consider the interplay of
multiple nonsymmorphic operations~\cite{PTAV2013}.
The derivation of
the tightest bound for all the 157 nonsymmorphic space groups with an
$\mathrm{SU}(N)$ symmetry is outside of the scope of this thesis.
As we will discuss \textit{e.g.} in Appendix~\ref{psg}, a nonsymmorphic symmetry sometimes exchanges
the bond label, and then it only exists when $J_\gamma$ obeys some
condition.  In this limited case, the generalized LSMA theorem only
applies in some parameter region defined by this condition.

\chapter{Basic theory for crystalline spin liquids in Kitaev spin liquids}\label{psg}

Since in the main text we have treated the notion of crystalline spin (or
spin-orbital) liquids in an abstract way, we would like to review how
it is materialized in real models.  We only give one typical example of crystalline
spin liquids, a Kitaev model on the $8^2.10$-$a$ lattice. In this case a nonsymmorphic
symmetry of the lattice space group protects the existence of fourfold degeneracy\footnote{In a correct sense there is twofold redundancy coming from the Majorana property.
From now on, we ignore this subtlety and regard it as fourfold degeneracy.}
and the emergence of a 3D Dirac cone in the Majorana spectrum, which is
impossible in the original classification of 3D Kitaev models based on time-reversal
and inversion symmetries~\cite{Obrien2016}. The difference between topological crystalline insulators and
crystalline spin liquids (XSLs) lies in how the space group symmetry acts on quasiparticle
excitations. Projective representations are allowed in spin liquids.
In some sense, it can be regarded as a gapless version of
symmetry-enriched topological (SET) phases.
The discussion here follows Ref.~\citenum{Yamada2017XSL}.
The notation is slightly different from the original one in the main text.
We use the standard $xyz$-notation instead of the $abc$-notation.
We note that we only solve pure Kitaev models and ignore any kinds of interactions,
although it has to be discussed if we wish to claim the phase to be stable.

\section{Kitaev's solution to the Kitaev model}

\begin{figure}
\centering
\includegraphics[width=15cm]{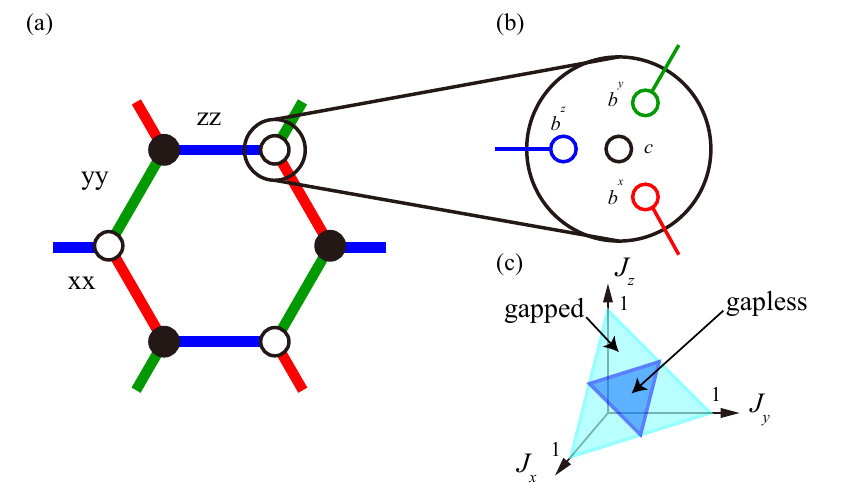}
\caption{Kitaev model on the honeycomb lattice. (a) Coloring of the honeycomb lattice
and the bond-dependent anisotropic interactions. Red, green, and blue bonds show
$x$-, $y$-, and $z$-directional anisotropy, respectively.  (b) Majorana representation of spin-1/2
degrees of freedom.  (c) Phase diagram of the Kitaev model on the honeycomb lattice.}
\label{kitaev}
\end{figure}

The construction of the Kitaev (honeycomb) model is based on exchange frustration.
It has a bond-dependent anisotropic interactions
between spin-1/2 degrees of freedom.  The Kitaev model can host both a gapless spin liquid
phase and a gapped $Z_2$ spin liquid phase,
which is related to the toric code~\cite{Kitaev2003}.
This section follows Ref.~\citenum{MT}.

The Kitaev model on the honeycomb lattice is defined as follows.
\begin{align}
        H_\textrm{Kitaev} &=K_x\sum_{\langle jk\rangle \in \textcolor{red}{x}} S_j^x S_k^x+K_y\sum_{\langle jk\rangle \in \textcolor{darkgreen}{y}} S_j^y S_k^y+K_z\sum_{\langle jk\rangle \in \textcolor{blue}{z}} S_j^z S_k^z \nonumber \\
        &=-J_x\sum_{\langle jk\rangle \in \textcolor{red}{x}} \sigma_j^x \sigma_k^x-J_y\sum_{\langle jk\rangle \in \textcolor{darkgreen}{y}} \sigma_j^y \sigma_k^y-J_z\sum_{\langle jk\rangle \in \textcolor{blue}{z}} \sigma_j^z \sigma_k^z,
\end{align}
where $\langle jk\rangle \in \alpha$ means that a nearest neighbor bond $\langle jk\rangle$
belongs to the $\alpha$-directional bond with the same color as shown
in Fig.~\ref{kitaev}(a), $K_x,\,K_y,\,K_z$ are real parameters, and $J_\alpha=|K_\alpha|/4$
for each $\alpha = x,\,y,\,z.$
This model is known to have the properties of quantum spin liquids with any nonzero parameters
$K_x,\,K_y,\,K_z,$ but we here concentrate on the ferromagnetic case where $K_x<0,\,K_y<0,$
and $K_z<0,$ for simplicity.\footnote{This is because the sign change can always be compensated
by the gauge transformation.}
In this case, the exchange frustration is clear because the red (resp. green and blue) bonds
want to align spin in the $x$- (resp. $y$- and $z$-) direction, and these conditions
cannot be met simultaneously for the classical spin.

The ground state is exactly solved by introducing a so-called Majorana representation
of the spin operators and mapping the problem to finding a correct flux sector.
For each site $j,$ we introduce four anticommuting real Majorana fermions
$b_j^x,\,b_j^y,\,b_j^z,$ and $c_j,$ as shown in Fig.~\ref{kitaev}(b).  From the anticommutation relations like
$\{b_j^\alpha,b_k^\beta\}=2\delta_{jk} \delta^{\alpha\beta},$ it is easy to show
that the spin-1/2 Pauli matrices can be represented as $\sigma_j^\alpha=ib_j^\alpha c_j.$
Even in the minimal representations for these Majorana operators, the Hilbert space
is expanded from the original spin-1/2 space (with 2 dimensions per site).
Therefore, the Hilbert space must be projected out from the Fock space $\tilde{\mathcal{L}}$
(with 4 dimensions per site because two Majorana fermions become one complex fermion)
to the physical subspace $\mathcal{L}$ to go back to the original spin representation.
The physical subspace $\mathcal{L}$ is defined by the condition $\ket{\xi} \in \mathcal{L}$
iff $D_j\ket{\xi}=\ket{\xi}$ for all $j,$ where $D_j=ib_j^x b_j^y b_j^z c_j.$
It is really physical because every algebra of Pauli matrices like
$\sigma_j^x\sigma_j^y\sigma_j^z=i$ is reproduced in this subspace.

If we define a $Z_2$ gauge flux
$\hat{u}_{jk}$ by $\hat{u}_{jk}=ib_j^{\alpha_{jk}}b_k^{\alpha_{jk}},$
where $\alpha_{jk}$ is the bond direction
between $j$ and $k,$ then the Hamiltonian in $\tilde{\mathcal{L}}$ becomes
$\tilde{H} = \frac{i}{2}\sum_{\langle jk \rangle} J_{\alpha_{jk}} \hat{u}_{jk} c_j c_k,$
where each bond $\langle jk \rangle$ is counted twice with $\hat{u}_{kj} = -\hat{u}_{jk}.$
This is nothing but a $Z_2$ lattice gauge theory for Majorana fermions $c_j$ with an external
magnetic field with a $Z_2$ gauge field defined by $\hat{u}_{jk}.$
Because $\hat{u}_{jk}$s all commute with $\tilde{H},$ after defining the eigenstates
of $\hat{u}_{jk}$s and replacing them by $c$-variables $u_{jk},$ we can diagonalize
the quadratic Hamiltonian $\frac{i}{2}\sum_{\langle jk \rangle} J_{\alpha_{jk}} u_{jk} c_j c_k$
for itinerant $c_j$ fermions to get the ground state for each flux sector by
applying projection operators $(1+D_j)/2$ to $\mathcal{L}.$
Given a magnetic flux $w_p$ for each hexagonal plaquette $p,$
this uniquely determines the ground state spectrum by diagonalizing a one-particle
Hamiltonian for $c_j$ fermions.  The ground state of this free model
$\ket{\tilde{\Psi}} \in \tilde{\mathcal{L}}$ can be projected onto the physical subspace
$\mathcal{L}$ by
\begin{equation}
    \ket{\Psi} = \prod_j \frac{1+D_j}{2} \ket{\tilde{\Psi}} \in \mathcal{L}.
\end{equation}

Therefore, the only thing left is to determine the correct flux sector including
the exact ground state of the original Hamiltonian.  From Lieb's beautiful theorem on the
flux sector with the lowest energy~\cite{Lieb1994},
we can rigorously conclude that the answer is the flux sector with zero magnetic flux.
Therefore, we can replace $u_{jk}$ by 1 and the ground state spectrum completely
becomes a Majorana version of the nearest-neighbor honeycomb tight-binding model.
In the case $J_x=J_y=J_z,$ this is the same model as that for graphene and
it is a well-known fact that there are Dirac cones at K and K$^\prime$ points
in the Brillouin zone [see also Fig.~\ref{lieb}(b)],
and we can conclude that the ground state is a gapless spin liquid.
From triangular inequalities, we can determine the
region where the Majorana spectrum is gapless, \textit{i.e.} the one-particle Hamiltonian
has a zero eigenstate as $J_\alpha \leq J_\beta+J_\gamma,$
where $\alpha,\,\beta,\,\gamma$ is a permutation of $x,\,y,\,z.$
The phase diagram (gapless or gapped) on the plane $J_x+J_y+J_z=1$
is shown in Fig.~\ref{kitaev}(c).
Dirac cones in the gapless region are protected by the time-reversal
symmetry and the topological nature of the vector bundle of the wavefunction,
which will be discussed in the next section again.

\section{Lieb's theorem and ground state flux sectors}

\begin{figure}
\centering
\includegraphics[width=12cm]{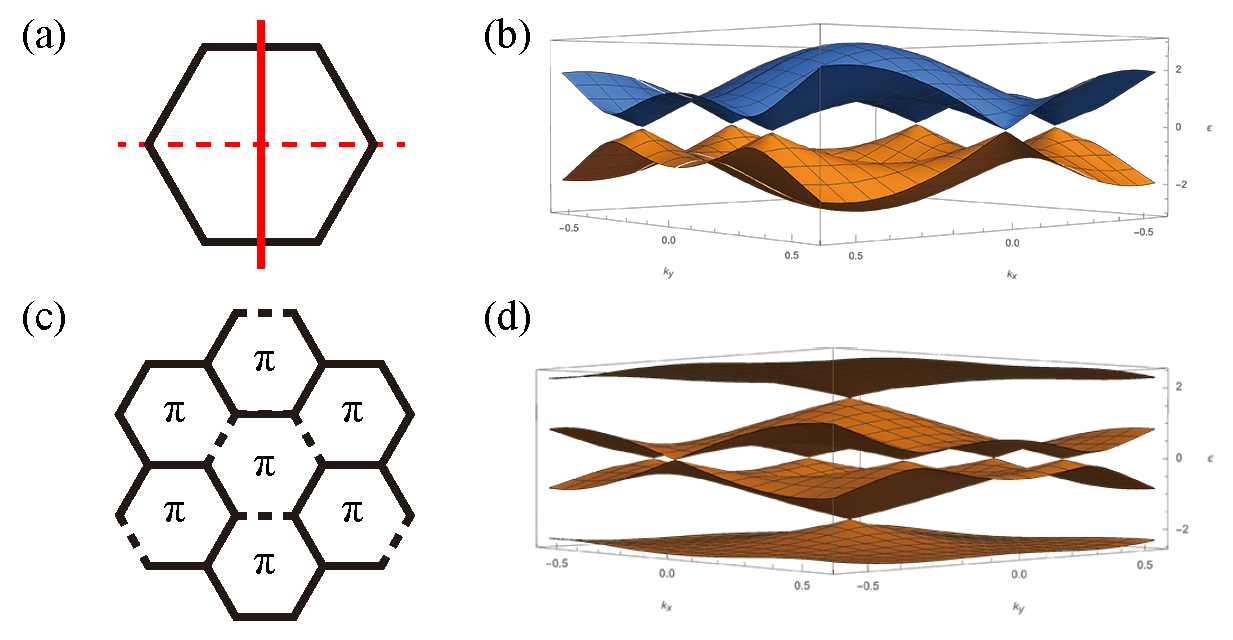}
\caption{(a) Mirror planes of the honeycomb lattice.  A solid red line shows a
mirror plane used in Lieb's theorem, and a dashed red line shows an irrelevant one.
(b) Band structure for the 0-flux state.  At half filling, there are six Dirac cones
shown in the figure.  Only two of them are independent, located at K and K$^\prime$ points.
(c) ``Benzene'' gauge for the $\pi$-flux state.  $-1$ bonds are shown by dashed lines.
(d) Band structure for the $\pi$-flux state.  At quarter filling it looks like
there is only one Dirac cone at $\Gamma$ point, but it is actually doubly degenerate
because the enlarged unit cell of the benzene gauge has redundancy, being twice larger than
that of the minimal one.  At half filling, the spectrum is similar to (b).}
\label{lieb}
\end{figure}

In the case of the honeycomb lattice, Lieb's theorem~\cite{Lieb1994}
is applicable to the flux problem when $J_x=J_y,$ but it requires a specific
reflection (mirror) symmetry on the lattice and
it will not apply to most of the generalized 3D Kitaev models.
Let us quickly review the claim of Lieb's theorem without giving a proof.
Assuming the existence of a translation symmetry and a reflection symmetry whose mirror plane
cuts bonds, not sites [see Fig.~\ref{lieb}(a) for comparison], we can prove the following
theorem.

\begin{thm}
For any periodic bipartite lattices, the flux problem for the half-filled
tight-binding (Hubbard) model is solved for each plaquette of length $l$ containing
a cutting mirror plane as follows.
\begin{enumerate}
\item A plaquette $C$ will carry zero flux, \textit{i.e.} $\prod_{\langle jk \rangle \in C} u_{jk} = 1,$ when $l \equiv 2 \mod 4.$
\item A plaquette $C$ will carry $\pi$ flux, \textit{i.e.} $\prod_{\langle jk \rangle \in C} u_{jk} = -1,$ when $l \equiv 0 \mod 4.$
\end{enumerate}
\end{thm}

We note that this theorem is generic not only for the free model, but also
for interacting models.  In Ref.~\citenum{Lieb1994} many types of reflection-positive
interactions are considered and Lieb's theorem applies to many interacting fermion
models.  Since this theorem is very generic, only involving a periodic array of mirror
planes, it is applicable to any dimensions and, if we could solve a flux problem for
every elementary loop, we can decide the correct ground state for the underlying
Kitaev model.  Otherwise, a numerical simulation is always necessary to determine
the ground state flux sector.

If this theorem is applied to the square lattice, as originally proposed by Lieb~\cite{Lieb1994},
a $\pi$ flux for each square plaquette should be optimal in accordance with
Affleck-Marston's theory~\cite{Affleck1988am}.  In the case of the honeycomb lattice,
it becomes 0-flux instead.\footnote{We note that in the honeycomb lattice
only the mirror planes cutting bonds work and such planes exist only when
$J_x=J_y,$ $J_y=J_z,$ or $J_z=J_x.$}
In this 0-flux case, the ground state Majorana spectrum of the Kitaev model is the same
as that of graphene.  We quickly review a one-body band structure of graphene to
solve the nearest-neighbor tight-binding model for the Kitaev honeycomb model.

For simplicity, we focus on the gapless phase of the 0-flux Kitaev model, and assume
$J_x = J_y = J_z \equiv J.$  Then, the ground state can be constructed from the
half-filled Fermi sea of the tight-binding model.  As already explained, it has
two Dirac cones at K and K$^\prime$ points.  The spectrum is conical and is effectively
described by a relativistic theory.  Because of the Nielsen-Ninomiya-type
theorem~\cite{Hatsugai2011}, the spectrum cannot be gapped unless two Dirac cones
collide by a nonperturbative threefold rotation symmetry breaking term.
The overview of the spectrum in the hexagonal Brillouin zone is shown in
Fig.~\ref{lieb}(b).

Just for a comparison, we also review the spectrum of the $\pi$-flux honeycomb lattice.
The most symmetric view for the band structure of the $\pi$-flux model can be
achieved by taking a ``benzene'' gauge, where only double bonds of benzene are set
to have $-1,$ as shown in Fig.~\ref{lieb}(c).  We again assume $J_x = J_y = J_z \equiv J.$
The spectrum has Dirac cones not only at half filling but also at 1/4 and 3/4
fillings [see Fig.~\ref{lieb}(d)].\footnote{This is why this $\pi$-flux state is
important not in half-filled Majorana models, but in $\mathrm{SU}(4)$ models discussed
in the main text.}  According to Kitaev~\cite{Kitaev2006},
the energy difference between the 0-flux and $\pi$-flux sectors is $0.067J$ per hexagon
at half filling.  We note that an isolated vison (flux) excitation from the
0-flux ground state has an energy $\Delta E \sim 0.27J.$

\section{Classification of Kitaev models by internal symmetries}

\begin{figure}
\centering
\includegraphics[width=10cm]{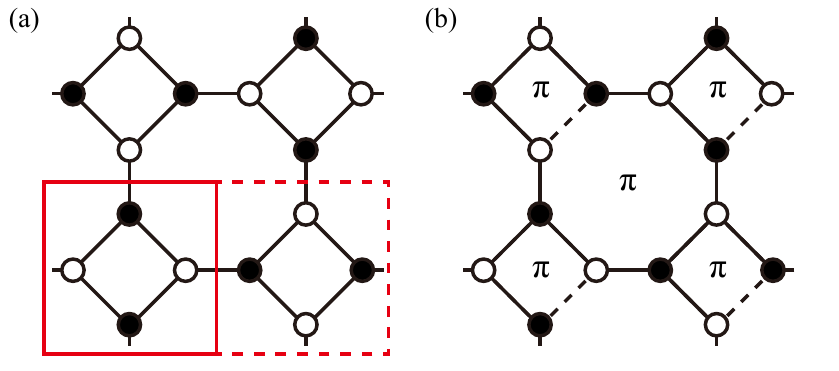}
\caption{Kitaev model on the squareoctagon lattice. (a) The squareoctagon
lattice and its sublattice labels (white and black circles).
The unit cell is shown by a red solid line and the neighboring one is shown
by a red dashed line. (b) One translation-symmetric gauge for the $\pi$-flux
state.  $-1$ bonds are shown by dashed lines.}
\label{squareoctagon}
\end{figure}

The general discussion for the classification of 3D Kitaev models is complicated,
so we just give known results on the classification, and present one 2D example, which
is more intuitive for most readers than 3D systems, in order to
show how internal symmetries are implemented projectively in Kitaev models.
This section follows Ref.~\citenum{Obrien2016} and thus include the inversion symmetry
in the set of ``internal'' symmetries for simplicity.\footnote{In a correct sense,
the inversion symmetry is also a lattice (spatial) symmetry.}

The classification of the Kitaev model is not as straightforward as the free-fermion
topological periodic table~\cite{Kitaev2009}.  In order to see this we first take up
the Kitaev model on the 2D squareoctagon lattice~\cite{Yang2007,Baskaran2009,Kells2011}.
The squareoctagon lattice is shown in Fig.~\ref{squareoctagon}(a).
This lattice is related to the 3D hyperoctagon lattice, but has a better property
because Lieb's theorem is applicable.\footnote{The 3D hyperhexagon lattice has
a nice property, too, but not suitable for our purpose here.} 
They share the same property that the projective implementation of the time-reversal
symmetry plays an important role.

In the same way as Affleck-Marston's ansatz~\cite{Affleck1988am} discussed in the main text,
the ground state of the squareoctagon lattice is described by a $\pi$-flux state.  Both
square and octagon plaquettes are likely to bind a $\pi$ flux in accordance with Lieb's theorem~\cite{Lieb1994}.
The translation is implemented trivially in contrast to Affleck-Marston's ansatz~\cite{Affleck1988am},
but other lattice symmetries are implemented projectively in a way similar to
Affleck-Marston's [see Fig.~\ref{squareoctagon}(b)].
Due to the property of the Kiteav model (or the Majorana representation),
there is an additional feature nonexistent in other complex fermion models.
Especially, the implementation of the time-reversal symmetry is the most exotic one
specific to the Kitaev model, so we only review its projective property in this appendix.

The time-reversal symmetry is different not only because it is antiunitary, but also
because it involves the sublattice symmetry of the lattice.  As already discussed by
Kitaev~\cite{Kitaev2006}, if the lattice is not bipartite (\textit{i.e.} has a loop of
an odd length), the time-reversal symmetry is spontaneously broken leading to degenerate
chiral spin liquid (CSL) states at the ground state, as is the case with the Kitaev models \textit{e.g.} on
the (9,3) lattices. Fortunately, the squareoctagon lattice is bipartite,
so the time-reversal symmetry is preserved.
However, as shown in Fig.~\ref{squareoctagon}(a), the sublattice symmetry is not compatible
with translation, so the time-reversal symmetry, too, becomes projective.

In the squareoctagon lattice, the translation along the $x$- or $y$-axis changes
the sublattice parity, so the projective symmetry group (PSG) of the time-reversal operation $T$ always involves a
gauge transformation extended across the unit cell to repair the sublattice.
Na\"ively, the action of $T$ is
$Tc_j T^{-1} = c_j$ and $Tb_j^\alpha T^{-1} = b_j^\alpha.$  Thus,
$T\hat{u}_{jk} T^{-1} = -\hat{u}_{jk}.$  However, since Kitaev's Majorana Hamiltonian
itself is defined from the sublattice index for bipartite lattices, the gauge
transformation is determined very easily.  If we define a sublattice parity
as $(-1)^j$ for the $j$th site, then the time-reversal operation supplemented by
a gauge transformation $\tilde{T}$ can only act like
$\tilde{T}c_j \tilde{T}^{-1} = (-1)^j c_j$ and $\tilde{T}b_j^\alpha \tilde{T}^{-1} = (-1)^j b_j^\alpha.$
Thus, because this gauge transformation is possible only in the extended
unit cell (both solid and dashed red unit cells are necessary in Fig.~\ref{squareoctagon}(a))
the action of $\tilde{T}$ requires a supplemental translation in the $k$-space.
In particular, in the case of the square/squareoctagon lattice
$\bm{k}_0 = (\pi,\,\pi)^t$ is necessary to maintain the gauge
transformation~\cite{Obrien2016}.  Since $T$ (or $\tilde{T}$) is antiunitary, the final form
of the implementation of the time-reversal symmetry becomes
\begin{align}
    \hat{h}(\bm{k}) &= U_T \hat{h}^*(-\bm{k}+\bm{k}_0)U_T^{-1}, \\
    \varepsilon(\bm{k}) &= \varepsilon(-\bm{k}+\bm{k}_0),
\end{align}
where $\hat{h}(\bm{k})$ is a (one-body) Bloch Hamiltonian for itinerant Majorana
fermions, and $\varepsilon(\bm{k})$ is its spectrum.  $U_T$ is a unitary matrix
defined from the action of $\tilde{T}.$

As for 2D lattices, the classification is almost completed:
when $\bm{k}_0 = 0,$ a stable object is a Dirac cone at the Fermi level if gapless,
and when $\bm{k}_0 \neq 0,$ a stable object is a Fermi surface if gapless, where
each Fermi surface may be related by the reciprocal lattice vector $\bm{k}_0.$
In the 3D case, it is more complicated because the (projective) inversion symmetry
may require a different $k$-space translation $\bm{k}_0^\prime$ depending on its
flux sector, even if gapless.  If $\bm{k}_0 = 0,$ a stable gapless object is a nodal line,
which is a natural generalization of a Dirac cone.  If $\bm{k}_0 \neq 0,$ the inversion
symmetry becomes important in the 3D case.  If $\bm{k}_0 \neq 0$ and $\bm{k}_0^\prime = 0,$
the inversion symmetry is implemented trivially and a stable object becomes a Weyl node.
Otherwise, $\hat{h}(\bm{k})$ has no trivially-implemented internal symmetry, and the
only remaining gapless object is a Fermi surface.\footnote{This type of Fermi surfaces is
usually unstable with interaction~\cite{Hermanns2015BCS}.}
The classification becomes rich if we break one of these symmetries
explicitly/spontaneously~\cite{Hermanns2015Weyl}, but this already completed
the classification based on the internal
symmetries~\cite{Obrien2016}.\footnote{It seems that a possibility that
the time-reversal and inversion symmetries share the same $\bm{k_0}$ is ignored,
but this never happens in the examples discussed in Ref.~\citenum{Obrien2016}.}
One interesting thing in this classification is that a Weyl semimetal of Majorana
fermions is possible even with time-reversal and inversion symmetries due to
their projective nature.

While from the symmetry analysis the spectrum of the squareoctagon lattice
has a Fermi surface if gapless, the $\pi$-flux state is actually gapped.
However, the 0-flux sector is known to have Fermi surfaces (or more accurately Fermi circles),
and would be stabilized by an additional flux stabilization term~\cite{Baskaran2009}.
In this 0-flux sector, the phase diagram is similar to the honeycomb case
[see Fig.~\ref{kitaev}(c)], and in the gapless regime Fermi surfaces are always stable.
Anyway, the Kitaev model on the squareoctagon lattice shows a rich variety of phases
depending on its flux sector, while it is not a main topic of this appendix.
Even though in both 0-flux and $\pi$-flux sectors translation is implemented trivially,
the time-reversal symmetry still plays an important role in the Kitaev model.
A take-home message here is that the projective implementation (PSG) of the time-reversal/inversion
symmetry topologically determines the spectrum in gapless phases in most 2D cases,
and the same is true for the 3D hyperoctagon lattice, hyperhexagon lattice, etc.

\section{Lieb's flux sector and Majorana spectrum}

\begin{figure}
\centering
\includegraphics[width=\textwidth]{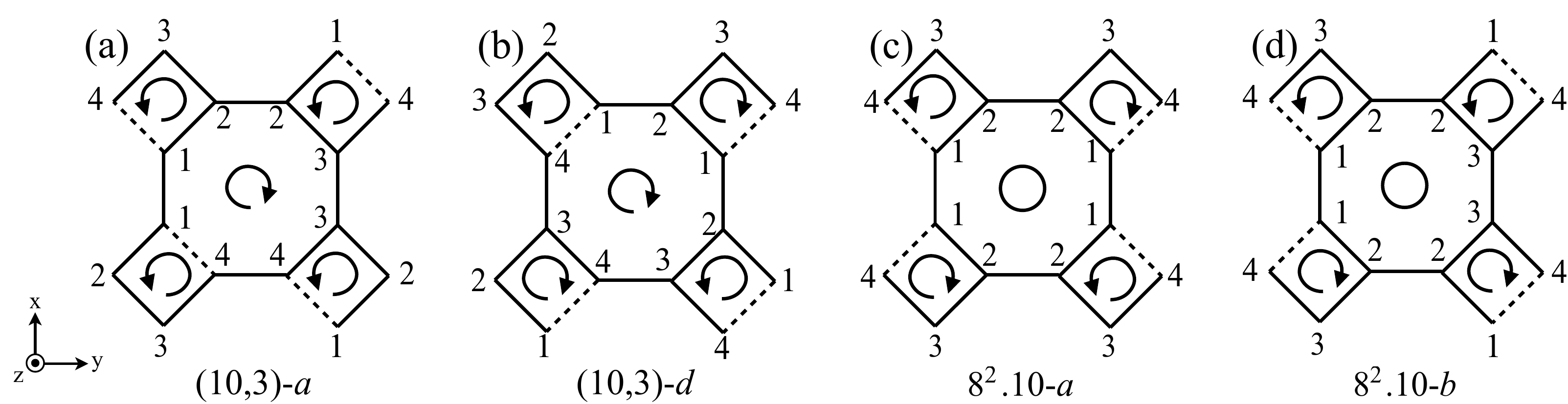}
\caption{3D lattices constructed from the squareoctagon lattice.  Every lattice is
constructed by extending the squareoctagon lattice along the $z$-direction.
(a) (10,3)-$a.$ (b) (10,3)-$d.$ (c) $8^2.10$-$a.$ (d) $8^2.10$-$b.$
Reprinted figure with permission from [\href{https://dx.doi.org/10.1103/PhysRevB.96.155107}{M.~G.~Yamada, V.~Dwivedi, and M.~Hermanns, Phys. Rev. B, \textbf{96}, 155107 (2017).}] Copyright 2017 by the American Physical Society.}
\label{comparison}
\end{figure}

The general classification discussed above for the 3D Kiteav model does not apply
to the $8^2.10$-$a$ lattice, and thus it is meaningful to solve this model explicitly.
Fortunately, Lieb's theorem is applicable to $8^2.10$-$a$ when $J_x=J_y.$
Thus, we mostly concentrate on the case $J_x=J_y,$ but we assume the same
flux configuration even for other parameters.  From the statement of Lieb's theorem
8-loops have to have a $\pi$ flux and 10-loops have to have a zero flux in the
ground state.  This completely determines the flux configuration of
$8^2.10$-$a.$\footnote{The naming of $8^2.10$-$a$ itself comes from the fact that it
is constructed from elementary 8-loops and 10-loops~\cite{Wells1977}.}

In $8^2.10$-$a,$ the time-reversal symmetry is implemented trivially, and
thus the stable object should be a nodal line.  However, it is not the case
for $J_x=J_y.$  When $J_x=J_y,$ the nodal lines get degenerate, reduced to
two gapless points with fourfold degenerate 3D Dirac cones.  This is beyond
the classification based on the internal symmetries.

\newcommand{\vR}{\bm{R}}
\newcommand{\va}{\bm{a}}
\newcommand{\vq}{\bm{q}}

\begin{figure}
\centering
\includegraphics[width=\textwidth]{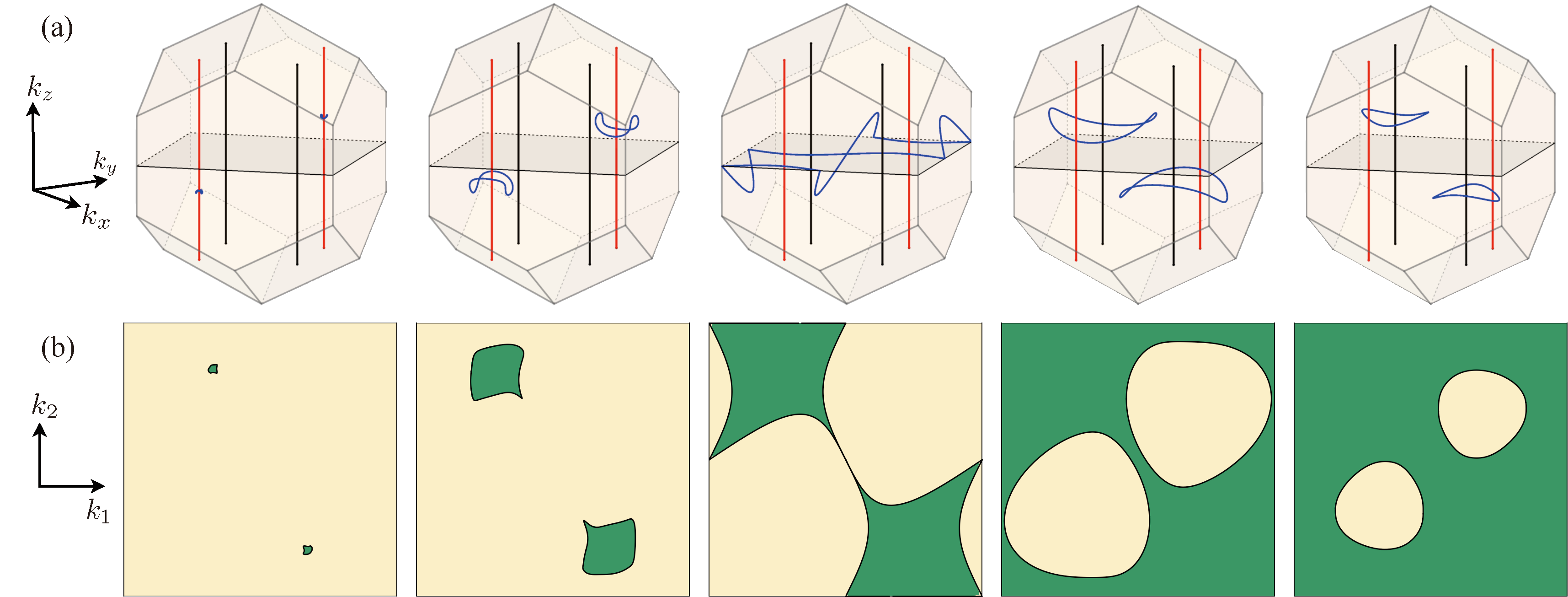}
\caption{(a) Zero-energy locus (blue) in the Brillouin zone (yellow) for the $8^2.10$-$a$ lattice
for parameters $J_x = \frac{1}{3} + \Delta J, J_y = J_z = \frac{1}{3} - \frac{1}{2} \Delta J,$
with $ \Delta J = 0.01, 0.05, 0.0808, 0.10, 0.12$ (from left to right).
Red solid lines are invariant under only the fourfold screw symmetry,
while black solid lines are invariant under the twofold and fourfold screw symmetries.
(b) Chiral invariant $\theta$ computed for the loops along the $k_3$ axis
as a function of $(k_1, k_2)$ for a corresponding parameter, where the values 0,
and $-1$ are represented by yellow and green, respectively. The black solid line
depicts the projection of the bulk nodal line along $k_3.$
Reprinted figure with permission from \cite{Yamada2017XSL} Copyright 2017 by the American Physical Society.}
\label{locus}
\end{figure}

The lattice structure of $8^2.10$-$a$ is schematically shown in Fig.~\ref{comparison}(c),
but the number labelled in the figure is just to show the height along the $z$-axis
in 2D.  The real site numbering used in the following discussion is included
in Ref.~\citenum{Yamada2017XSL}, instead.  The coloring of bonds and the flux sector are
accidentally the same as those discussed for the $\mathrm{SU}(4)$ model
[see Subsec.~\ref{810acolor}].\footnote{As for the $\mathrm{SU}(4)$ model, the flux sector
here means the embedded flux configuration.}
Thus, we omit a detailed description of the lattice structure and directly move on to
the construction of a Hamiltonian.  Using the crystallographic axes, $\va_1,$ $\va_2,$ and
$\va_3$ are defined as lattice vectors, and $\vR$ spans every lattice point, \textit{i.e.}
$\vR \in \mathbb{Z}\va_1 + \mathbb{Z}\va_2 + \mathbb{Z}\va_3.$

\begin{align}
    H_\textrm{XSL} = - & \sum_{\vR} \left\{ J_x \left[ \sigma_1^x(\vR) \sigma_2^x(\vR) + \sigma_3^x(\vR) \sigma_4^x(\vR) \right. \right. \nonumber \\ 
  & \left. \;\; \left. + \sigma_5^x(\vR) \sigma_8^x(\vR+\va_3)  + \sigma_6^x(\vR) \sigma_7^x(\vR) \right] \right. \nonumber \\ 
  & \left. + J_y \left[ \sigma_1^y(\vR) \sigma_4^y(\vR-\va_3) + \sigma_2^y(\vR) \sigma_3^y(\vR) \right. \right. \phantom{\sum} \nonumber \\ 
  & \left. \;\; \left. + \sigma_5^y(\vR) \sigma_6^y(\vR) + \sigma_7^y(\vR) \sigma_8^y(\vR) \right] \right. \nonumber \\ 
  & \left. + J_z \left[ \sigma_1^z(\vR) \sigma_6^z(\vR+\va_2-\va_3) \right. \right. \phantom{\sum} \nonumber \\ 
  & \left. \;\; \left. + \sigma_2^z(\vR) \sigma_7^z(\vR-\va_1+\va_2)  \right. \right. \nonumber \\ 
  & \left. \;\; \left. + \sigma_3^z(\vR) \sigma_8^z(\vR-\va_1+\va_3) + \sigma_4^z(\vR) \sigma_5^z(\vR) \right] \right\}
\end{align}
This can be solved simply by introducing Majorana fermions.  The desired flux configuration
can be realized by setting the bond operators $u_{jk}=+1$ (resp. $-1$) if $j$ is
odd (resp. even), except for the 1-6 bond where $u_{16}=-1.$
The resulting one-body Majorana Hamiltonian $\hat{h}(\bm{k})$ after a Fourier transformation
looks like
\begin{equation}
\hat{h}(\bm{k}) = 
  \begin{pmatrix}
   0 & A(\bm{k}) \\ 
   A^\dagger(\bm{k}) & 0 
  \end{pmatrix}.
\end{equation}
As before the spectrum of $\hat{h}(\bm{k})$ is $\varepsilon(\bm{k}).$
$A(\bm{k})$ is defined by
\begin{equation}
  A(\bm{k}) = i \begin{pmatrix}
   0 & 				-J_z e^{2\pi ik_{23}} & 	J_x & 			J_y e^{-2\pi ik_3} 	\\
   J_z e^{2\pi i k_{31}} & 	0 & 			J_y & 			J_x 		\\
   J_x e^{2\pi ik_3} & 		J_y & 			0 & 			J_z 	  	\\ 
   J_y & 			J_x & 			J_z e^{2\pi ik_{12}} & 	0 			 
  \end{pmatrix},
\end{equation}
where $\bm{k}$ is a reciprocal vector normalized by $k_i \in [0,1),$ and
$k_{mn}=k_m-k_n.$  We note that the matrix index is reordered to make it symmetric.
Just by diagonalizing this Hamiltonian, we can see that in a gapless region with $J_x=J_y$
the spectrum of Majorana fermions is described by 3D Dirac cones with fourfold degeneracy
at some points in the Brillouin zone.  These Dirac points actually lie on the invariant line
of the fourfold screw symmetry.  Fig.~\ref{locus} clearly shows how the zero-energy object
evolves from Dirac cones to nodal lines by changing the parameters from $J_x=J_y$
to $J_x \neq J_y.$  Detailed description is included in Ref.~\citenum{Yamada2017XSL}.

A chiral invariant $\theta$ in Fig.~\ref{locus}(b) is defined by $A(\bm{k})$ as
\begin{equation}
    \theta = \frac{1}{4\pi i} \oint_{\mathcal{C}} \textrm{tr}\left\{ A^{-1} dA - \left( A^\dagger \right)^{-1} d A^\dagger \right\},
\end{equation}
for a contour $\mathcal{C}$ in the $k$-space.

\section{Physics of crystalline spin liquids}

We claim it to be a crystalline phase because the fourfold degeneracy can
only be protected under the space group symmetry in Majorana systems,
while 2D Dirac cones are protected in the Kitaev model on the honeycomb lattice
just by the time-reversal symmetry, and breaking the lattice symmetry just moves
the location of the Dirac cones in the Brillouin zone.  In order to see
this we will check PSG of the screw symmetry of this lattice.

As easily seen from the spiral lattice structure [see Fig.~\ref{comparison}],
there is a fourfold screw rotation symmetry, and we name it $S_4$ and its subgroup twofold
screw rotation $S_2.$  Of course, four $S_4$ and two $S_2$ operations
are reduced to translation along the $z$-direction.\footnote{This Cartesian $xyz$-axis
is different from the one used to derive the Jackeli-Khaliullin mechanism.}
The importance of this symmetry is clearly shown in Fig.~\ref{locus}(a) because
the screw-invariant lines are always surrounded by the zero-energy objects and
at the center of the spectrum.  We note that all invariant lines do not pass through
the origin because the symmetry is implemented projectively and the screw rotation
involves the reciprocal lattice translation in the $k$-space.  In Fig.~\ref{locus}(a),
those lines look like separated, but the ones in the same color are actually a single
line connected on the boundary.

\begin{figure}
\centering
\includegraphics[width=8cm]{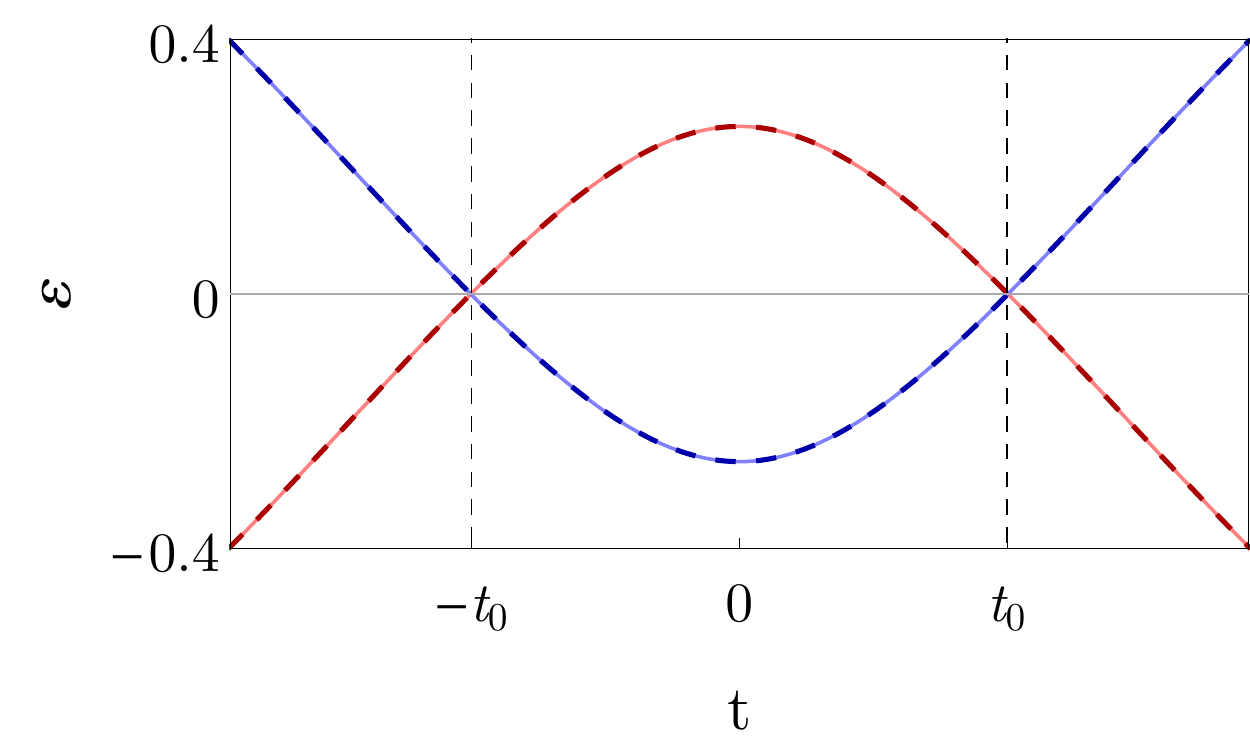}
\caption{Spectrum of the four bulk bands that are closest to $\varepsilon(t)=0$ and
    that form the Dirac nodes, plotted along the screw symmetric line Eq.~\eqref{scr4}.
The parameters are given by $J_x=J_y=0.37,$ $J_z=0.26,$
with the Dirac nodes corresponding to $t_0 \sim \pm 0.21,$ solutions of $\varepsilon(t)=0.$
The bands are labeled by their screw eigenvalues
$\rho_n e^{- 3 \, i\pi t}$ with $\rho_n = e^{ i (2n+1)\pi / 4}.$
The bands with eigenvalues $\rho_0, \rho_1$ are denoted by red solid and dashed lines,
and those with eigenvalues $\rho_2, \rho_3$ by blue solid and dashed lines, respectively. 
Reprinted figure with permission from \cite{Yamada2017XSL} Copyright 2017 by the American Physical Society.}
\label{dirac}
\end{figure}

When $J_x = J_y,$ the action of the fourfold screw symmetry is as follows.
\begin{equation}
    \hat{h}(S_4\bm{k}) = \mathcal{U}_{S4}(\bm{k}) \hat{h}(\bm{k}) \mathcal{U}_{S4}^\dagger(\bm{k}),
\end{equation}
where
\begin{equation}
S_4\left( k_1,k_2,k_3 \right) = \left( k_3-k_2, k_1+\frac{1}{2}, k_3 \right),
\end{equation}
and the unitary matrix $\mathcal{U}_{S4}(\bm{k})$ depends on the rotation axis.
Though we will not show an explicit form of the matrix representation,
an important fact is that there is an invariant line where a Majorana Hamiltonian
has some commuting unitary matrix.  The invariant line can be parametrized as 
\begin{equation}
    \bm{\gamma}_t = t \, \vq_1 + \left( t + \frac{1}{2} \right) \vq_2 + \left( 2t + \frac{1}{2} \right) \vq_3,\label{scr4}
\end{equation}
where $\vq_i$ are reciprocal lattice vectors obeying $\vq_i \cdot \va_j = 2\pi \delta_{ij},$
and is periodic under $t \mapsto t + 1.$  This line is shown in red in Fig.~\ref{locus}(a).
Explicitly, we note that for any momentum $\bm{\gamma}_t,$
\begin{equation}
    [\mathcal{U}_{S4}(\bm{\gamma}_t), \hat{h}(\bm{\gamma}_t)] = 0,
\end{equation}
and $\left[ \mathcal{U}_{S4}(\bm{\gamma}_t) \right]^4 = -e^{- 12 \, i \pi t} I_8.$
Thus, the eigenstates $\ket{\varphi_t}$ of $\hat{h}(\bm{\gamma}_t)$ satisfy
\begin{equation}
  \mathcal{U}_{S4}(\bm{\gamma}_t) \ket{\varphi_t} = \rho_n e^{- 3 \, i\pi t} \ket{\varphi_t},
\end{equation}
with $\rho_n = \exp \left(i \frac{2n+1}{4} \pi \right); \, n = 0,1,2,$ or $3,$ and to each bulk band
we associate a screw eigenvalue corresponding to $\rho_n.$

Finally, we can plot the value of $\rho_n$ for each band on the line defined by
$\bm{\gamma}_t.$  This is shown around $\varepsilon(t)=0$ in Fig.~\ref{dirac},\footnote{$\varepsilon(t)$ is defined as eigenvalues of $\hat{h}(\bm{\gamma}_t).$}
and all the four bands consisting of Dirac cones have a different quantum number.
Thus, we have proven that the fourfold degeneracy is indeed protected by the
screw symmetry in a projective form, and breaking the fourfold screw or time-reversal symmetry
will result in a gap opening of the Dirac cones.  All of these phenomena are
beyond the previous study, and in this sense we can regard it as a new crystalline phase.
We note that the double degeneracy of bulk bands shown in Fig.~\ref{dirac} is protected by
anticommutivity of screw and glide symmetries~\cite{Yamada2017XSL}.

In summary, the role of symmetry in spin liquids is still not fully understood,
and there should be a rich variety of new exotic phases.  Especially, the classification
of crystalline phases is completely beyond the classification in free-fermion systems,
and almost nothing is known for this huge iceberg, except for a small number of
exactly solvable models.  This is because we still do not have
a systematic theory which can treat the projective implementation of the lattice symmetry
beyond the mean-field approximation.  Interacting systems, especially in the gapless case,
are difficult in many senses.  For example, it is difficult to include a gauge fluctuation
correctly, and we have ignored it even in this thesis in most spin liquids.
We believe that this analysis is complementary to Appendix~F of Ref.~\citenum{Kitaev2006}
because we discussed the importance of symmetry fractionalization, which was mostly
discussed in gapped systems so far, in gapless Kitaev spin liquids.

\bibliographystyle{oreno2} 
\bibliography{paper}

\chapter*{Acknowledgement \markboth{Acknowledgement}{}}
We thank Arash~Banisafar, Kelsey~Collins,
Kedar~Damle, Eugene~Demler, Vatsal~Dwivedi, Shu~Ebihara, Tim~Eschmann, Danna~E.~Freedman,
Liang~Fu, Yohei~Fuji, Hiroyuki~Fujita, Bertrand~I.~Halperin,
Maria~Hermanns, Hosho~Katsura, Giniyat~Khaliullin, Daniel~I.~Khomskii, 
Ryohei~Kobayashi, Mikl\'os~Lajk\'o, Linhao~Li, Fr\'ed\'eric~Mila, Yuki~Nagai,
Yuji~Nakagawa, 
Judit~Romh\'anyi, Ryoya~Sano, Constantin~Schrade,
Kirill~Shtengel, Andrew~Smerald, Tomohiro~Soejima, Yasuhiro~Tada, Hidenori~Takagi,
Tomohiro~Takayama, T.~Senthil, Ryo~Takahashi, Simon~Trebst, Shinji~Tsuneyuki, and
especially Itamar~Kimchi for helpful comments.
M.G.Y. thanks George~Jackeli, Masaki~Oshikawa, and personally Atsuko~Tsuji for effortless
help to write a thesis.
The crystal data included in this work have been taken from Materials Project~\cite{osti_1275075}.
M.G.Y. is supported by the Materials Education program for the future leaders in Research, Industry, and Technology (MERIT), and by JSPS.
This work was supported by JSPS KAKENHI Grant Numbers JP15H02113, JP17J05736, and JP18H03686, and by JSPS Strategic International Networks Program No. R2604 ``TopoNet''.
M.G.Y. acknowledges the support of the Max-Planck-UBC-UTokyo Centre for Quantum Materials.
M.G.Y. also acknowledges the Quantum Materials Department at MPI-FKF, Stuttgart for 
kind hospitality during his visits, and the Department of Physics, MIT.
This research was supported in part by the National Science Foundation under Grant No. NSF PHY-1748958.
\end{document}